\documentclass[reprint,aps,prb,longbibliography,amssymb,amsmath,amsfonts,superscriptaddress,nofootinbib,tikz]{revtex4-2}
\usepackage{graphicx,indentfirst,dsfont,bbold,tikz,mathtools,color,xcolor,makecell,tabularx,multirow}
\usepackage[caption=false]{subfig}
\usepackage{yfonts}
\usepackage[colorlinks,
linkcolor=blue,
anchorcolor=red,
citecolor=blue,
urlcolor=blue]{hyperref}
\usetikzlibrary{patterns}
\definecolor{purplish}{rgb}{0.8, 0, 0.8}

\begin{document}

\title{Boundary anomaly detection in two-dimensional subsystem symmetry-protected topological phases}
\author{Ke Ding}
\affiliation{State Key Laboratory of Low-Dimensional Quantum Physics, Department of Physics, Tsinghua University, Beijing 100084, China}
\author{Hao-Ran Zhang}
\affiliation{State Key Laboratory of Low-Dimensional Quantum Physics, Department of Physics, Tsinghua University, Beijing 100084, China}
\author{Bai-Ting Liu}
\affiliation{State Key Laboratory of Low-Dimensional Quantum Physics, Department of Physics, Tsinghua University, Beijing 100084, China}
\author{Shuo Yang}
\email{shuoyang@tsinghua.edu.cn}
\affiliation{State Key Laboratory of Low-Dimensional Quantum Physics, Department of Physics, Tsinghua University, Beijing 100084, China}
\affiliation{Frontier Science Center for Quantum Information, Beijing, China}
\affiliation{Hefei National Laboratory, Hefei 230088, China}

\begin{abstract}
We generalize the topological response theory to detect the boundary anomalies of linear subsystem symmetries.
This approach allows us to distinguish different subsystem symmetry-protected topological (SSPT) phases and uncover new ones. We focus on the cases where the mixed anomaly exists within the adjacent subsystems.
Using numerical simulations, we demonstrate the power of this method by identifying strong and weak $Z_2^\tau\times Z_2^\sigma$ SSPT phases in a tunable tensor network state. 
Our analysis reveals an intrinsic $Z_2$ SSPT phase characterized by its degenerate entanglement spectrum.
Furthermore, we extend the anomaly indicator to mixed-state density matrices and show that quantum anomalies of subsystem symmetry can persist under both uniform and alternating disorders.
This finding establishes a connection between boundary quantum anomalies in pure and mixed states.
Our work provides a numerical method to detect quantum anomalies of subsystem symmetries, offering new insights into the study of topological quantum phases.
\end{abstract}

\maketitle
\tableofcontents

\section{Introduction}
Global symmetry plays a crucial role in the classification of topological quantum matter, enabling even short-range entangled (SRE) wave functions to exhibit nontrivial topological features \cite{RevModPhys.89.041004,PhysRevB.80.155131,PhysRevB.87.155114,doi:10.1126/science.1227224,Chen2014,PhysRevB.83.035107,PhysRevB.84.165139}. 
A prominent example is the spin-$1$ Haldane chain, where global time-reversal symmetry protects its boundary degeneracy \cite{AKLT}.
Similarly, gapless edge modes in topological insulators are protected by both global time-reversal and charge-conservation symmetries \cite{RevModPhys.83.1057,PhysRevX.3.011016}. 
These novel features reflect the 't Hooft anomaly in the boundary Hilbert space \cite{PhysRevB.84.235141,PhysRevB.86.115109,PhysRevB.104.115156,GarreRubio2023classifyingphases}. 
The Li-Haldane conjecture \cite{PhysRevLett.101.010504} further shows that the entanglement properties of the bulk state lead to quantum anomalies on the boundary. 

Given that tensor networks capture the entanglement features of quantum many-body wave functions \cite{PhysRevB.81.064439,PhysRevB.83.245134,PhysRevB.91.224431,PhysRevLett.110.236801,PhysRevLett.111.090501,PhysRevLett.112.036402,PhysRevLett.114.106803,Jiang2012,PhysRevB.94.075151,PhysRevLett.117.131602}, they are naturally used to characterize boundary quantum anomalies \cite{PhysRevB.90.205114,PhysRevB.90.235113,PhysRevB.93.155163,PhysRevLett.120.156601,Shiozaki2017,PhysRevB.96.075125,xu2024entanglementpropertiesgaugetheories}. 
The tensor equation of the symmetric wave function provides the action of symmetry groups in the virtual space, allowing for direct identification of the boundary quantum anomalies.
Tensor networks offer a powerful platform for detecting and classifying topological quantum phases \cite{PhysRevB.84.165139,SCHUCH20102153,PhysRevB.86.125441,PhysRevB.94.205150,PhysRevB.95.075108,PhysRevB.95.125107,PhysRevLett.118.110504,Sahinoglu2021,PhysRevLett.132.126504,Xu_2024,guo2024locally} as well as characterizing phase transitions between different quantum phases \cite{PhysRevB.97.195124,PhysRevX.11.041014,PhysRevLett.102.255701,PhysRevLett.115.116803,PhysRevLett.119.070401,PhysRevLett.132.086503,PhysRevResearch.5.043078}.

Recent research has revealed that 't Hooft anomalies also arise in quantum many-body wave functions with subsystem symmetries \cite{PhysRevB.98.035112,PhysRevB.98.235121,PhysRevResearch.2.012059,PhysRevB.106.085113}. 
These subsystem symmetries are characterized by charges localized within rigid subregions of the lower-dimensional systems. 
Investigations from both quantum field theory \cite{10.21468/SciPostPhys.8.4.050,PhysRevB.103.245128,PhysRevB.106.085113,PhysRevResearch.2.012059,casasola2024fractalsubsystemsymmetriesanomalies,Ebisu2024} and lattice models \cite{PhysRevB.98.035112,PhysRevB.98.235121,PhysRevB.100.115112,PhysRevB.106.085104,PhysRevResearch.2.033331,PhysRevB.103.035148} have explored the quantum anomalies of these subsystem symmetries, uncovering their potential to construct higher-order gapless modes \cite{PhysRevB.98.235102,PhysRevB.105.245122,PhysRevB.108.045133,PhysRevB.108.155123,PhysRevResearch.2.012059}. 
Furthermore, studies on the two-dimensional (2D) cluster state, which is characterized by a $Z_2^\tau\times Z_2^\sigma$ subsystem symmetry, have demonstrated its ability to host universal computational resources \cite{PhysRevLett.86.5188,Briegel_2009,PhysRevLett.122.090501,Stephen2019subsystem,Stephen2019subsystem}. 
This finding has significantly improved our understanding of measurement-based quantum computation (MBQC) \cite{PhysRevA.98.022332}.

Despite this progress, detecting boundary anomalies of subsystem symmetries in general wave functions remains an unresolved problem. 
Although spurious topological entanglement entropy can identify nontrivial subsystem symmetry-protected topological (SSPT) phases \cite{PhysRevB.94.075151,PhysRevLett.122.140506,PhysRevB.100.115112}, it is insufficient to distinguish different nontrivial SSPT phases. 
A detailed analysis of the subsystem symmetry operators within the boundary space is required to make such distinctions \cite{PhysRevB.98.035112,PhysRevB.98.235121}.

To address this issue, we extend the topological response theory based on matrix product state (MPS) \cite{Shiozaki2017,PhysRevB.96.075125,deGroot2022symmetryprotected} to subsystem symmetries. 
The topological invariant of the SSPT phase can be obtained by calculating the subsystem symmetry charge of the twisted sector state.
This approach enables us to characterize the mixed anomaly of strong $Z_2^\tau\times Z_2^\sigma$ SSPT phases, which is consistent with the boundary anomaly characterized by the effective boundary Hamiltonian. 
We further identify an intrinsic $Z_2$ SSPT phase and detect its mixed anomaly from both the bulk invariant and boundary spectrum perspectives. 
Our method is also extended to mixed-state density matrices with average subsystem symmetries. 
By calculating the anomaly indicator for various lattice models, we uncover the connection between boundary anomalies in closed and open quantum systems.

This paper is organized as follows. 
In Sec. \ref{sec detect}, we revisit the SSPT phases and propose different methods to detect boundary anomalies of subsystem symmetries. 
In Sec. \ref{sec cluster}, we construct a tunable tensor to detect the quantum anomaly of the strong and weak $Z_2^\tau\times Z_2^\sigma$ SSPT phases. 
In Sec. \ref{sec z2}, we identify an intrinsic SSPT phase without a weak counterpart and characterize this topological phase through the anomaly indicator and the entanglement spectrum.
In Sec. \ref{sec disordered}, we find two distinct types of average subsystem symmetry-protected topological (ASSPT) phases and discuss how to extract their mixed-state quantum anomalies of average subsystem symmetry. 
Our research provides a numerical method to detect topological quantum phases protected by subsystem symmetries.

\section{Boundary anomaly detection of subsystem symmetries}\label{sec detect}

This section provides a concise review of SSPT phases and presents our approach to detecting boundary anomalies from the bulk topological invariant and the boundary entanglement spectrum. 
We also propose that the subsystem symmetry charge of the twisted sector state can serve as an anomaly indicator for these SSPT phases.

\subsection{Boundary anomaly of SSPT phases}
\begin{figure}
    \centering
    \begin{tikzpicture}
    \draw[thick] 
    (-0.5,0) -- (3.5,0) 
    (0,-0.5) -- (0,3.5)
    (-0.5,1) -- (3.5,1) 
    (1,-0.5) -- (1,3.5)
    (-0.5,2) -- (3.5,2) 
    (2,-0.5) -- (2,3.5)
    (-0.5,3) -- (3.5,3)
    (3,-0.5) -- (3,3.5);
    
    \filldraw[purplish,opacity=0.2] (-0.5,1.75) rectangle ++(4,0.5);
    \filldraw[purplish,opacity=0.2] (1.75,-0.5) rectangle ++(0.5,4);
    \draw[purplish] (4.1,2.2) node {$S^h(g^{[y]})$};
    \draw[purplish] (2,3.75) node {$S^v(g^{[x]})$};
    \draw[black] (2.5,1.75) node {$(x,y)$};    
    \foreach \x in {0,1,2,3}{
    \foreach \y in {0,1,2,3}{
    \filldraw[purplish] (\x,\y) circle[radius=0.13];
    }}
    \end{tikzpicture}
\caption{Linear subsystem symmetry transformations $S^v(g^{[x]})$ and $S^h(g^{[y]})$ on a 2D square lattice.
}\label{fig subsystem symmetry}
\end{figure}

A nontrivial symmetry-protected topological (SPT) phase is an SRE state that cannot be adiabatically deformed into a trivial product state through $G$-symmetric finite-depth local unitary circuits \cite{PhysRevB.87.155114}.
The classification of $d$-dimensional SPT phases protected by bosonic symmetry $G$ is given by the group cohomology $\mathcal{H}^{d+1}[G,U(1)]$ \cite{PhysRevB.87.155114}.
This classification aligns with the classification of quantum anomalies present in the $d$-dimensional boundary Hilbert space, which has been established through the \textit{anomaly inflow} mechanism. 
This mechanism has been generalized to systems with subsystem symmetry in Ref. \cite{PhysRevB.106.085113}, where the 't Hooft anomaly of a subsystem symmetry is canceled by the bulk theory of a nontrivial SSPT phase in one higher dimension. 

We aim to develop a numerical method to detect the boundary anomaly of subsystem symmetries directly from the bulk SSPT wave function.
To this end, we consider an SSPT state on a $L_x\times L_y$ square lattice with open boundary condition, as shown in Fig. \ref{fig subsystem symmetry}. 
The total linear subsystem symmetries are denoted by the large group $G = G_v\times G_h = \prod_{x=1}^{L_x}G_s^{[x]}\times \prod_{y=1}^{L_y}G_s^{[y]}$, where the subsystem symmetries are uniformly described by the group $G_s=\{g,f,\cdots\}$. 
We denote the group elements associated with the $x$-th column and $y$-th row as $g^{[x]}\in G_s^{[x]}$ and $g^{[y]}\in G_s^{[y]}$, respectively.
As illustrated in Fig. \ref{fig subsystem symmetry}, the subsystem symmetry transformations labeled by $g^{[x]}$ and $g^{[y]}$ are defined within the physical space by unitary operators
\begin{align}
\begin{aligned}
    S^v(g^{[x]}) = \prod_y U_{x,y}(g^{[x]}),\quad
    S^h(g^{[y]}) = \prod_x U_{x,y}(g^{[y]}).
\end{aligned}
\end{align}
Here, $U_{x,y}$ is a local unitary operator within the physical local Hilbert space $\mathcal{H}_{x,y}$. 
In the following, we refer to $S^v(g^{[x]})$ and $S^h(g^{[y]})$ as vertical and horizontal linear subsystem symmetry transformation operators. 
To investigate mixed quantum anomalies between subsystem symmetries, we propose that the subsystem symmetry operators decompose into a tensor product of two local boundary operators upon projection into the boundary Hilbert space. 
This is the simplest case where mixed anomalies exist only between adjacent subsystems.

To investigate the boundary anomaly of the SSPT phase, we construct a projected entangled-pair state (PEPS) representation on the $L_x\times L_y$ square grid, with periodic boundary identification between $y=0$ and $y=L_y$ as illustrated by
\begin{equation*}
    \begin{tikzpicture}
        \draw[white,fill=white,thick,dotted] (0,0) ellipse (0.5 and 1);
        \draw[black,ultra thick,dashed] (0,0) ellipse (0.5 and 1);
        \draw[black,ultra thick,dashed] (4,1) arc [start angle=90, end angle=-90, x radius=0.5, y radius=1];
        \draw[black,thick,dotted] (0,1) -- (4,1) (4,-1) -- (0,-1);
        \draw[black,thick] (-0.5,0) -- (0.5,0) (-0.3,0.75) -- (0.3,0.75) (-0.3,-0.75) -- (0.3,-0.75);
        \draw[black,thick] (0.41,0.5) -- (4.41,0.5) (0.41,-0.5) -- (4.41,-0.5) (1,1) arc[start angle=90, end angle=-90, x radius=0.5, y radius=1] (2,1) arc[start angle=90, end angle=-90, x radius=0.5, y radius=1] (3,1) arc[start angle=90, end angle=-90, x radius=0.5, y radius=1] (5.5,1) arc[start angle=90, end angle=20, x radius=0.5, y radius=1] (5.5,-1) arc[start angle=-90, end angle=-20, x radius=0.5, y radius=1] (0,-1.4) -- (1.7,-1.4) (2.3,-1.4) -- (4,-1.4);
        \draw[ultra thick,purplish] (1.15,0.8) -- (1.45,0.5) (2.15,0.8) -- (2.45,0.5) (3.15,0.8) -- (3.45,0.5) (1.15,-0.2) -- (1.45,-0.5) (2.15,-0.2) -- (2.45,-0.5) (3.15,-0.2) -- (3.45,-0.5);
        \draw (6,0) node {$L_y$} (2,-1.4) node {$L_x$};
        \draw (5.1,0) node {$\mathcal{H}_{\mathrm{edge}}^{\mathrm{Right}}$} (-1,0) node {$\mathcal{H}_{\mathrm{edge}}^{\mathrm{Left}}$};
    \end{tikzpicture}\label{fig cylinder}
\end{equation*}
Since the bulk wave function is symmetric, we map the subsystem symmetry operator $S^h(g^{[y]})$ with $g^{[y]}\in G_s^{[y]}$ to the boundary Hilbert space $\mathcal{H}_{\mathrm{edge}}^{\mathrm{Left}}\otimes\mathcal{H}_{\mathrm{edge}}^{\mathrm{Right}}$ as
\begin{equation}
    S^h(g^{[y]}) |\psi\rangle = W^{\mathrm{Left}}(g^{[y]})W^{\mathrm{Right}}(g^{[y]})|\psi\rangle.\label{eq nonlocal sym}
\end{equation}
We note that $W^{\mathrm{Right}}(g^{[y]})$ ($W^{\mathrm{Left}}(g^{[y]})$) forms a group representation of $G_h$ within the right (left) boundary Hilbert space $\mathcal{H}_{\mathrm{edge}}^{\mathrm{Right}}$ ($\mathcal{H}_{\mathrm{edge}}^{\mathrm{Left}}$), which is denoted by the thick dashed line of the cylinder. 
The boundary anomaly associated with $G_h$ is encoded within the factor system $\omega(g^{[y_1]},f^{[y_2]})$ of the boundary symmetry operators defined as \cite{PhysRevB.83.035107,PhysRevB.84.165139,PhysRevB.98.235121}
\begin{align}
    &W^{\mathrm{Right}}(g^{[y_1]})W^{\mathrm{Right}}(f^{[y_2]}) \nonumber \\
    =&\omega(g^{[y_1]},f^{[y_2]})W^{\mathrm{Right}}(g^{[y_1]}f^{[y_2]}).\label{eq boundary tensor eq1}
\end{align}
By reordering the symmetry operators associated with $g^{[y_1]}$ and $f^{[y_2]}$, we have
\begin{align}
    &W^{\mathrm{Right}}(f^{[y_2]})W^{\mathrm{Right}}(g^{[y_1]}) \nonumber \\
    =&\omega(f^{[y_2]},g^{[y_1]})W^{\mathrm{Right}}(f^{[y_2]}g^{[y_1]}).\label{eq boundary tensor eq2}
\end{align}
Considering $G_s$ as a finite Abelian group, the commutativity between $f^{[y_2]}$ and $g^{[y_1]}$ ensures that
\begin{equation}
    W^{\mathrm{Right}}(g^{[y_1]}f^{[y_2]}) = W^{\mathrm{Right}}(f^{[y_2]}g^{[y_1]}).\label{eq boundary tensor eq3}
\end{equation}
Combining Eqs. (\ref{eq boundary tensor eq1})-(\ref{eq boundary tensor eq3}), we have
\begin{align}
\begin{aligned}
    &W^{\mathrm{Right}}(g^{[y_1]})W^{\mathrm{Right}}(f^{[y_2]})\\ =& \phi(g^{[y_1]},f^{[y_2]})W^{\mathrm{Right}}(f^{[y_2]})W^{\mathrm{Right}}(g^{[y_1]}),\label{eq boundary tensor eq4}
\end{aligned}
\end{align}
where the phase factor
\begin{equation}
    \phi(g^{[y_1]},f^{[y_2]}) = \frac{\omega(g^{[y_1]},f^{[y_2]})}{\omega(f^{[y_2]},g^{[y_1]})}\label{eq boundary tensor eq5}
\end{equation}
represents the topological invariant of the wave function \cite{PhysRevB.86.125441,deGroot2022symmetryprotected}.

Based on our assumption of the boundary subsystem symmetry operators, we introduce the following two constraints:
\begin{itemize}
    \item The boundary operator $W^{\mathrm{Left}/\mathrm{Right}}(g^{[y]})$ at $y$-th row acts solely within the adjacent local boundary spaces $\mathcal{H}_{\mathrm{edge},y}^{\mathrm{Left}/\mathrm{Right}}\otimes\mathcal{H}_{\mathrm{edge},y+1}^{\mathrm{Left}/\mathrm{Right}}$.
    \item The boundary operator can be decomposed into the tensor product of the operators in two local boundary spaces, i.e., $W^{\mathrm{Right}}(g^{[y]})=V_y(g^{[y]})V_{y+1}(g^{[y]})$ and $W^{\mathrm{Left}}(g^{[y]})=V_y^{-1}(g^{[y]})V_{y+1}^{-1}(g^{[y]})$.
\end{itemize}
In this case, mixed quantum anomalies exist only in adjacent subsystems.
Namely, $\phi(g^{[y_1]},f^{[y_2]}) = 1$ if $|y_1-y_2|>1$. 
In Appendix \ref{app mixed anomaly}, we discuss the specific forms of $\phi(g^{[y_1]},f^{[y_2]})$ for the cases with $|y_1-y_2|>1$, $|y_1-y_2|=0$, and $|y_1-y_2|=1$. By expanding the region of the investigation while using a similar approach, the analysis is also applicable to more complex cases.
Given the system is translationally invariant in both $x$ and $y$ directions, Eq. (\ref{eq nonlocal sym}) can be expressed as the following tensor equation
\begin{equation}
\begin{tikzpicture}
    \draw[thick] (0,0.8) -- (0,-0.8) (-0.5,0.4) -- (0.5,0.4) (-0.5,-0.4) -- (0.5,-0.4);
    \draw[thick,fill=white] (-0.25,0.15) rectangle ++(0.5,0.5) (-0.25,-0.65) rectangle ++(0.5,0.5);
    \draw[thick] (0,0.8) arc[start angle=180, end angle=0, x radius=0.075, y radius=0.3]
    (0,-0.8) arc[start angle=-180, end angle=0, x radius=0.075, y radius=0.3];
    \draw[line width=0.1cm,white] (0,0.4) -- (-0.3,0.7) (0,-0.4) -- (-0.3,-0.1);
    \draw[ultra thick,purplish] (0,0.4) -- (-0.3,0.7) (0,-0.4) -- (-0.3,-0.1);
    \draw[purplish] (-0.6,0.1) node {$U_{x,y}$} (-0.5,0.8) node {$I$};
    \draw (1.25,0) node {$=W^{\mathrm{Left}}$} (2.6,0.4) node {$V^{-1}_{y+1}$} (2.6,-0.4) node {$V^{-1}_{y}$};
    \draw (5.75,0) node {$W^{\mathrm{Right}}$} (4.5,0.4) node {$V_{y+1}$} (4.35,-0.4) node {$V_{y}$};
    \node[rotate=270] at (2.15,0) {$\underbrace{\hspace{1.4cm}}$};
    \node[rotate=-270] at (4.95,0) {$\underbrace{\hspace{1.4cm}}$};
    \draw[thick] (3.5,0.8) -- (3.5,-0.8) (3,0.4) -- (4,0.4) (3,-0.4) -- (4,-0.4);
    \draw[thick,fill=white] (3.25,0.15) rectangle ++(0.5,0.5) (3.25,-0.65) rectangle ++(0.5,0.5);
    \draw[thick] (3.5,0.8) arc[start angle=180, end angle=0, x radius=0.075, y radius=0.3]
    (3.5,-0.8) arc[start angle=-180, end angle=0, x radius=0.075, y radius=0.3];
    \draw[line width=0.1cm,white] (3.5,0.4) -- (3.2,0.7) (3.5,-0.4) -- (3.2,-0.1);
    \draw[ultra thick,purplish] (3.5,0.4) -- (3.2,0.7) (3.5,-0.4) -- (3.2,-0.1);
\end{tikzpicture},\label{eq tensor eq}
\end{equation}
where $(g^{[y]})$ is omitted for simplicity.
The quantum anomaly $\phi(g^{[y_1]},f^{[y_2]})$ can be extracted from the group representation $W^{\mathrm{Right}}$ using Eq. (\ref{eq boundary tensor eq4}) and Eq. (\ref{eq tensor eq}).
Given the extensibility of subsystem symmetries, the number of equivalence classes in $\mathcal{H}^2[G_h,U(1)]$ may increase with system size $L_y$.
To eliminate redundancy in classification, Ref. \cite{PhysRevB.98.235121} introduces an additional equivalence relation.
Specifically, SSPT phases that can be deformed into each other via stacking of lower-dimensional SPT phases are considered to be in the same equivalence class. 
Based on this definition, SSPT phases are categorized into two distinct types: \textit{weak SSPT phases} and \textit{strong SSPT phases}. 
Weak SSPT phases can be trivialized via stacking lower-dimensional SPT phases, with their boundary anomaly existing only within each subsystem labeled by $\phi(g^{[y]},f^{[y]})$. 
The strong SSPT phase is characterized by the mixed quantum anomaly of two adjacent subsystem symmetries, labeled by $\phi(g^{[y]},f^{[y+1]})$. 

In this paper, we introduce the concept of an \textit{intrinsic SSPT phase} to denote the strong SSPT phase that lacks a weak counterpart. 
In Sec. \ref{sec z2}, we construct and investigate the properties of intrinsic SSPT phases, taking $Z_2$ subsystem symmetry as an example.

\subsection{Anomaly indicator of subsystem symmetries}
\begin{table*}[htbp]
\centering
\setlength{\arrayrulewidth}{1pt}
\setlength{\tabcolsep}{15pt}
\setlength{\extrarowheight}{5pt}
\begin{tabularx}{\textwidth}{cccc}

\hline
\hline
Topological phase & Symmetry group & Topological invariant & Anomaly indicator \\ 
\hline
1D SPT & $G$ & $\phi(g,f)$ & $Q(g,f)$ \\ \\

1D ASPT & $\mathcal{G}\times \mathcal{K}$ & \makecell{$\phi(g,\Tilde{g})$ \\ $\phi(k,\Tilde{g})$} & \makecell{$Q^{\mathrm{ave}}(g,\Tilde{g})$ \\ $Q^{\mathrm{exa}}(k,\Tilde{g})$}  \\ \\

2D SSPT & $G_h=\prod_y G_s^{[y]}$ & \makecell{$\phi(g^{[y]},f^{[y+1]})$ \\ $\phi(g^{[y]},f^{[y]})$} & \makecell{$Q(g^{[y]},f^{[y+1]})$ \\ $Q(g^{[y]},f^{[y]})$} \\ \\

2D uniform ASSPT & $\mathcal{G}_h=\prod_y \mathcal{G}_s^{[y]}\times\mathcal{K}_s^{[y]}$ & \makecell{$\phi(g^{[y]},\Tilde{g}^{[y+1]})$ \\ $\phi(k^{[y]},\Tilde{g}^{[y+1]})$} & \makecell{$Q^{\mathrm{ave}}(g^{[y]},\Tilde{g}^{[y+1]})$ \\ $Q^{\mathrm{exa}}(k^{[y]},\Tilde{g}^{[y+1]})$}  \\ \\

2D alternating ASSPT & $\mathcal{G}_h=\prod_y \mathcal{G}_s^{[2y]}\times \mathcal{K}_s^{[2y+1]}$ & \makecell{$\phi(g^{[2y]},k^{[2y+1]})$ \\ $\phi(k^{[2y-1]},g^{[2y]})$} & \makecell{$Q^{\mathrm{ave}}(g^{[2y]},k^{[2y+1]})$ \\ $Q^{\mathrm{exa}}(k^{[2y-1]},g^{[2y]})$} \\
\hline
\hline
\end{tabularx}
\caption{
Anomaly indicator of distinct SRE topological phases. 
$G_s^{[y]}$ represents the subsystem symmetry group for the $y$-th row. 
Mixed-state symmetries are denoted in calligraphic font, while pure-state symmetries are in italic font. 
The average symmetry is labeled by $\mathcal{G}$ and exact symmetry by $\mathcal{K}$. 
General group elements of ASPT phases are represented as $\Tilde{g}=(g,k)$ in $\mathcal{G}\times\mathcal{K}$. 
The forms of $Q^{\mathrm{ave}}$ and $Q^{\mathrm{exa}}$ for the mixed-state density matrices are given in Eqs. (\ref{eq transfer 1}) and (\ref{eq transfer 2}).
}
\label{tab anomaly detection}
\end{table*}
In this section, we extend the anomaly indicator for global symmetries, which is formulated using MPS representation \cite{Shiozaki2017,PhysRevB.96.075125,deGroot2022symmetryprotected}, to systems with linear subsystem symmetries. 
We then discuss how to apply this approach to identify mixed anomalies in strong SSPT phases.

Let us start with the SSPT tensor wave function $|\psi\rangle$ on a cylinder with $L_y=2$. 
We construct a twisted sector state $|\psi_{f^{[y+1]}}\rangle$ by inserting a zero-dimensional symmetry defect $V_{y+1}(f^{[y+1]})$ at the virtual bond of the tensor wave function with periodic boundary condition
\begin{equation}
\begin{tikzpicture}[>=stealth]
    \draw (-0.25,0) node {$|\psi_{f^{[y+1]}}\rangle=$}
    (3.5,0.4) node {$\cdots$} (3.5,-0.4) node {$\cdots$};
    \draw[thick] (1,0.4) -- (3,0.4) (4,0.4) -- (5.5,0.4) (1,-0.4) -- (3,-0.4) (4,-0.4) -- (5.5,-0.4) (1.5,0.8) -- (1.5,-0.8) (2.5,0.8) -- (2.5,-0.8) (4.5,0.8) -- (4.5,-0.8) (1,0.4) arc[start angle=270, end angle=90, x radius=0.25, y radius=0.075] (1,-0.4) arc[start angle=270, end angle=90, x radius=0.25, y radius=0.075] (5.5,0.4) arc[start angle=-90, end angle=90, x radius=0.25, y radius=0.075] (5.5,-0.4) arc[start angle=-90, end angle=90, x radius=0.25, y radius=0.075];
    \draw[thick,fill=white] (1.25,0.15) rectangle ++(0.5,0.5) (1.25,-0.65) rectangle ++(0.5,0.5) (2.25,0.15) rectangle ++(0.5,0.5) (2.25,-0.65) rectangle ++(0.5,0.5) (4.25,0.15) rectangle ++(0.5,0.5) (4.25,-0.65) rectangle ++(0.5,0.5);
    \draw[line width=0.1cm,white] (1.5,0.4) -- (1.2,0.7) (1.5,-0.4) -- (1.2,-0.1) (2.5,0.4) -- (2.2,0.7) (2.5,-0.4) -- (2.2,-0.1) (4.5,0.4) -- (4.2,0.7) (4.5,-0.4) -- (4.2,-0.1);
    \draw[ultra thick,purplish] (1.5,0.4) -- (1.2,0.7) (1.5,-0.4) -- (1.2,-0.1) (2.5,0.4) -- (2.2,0.7) (2.5,-0.4) -- (2.2,-0.1) (4.5,0.4) -- (4.2,0.7) (4.5,-0.4) -- (4.2,-0.1);
    \draw[thick] (1.5,0.8) arc[start angle=180, end angle=0, x radius=0.075, y radius=0.3]
    (1.5,-0.8) arc[start angle=-180, end angle=0, x radius=0.075, y radius=0.3] (2.5,0.8) arc[start angle=180, end angle=0, x radius=0.075, y radius=0.3]
    (2.5,-0.8) arc[start angle=-180, end angle=0, x radius=0.075, y radius=0.3] (4.5,0.8) arc[start angle=180, end angle=0, x radius=0.075, y radius=0.3]
    (4.5,-0.8) arc[start angle=-180, end angle=0, x radius=0.075, y radius=0.3];
    \filldraw (5.25,0.4) circle[radius=0.06];
    \draw (5.7,0.9) node {$V_{y+1}(f^{[y+1]})$};
\end{tikzpicture}.\label{eq single insertion}
\end{equation}
The mixed anomaly between the adjacent subsystem symmetries $G_s^{[y]}\times G_s^{[y+1]}$ for $|\psi\rangle$ is detected using the symmetry charges of the twisted sector state 
\begin{align}
\begin{aligned}
    Q(g^{[y]},f^{[y+1]}) = \frac{\langle\psi_{f^{[y+1]}}|S^h(g^{[y]})|\psi_{f^{[y+1]}}\rangle}{\langle\psi_{f^{[y+1]}}|\psi_{f^{[y+1]}}\rangle}.\label{eq topo response}
\end{aligned}
\end{align}
We calculate $Q(g^{[y]},f^{[y+1]})$ via the spectra of the transfer matrices with $f^{[y+1]}$ symmetry defect
\begin{equation}
\begin{tikzpicture}
    \draw (5,0.2) node {$\mathbb{T}(g^{[y]},f^{[y+1]})=$};
    \draw[thick] (7.5,0.8) -- (7.5,-0.8) (7,0.4) -- (8.1,0.4) (7,-0.4) -- (8.1,-0.4)
    (6.85,1.15) -- (6.85,-0.45) (6.35,0.75) -- (7.45,0.75) (6.35,-0.05) -- (7.45,-0.05);
    \draw[thick,fill=white] (7.25,0.15) rectangle ++(0.5,0.5) (7.25,-0.65) rectangle ++(0.5,0.5);
    \draw[thick] (7.5,0.8) arc[start angle=180, end angle=0, x radius=0.075, y radius=0.3]
    (7.5,-0.8) arc[start angle=-180, end angle=0, x radius=0.075, y radius=0.3] (6.85,1.15) arc[start angle=180, end angle=0, x radius=0.075, y radius=0.3]
    (6.85,-0.45) arc[start angle=-180, end angle=0, x radius=0.075, y radius=0.3];
    \draw[line width=0.1cm,white] (7.5,0.4) -- (6.85,0.75) (7.5,-0.4) -- (6.85,-0.05);
    \draw[ultra thick,purplish] (7.5,0.4) -- (6.85,0.75) (7.5,-0.4) -- (6.85,-0.05);
    \draw[thick,fill=white] 
    (6.6,0.5) rectangle ++(0.5,0.5) (6.6,-0.3) rectangle ++(0.5,0.5);
    \filldraw[black] (7.9,0.4) circle[radius=0.075] (7.3,0.75) circle[radius=0.075];
    \filldraw[white] (7.25,-0.265) circle[radius=0.1];
    \filldraw[purplish] (7.25,-0.265) circle[radius=0.075];
    \draw[purplish] (6.75,-1) node {$U_{x,y}(g^{[y]})$};
    \draw (8.75,0.65) node {$V_{y+1}(f^{[y+1]})$} (8,1.3) node {$V^*_{y+1}(f^{[y+1]})$};
\end{tikzpicture}.
\end{equation}
The details of the calculation are provided in Appendix \ref{app anomaly detection numerical}. 
When the system has subsystem symmetries, according to the tensor equation (\ref{eq tensor eq}), we have the following transfer matrix equation
\begin{equation}
\begin{tikzpicture}
    \draw[thick] (0,0.8) -- (0,-0.8) (-0.5,0.4) -- (0.6,0.4) (-0.5,-0.4) -- (0.6,-0.4)
    (-0.65,1.15) -- (-0.65,-0.45) (-1.15,0.75) -- (-0.05,0.75) (-1.15,-0.05) -- (-0.05,-0.05);
    \draw[thick,fill=white] (-0.25,0.15) rectangle ++(0.5,0.5) (-0.25,-0.65) rectangle ++(0.5,0.5);
    \draw[thick] (0,0.8) arc[start angle=180, end angle=0, x radius=0.075, y radius=0.3]
    (0,-0.8) arc[start angle=-180, end angle=0, x radius=0.075, y radius=0.3] (-0.65,1.15) arc[start angle=180, end angle=0, x radius=0.075, y radius=0.3]
    (-0.65,-0.45) arc[start angle=-180, end angle=0, x radius=0.075, y radius=0.3];
    \draw[line width=0.1cm,white] (0,0.4) -- (-0.65,0.75) (0,-0.4) -- (-0.65,-0.05);
    \draw[ultra thick,purplish] (0,0.4) -- (-0.65,0.75) (0,-0.4) -- (-0.65,-0.05);
    \draw[thick,fill=white] 
    (-0.9,0.5) rectangle ++(0.5,0.5) (-0.9,-0.3) rectangle ++(0.5,0.5);
    \filldraw[white] (-0.25,-0.265) circle[radius=0.1];
    \filldraw[purplish] (-0.25,-0.265) circle[radius=0.075];
    \draw[purplish] (-0.8,-1) node {$U_{x,y}(g^{[y]})$};
    \draw (1.25,0) node {$=$} (4.5,0.75) node {$V_{y+1}(f^{[y+1]})$};
    \draw[purplish] (4,0.1) node {$V_{y+1}(g^{[y]})$} (4,-0.7) node {$V_{y}(g^{[y]})$};
    \draw[thick] (3,0.8) -- (3,-0.8) (2.4,0.4) -- (3.9,0.4) (2.4,-0.4) -- (3.9,-0.4)
    (2.35,1.15) -- (2.35,-0.45) (1.85,0.75) -- (2.95,0.75) (1.85,-0.05) -- (2.85,-0.05);
    \draw[thick,fill=white] (2.75,0.15) rectangle ++(0.5,0.5) (2.75,-0.65) rectangle ++(0.5,0.5);
    \draw[thick] (3,0.8) arc[start angle=180, end angle=0, x radius=0.075, y radius=0.3]
    (3,-0.8) arc[start angle=-180, end angle=0, x radius=0.075, y radius=0.3] (2.35,1.15) arc[start angle=180, end angle=0, x radius=0.075, y radius=0.3]
    (2.35,-0.45) arc[start angle=-180, end angle=0, x radius=0.075, y radius=0.3];
    \draw[line width=0.1cm,white] (3,0.4) -- (2.35,0.75) (3,-0.4) -- (2.35,-0.05);
    \draw[ultra thick,purplish] (3,0.4) -- (2.35,0.75) (3,-0.4) -- (2.35,-0.05);
    \draw[thick,fill=white] 
    (2.1,0.5) rectangle ++(0.5,0.5) (2.1,-0.3) rectangle ++(0.5,0.5);
    \filldraw[purplish] (3.4,-0.4) circle[radius=0.075]  (3.4,0.4) circle[radius=0.075] (2.6,-0.4) circle[radius=0.075]  (2.6,0.4) circle[radius=0.075];
    \draw[thick,fill=black] (3.7,0.4) circle[radius=0.075] (2.75,0.75) circle[radius=0.075] (0.4,0.4) circle[radius=0.075] (-0.25,0.75) circle[radius=0.075];
\end{tikzpicture}.
\end{equation}
We denote the group representation of $g^{[y]}$ and $f^{[y+1]}$ in the virtual space by purple and black dots. 
According to Eq. (\ref{eq single bond phi}), reordering the gauge transformation within the virtual bond
\begin{equation}
\begin{tikzpicture}
    \draw (3.5,0) node {$=\phi(g^{[y]},f^{[y+1]})\cdot$}  (1.5,0.4) node {$V_{y+1}(f^{[y+1]})$} (5.5,0.4) node {$V_{y+1}(f^{[y+1]})$}
    (7.1,0) node {$,$};
    \draw[purplish] (0.5,-0.4) node {$V_{y+1}(g^{[y]})$} (6.5,-0.4) node {$V_{y+1}(g^{[y]})$};
    \draw[thick] (0,0) -- (2.,0) (5,0) -- (7,0);
    \filldraw[purplish] (0.5,0) circle[radius=0.075] (6.5,0) circle[radius=0.075];
    \filldraw[black] (1.5,0) circle[radius=0.075] (5.5,0) circle[radius=0.075];
\end{tikzpicture}
\end{equation}
yields the following equation
\begin{align}
\begin{aligned}
        \mathbb{T}(g^{[y]},f^{[y+1]}) = &\phi(g^{[y]},f^{[y+1]})\biggl(I\otimes W^{\mathrm{Left}}(g^{[y]})\biggr)\\ \cdot &\mathbb{T}(e,f^{[y+1]}) \cdot \biggl(I\otimes W^{\mathrm{Right}}(g^{[y]})\biggr).\label{eq transfer equation}
\end{aligned}
\end{align}
According to Eq. (\ref{eq transfer equation}), the numerator of $Q(g^{[y]},f^{[y+1]})$ in Eq. (\ref{eq topo response}) can be evaluated with the expression
\begin{align}
\begin{aligned}
&\langle\psi_{f^{[y+1]}}|S^h(g^{[y]})|\psi_{f^{[y+1]}}\rangle\\
=&\mathrm{Tr}[\mathbb{T}(g^{[y]},e)\cdots\mathbb{T}(g^{[y]},e)\mathbb{T}(g^{[y]},f^{[y+1]})]\\
=&\phi(g^{[y]},f^{[y+1]})\cdot\mathrm{Tr}[\mathbb{T}(e,e)\cdots\mathbb{T}(e,e)\mathbb{T}(e,f^{[y+1]})]\\
=&\phi(g^{[y]},f^{[y+1]})\langle\psi_{f^{[y+1]}}|\psi_{f^{[y+1]}}\rangle.
\end{aligned}
\end{align}
This yields the relationship
\begin{equation}
    Q(g^{[y]},f^{[y+1]}) = \phi(g^{[y]},f^{[y+1]})
\end{equation}
for a symmetric subsystem state.
It indicates that $Q(g^{[y]},f^{[y+1]})$ serves as an \textit{anomaly indicator} of the nontrivial strong SSPT phases. 
In fact, this quantity exactly represents the topological response of the background gauge field associated with subsystem symmetries, as detailed in Appendix \ref{app inflow}. 
In practical calculations, we can obtain $|\psi_{f^{[y+1]}}\rangle$ by diagonalizing the Hamiltonian with gauge flux insertion.
When the subsystem of the wave function $|\psi\rangle$ experiences symmetry breaking, $S^h(g^{[y]})$ will flip $|\psi\rangle$ to a different symmetry-broken configuration. 
Given that distinct symmetry-broken states are mutually orthogonal, the anomaly indicator converges to $0$ in the thermodynamic limit as
\begin{equation}
    \lim_{L_x\rightarrow\infty}Q(g^{[y]},f^{[y+1]}) = 0.
\end{equation}

We further generalize this method to various SRE phases protected by linear subsystem symmetries in both closed and open quantum systems. 
Our results are summarized in Tab. \ref{tab anomaly detection}. 
We will delve into the details in the following sections. 
In Sec. \ref{sec cluster} and \ref{sec z2}, we focus on anomaly indicators in SSPT phases and numerically compare them with $\phi(g^{[y]},f^{[y+1]})$. 
In Sec. \ref{sec disordered}, we discuss mixed-state quantum anomalies in average symmetry-protected topological (ASPT) phases involving both average and exact symmetries, and identify two distinct types of ASSPT phases. 

\subsection{Entanglement spectrum and edge theory}
The boundary anomaly of the 1D SPT phases, labeled by the nontrivial elements in $\mathcal{H}^2[G,U(1)]$, can also be detected by the two-fold degeneracy of the entanglement spectrum \cite{PhysRevB.81.064439}.
Consequently, the entanglement spectrum also serves as a fingerprint for detecting the quantum anomaly of 2D strong SSPT phases in $\mathcal{H}^2[G_h,U(1)]$. 
We begin by revisiting the method for calculating the entanglement spectrum of 2D tensor wave functions, as proposed in Ref. \cite{PhysRevB.83.245134}. 
We consider a 2D PEPS on a cylinder with circumference $L_y$ and length $L_x$. 
The system is cut at the length $l$, and the tensor wave function is represented as
\begin{equation}
    |\psi\rangle = \sum_{I,J}\sum_{\alpha}L^I_\alpha R^J_\alpha |I,J\rangle.
\end{equation}
Here, $I=(i_1,\cdots,i_{lL_y})$ denotes the physical indices on the left side, and $J=(i_{lL_y+1},\cdots,i_{L_xL_y})$ represents the physical indices on the right side. 
The virtual indices along the cut line are represented by $\alpha=(\alpha_1,\cdots,\alpha_{L_y})$. 
The reduced density matrix on the left part of the cylinder is \cite{PhysRevB.83.245134} 
\begin{equation}
    \rho_l\propto \sqrt{\sigma^T_L}\sigma_R\sqrt{\sigma^T_L},
\end{equation}
where $\sigma_L$ and $\sigma_R$ denote the left and right fixed points of the transfer matrix on the cylinder, respectively. 
This enables us to characterize the energy spectrum of the effective boundary Hamiltonian from the entanglement spectrum of $\rho_l$ \cite{PhysRevLett.101.010504}. 
Since linear subsystem symmetry transformations act projectively within the boundary vector space of 2D nontrivial SSPT phases, it is expected that their boundary spectrum would exhibit a scale-dependent degeneracy.

\section{Mixed anomaly of strong SSPT phases}\label{sec cluster}
In this section, we apply our method to distinguish the boundary anomalies of strong and weak $Z_2^\tau\times Z_2^\sigma$ SSPT phases. 
We provide a brief overview of the 2D cluster state on the square lattice using the tensor network formalism.
Then, we construct a tunable tensor to explore the nontrivial strong and weak SSPT phases with respect to $Z_2^\tau\times Z_2^\sigma$ subsystem symmetry.

\subsection{$Z_2^\tau\times Z_2^\sigma$ SSPT phase}
\begin{figure}
    \centering
    \begin{tikzpicture}
        \draw[thick] 
        (0,0) -- (0.4,0) -- (1.1,0.7) -- (1.8,0) -- (1.1,-0.7) -- (0.4,0)
        (1.8,0) -- (2.2,0) -- (2.9,0.7) -- (3.6,0) -- (2.9,-0.7) -- (2.2,0)
        (4,0) -- (4.7,0.7) -- (5.4,0) -- (4.7,-0.7) -- (4,0)
        (0,1.8) -- (0.4,1.8) -- (1.1,2.5) -- (1.8,1.8) -- (1.1,1.1) -- (0.4,1.8)
        (1.8,1.8) -- (2.2,1.8) -- (2.9,2.5) -- (3.6,1.8) -- (2.9,1.1) -- (2.2,1.8)
        (3.6,1.8) -- (4,1.8) -- (4.7,2.5) -- (5.4,1.8) -- (4.7,1.1) -- (4,1.8)
        (3.6,0) -- (4,0) (1.1,-0.7) -- (1.1,-1.1) (2.9,-0.7) -- (2.9,-1.1)
        (5.4,0) -- (5.8,0) (4.7,-0.7) -- (4.7,-1.1) (5.4,1.8) -- (5.8,1.8)
        (1.1,0.7) -- (1.1,1.1) (2.9,0.7) -- (2.9,1.1) (4.7,0.7) -- (4.7,1.1)
        (1.1,2.5) -- (1.1,2.9) (2.9,2.5) -- (2.9,2.9) (4.7,2.5) -- (4.7,2.9);
        \filldraw[blue] 
        (1.1,0.7) circle[radius=0.15]
        (1.1,2.5) circle[radius=0.15]
        (2.9,0.7) circle[radius=0.15]
        (2.9,2.5) circle[radius=0.15]
        (4.7,0.7) circle[radius=0.15]
        (4.7,2.5) circle[radius=0.15];

        \filldraw[red] (1.65,-0.15) -- (1.65,0.15) -- (1.95,0.15) -- (1.95,-0.15) -- cycle
        (1.65,1.65) -- (1.65,1.95) -- (1.95,1.95) -- (1.95,1.65) -- cycle
        (3.45,-0.15) -- (3.45,0.15) -- (3.75,0.15) -- (3.75,-0.15) -- cycle
        (3.45,1.65) -- (3.45,1.95) -- (3.75,1.95) -- (3.75,1.65) -- cycle
        (5.25,-0.15) -- (5.25,0.15) -- (5.55,0.15) -- (5.55,-0.15) -- cycle
        (5.25,1.65) -- (5.25,1.95) -- (5.55,1.95) -- (5.55,1.65) -- cycle;
        \draw[very thick,white,fill=blue]
        (1.1,0.7) circle[radius=0.075]
        (1.1,2.5) circle[radius=0.075]
        (2.9,0.7) circle[radius=0.075]
        (2.9,2.5) circle[radius=0.075]
        (4.7,0.7) circle[radius=0.075]
        (4.7,2.5) circle[radius=0.075];
        \draw[very thick,white,fill=red]
        (1.8,0) circle[radius=0.075]
        (1.8,1.8) circle[radius=0.075]
        (3.6,0) circle[radius=0.075]
        (3.6,1.8) circle[radius=0.075]
        (5.4,0) circle[radius=0.075]
        (5.4,1.8) circle[radius=0.075];
        \draw[very thick,white,fill=white] 
        (0.75,-0.35) circle[radius=0.2]
        (1.45,-0.35) circle[radius=0.2]
        (2.55,-0.35) circle[radius=0.2]
        (3.25,-0.35) circle[radius=0.2]
        (4.35,-0.35) circle[radius=0.2]
        (5.05,-0.35) circle[radius=0.2]
        (0.75,0.35) circle[radius=0.2]
        (1.45,0.35) circle[radius=0.2]
        (2.55,0.35) circle[radius=0.2]
        (3.25,0.35) circle[radius=0.2]
        (4.35,0.35) circle[radius=0.2]
        (5.05,0.35) circle[radius=0.2]
        (0.75,1.45) circle[radius=0.2]
        (1.45,1.45) circle[radius=0.2]
        (2.55,1.45) circle[radius=0.2]
        (3.25,1.45) circle[radius=0.2]
        (4.35,1.45) circle[radius=0.2]
        (5.05,1.45) circle[radius=0.2]
        (0.75,2.15) circle[radius=0.2]
        (1.45,2.15) circle[radius=0.2]
        (2.55,2.15) circle[radius=0.2]
        (3.25,2.15) circle[radius=0.2]
        (4.35,2.15) circle[radius=0.2]
        (5.05,2.15) circle[radius=0.2];
        \draw 
        (0.75,-0.35) node {\small $H$}
        (1.45,-0.35) node {\small $H$}
        (2.55,-0.35) node {\small $H$}
        (3.25,-0.35) node {\small $H$}
        (4.35,-0.35) node {\small $H$}
        (5.05,-0.35) node {\small $H$}
        (0.75,0.35) node {\small $H$}
        (1.45,0.35) node {\small $H$}
        (2.55,0.35) node {\small $H$}
        (3.25,0.35) node {\small $H$}
        (4.35,0.35) node {\small $H$}
        (5.05,0.35) node {\small $H$}
        (0.75,1.45) node {\small $H$}
        (1.45,1.45) node {\small $H$}
        (2.55,1.45) node {\small $H$}
        (3.25,1.45) node {\small $H$}
        (4.35,1.45) node {\small $H$}
        (5.05,1.45) node {\small $H$}
        (0.75,2.15) node {\small $H$}
        (1.45,2.15) node {\small $H$}
        (2.55,2.15) node {\small $H$}
        (3.25,2.15) node {\small $H$}
        (4.35,2.15) node {\small $H$}
        (5.05,2.15) node {\small $H$};
        \draw[blue]
        (0.75,0.75) node {\small $i_\tau$};
        \draw[red]
        (1.9,-0.4) node {\small $i_\sigma$};
        \draw[dashed,very thick,black] (0.2,-0.9) rectangle ++(1.9,1.9);
    \end{tikzpicture}
    \caption{
    Tensor network representation of a 2D cluster state on the square lattice.
    The local tensor $T^{i_\tau i_\sigma}$ is depicted by the dashed square. 
    The vertex of the PEPS denotes $\delta_{i j_1\cdots j_3}$. 
    The ``$H$'' symbol represents the Hadamard gate of the virtual bond.
    }\label{fig 2D cluster PEPS}
\end{figure}

A typical example that incorporates the $Z_2^\tau\times Z_2^\sigma$ subsystem symmetry is the 2D cluster state on the square lattice.
We formulate the PEPS representation of the 2D cluster state wave function based on the domain wall decoration structure reviewed in Appendix \ref{app tn rep 2d cluster}. 
The PEPS representation is shown in Fig. \ref{fig 2D cluster PEPS}. 
The linear $Z_2^\tau\times Z_2^\sigma$ subsystem symmetry operators are defined as
\begin{align}
\begin{aligned}
    S^v(g^{[x]}_\tau) &= \prod_y \tau^x_{x,y},\ 
    S^v(g^{[x]}_\sigma) = \prod_y \sigma^x_{x,y},\\
    S^h(g^{[y]}_\tau) &= \prod_x \tau^x_{x,y},\ 
    S^h(g^{[y]}_\sigma) = \prod_x \sigma^x_{x,y}.
\end{aligned}
\end{align}  
The local tensor $T^{i_\tau i_\sigma}$ of the PEPS is provided by
\begin{equation}
\begin{tikzpicture}
    \draw[thick] (3.5,0) -- (5,0) (4.25,-0.75) -- (4.25,0.75);
    \draw (5.5,0) node {$=$};
    \draw[purplish] (3.5,0.8) node {$(i_\tau,i_\sigma)$};
    \draw[thick]
    (6,0) -- (6.4,0) -- (7.1,0.7) -- (7.8,0) -- (7.1,-0.7) -- (6.4,0)
    (7.1,0.7) -- (7.1,1.1) (7.1,-0.7) -- (7.1,-1.1) (7.8,0) -- (8.2,0);
    \draw[very thick,white,fill=white] (6.65,0.35) circle[radius=0.2]
    (7.45,0.35) circle[radius=0.2]
    (6.65,-0.35) circle[radius=0.2]
    (7.45,-0.35) circle[radius=0.2];
    \draw (6.65,0.35) node {\small $H$}
    (6.65,-0.35) node {\small $H$}
    (7.45,0.35) node {\small $H$}
    (7.45,-0.35) node {\small $H$};
    \draw[blue]
    (6.75,0.8) node {$i_\tau$};
    \draw[red]
    (7.9,-0.4) node {$i_\sigma$};
    \draw[thick,fill=white] (3.95,-0.3) -- (4.55,-0.3) -- (4.55,0.3) -- (3.95,0.3) -- cycle;
    \filldraw[blue] (7.1,0.7) circle[radius=0.15];

    \filldraw[red] (7.65,-0.15) -- (7.65,0.15) -- (7.95,0.15) -- (7.95,-0.15) -- cycle;
    \draw[white,line width=0.125cm] (3.75,0.5) -- (4.25,0);
    \draw[purplish,ultra thick] (3.75,0.5) -- (4.25,0);
    \draw[very thick,white,fill=blue]
    (7.1,0.7) circle[radius=0.075];
    \draw[very thick,white,fill=red]
    (7.8,0) circle[radius=0.075];
\end{tikzpicture},\label{eq local tensor 2d cluster}
\end{equation}
where $H$ represents the Hadamard gate 
\begin{equation}
    H = \frac{1}{\sqrt{2}}
    \begin{pmatrix}
        1 & 1\\
        1 & -1
    \end{pmatrix}.
\end{equation}
The blue (red) vertex depicts the delta function $\delta_{i_\tau j_1\cdots j_3}$ ($\delta_{i_\sigma j_1\cdots j_3}$). 
We derive the local tensor equations of $T^{i_\tau i_\sigma}$ in Appendix \ref{app tn rep 2d cluster} in detail. 
The explicit form of the boundary operators $W^{\mathrm{Right}}$ are given by
\begin{align}
    W^{\mathrm{Right}}(g_\tau^{[y]}) &= Z_{L_x,y}Z_{L_x,y+1},\nonumber \\
    W^{\mathrm{Right}}(g_\sigma^{[y]}) &= X_{L_x,y}.\label{eq w structure}
\end{align}
By substituting the specific form of boundary symmetry operators from Eq. (\ref{eq w structure}) to Eq. (\ref{eq boundary tensor eq4}), we obtain the mixed anomaly between adjacent subsystem symmetries
\begin{equation}
\phi(g_{\tau}^{[y]},g_\sigma^{[y+1]}) = \frac{\omega(g_\tau^{[y]},g_\sigma^{[y+1]})}{\omega(g_\sigma^{[y+1]},g_\tau^{[y]})} = -1.
\end{equation}
Similarly, the boundary anomaly within a single subsystem is encoded by the anti-commutativity
\begin{align}
\begin{aligned}
    &W^{\mathrm{Right}}(g_\tau^{[y]})W^{\mathrm{Right}}(g_\sigma^{[y]})\\ =& \phi(g_\tau^{[y]},g_\sigma^{[y]})W^{\mathrm{Right}}(g_\sigma^{[y]})W^{\mathrm{Right}}(g_\tau^{[y]}),
\end{aligned}
\end{align}
with $\phi(g_\tau^{[y]},g_\sigma^{[y]})=-1$.
Since the bulk wave function exhibits translational invariance, we interpret the boundary anomalies through the graphical representation \cite{PhysRevB.98.235121}
\begin{equation}
\begin{tikzpicture}
    \draw (-0.35,0) node {$g_{\tau}^{[y]}$} (-0.55,1) node {$g_{\tau}^{[y+1]}$} (1.6,1) node {$g_{\sigma}^{[y+1]}$} (1.4,0) node {$g_{\sigma}^{[y]}$};
    \filldraw[red] 
    (1,1) circle[radius=0.08] 
    (1,0) circle[radius=0.08];
    \filldraw[blue] 
    (0,0) circle[radius=0.08] 
    (0,1) circle[radius=0.08];
    \draw[very thick] (0.15,0) -- (0.85,0) (0.15,1) -- (0.85,1);
    \draw[very thick] (0.1,0.1) -- (0.9,0.9);
\end{tikzpicture}.
\end{equation}
The links represent the anti-commutativity between the group representations $W^{\mathrm{Right}}$ of distinct group elements in $G_s^{[y]}\times G_s^{[y+1]}$. 
Different SSPT phases are clearly distinguished by the graphical representations for their projective representations.

\subsection{Mixed anomaly detection}\label{sec mixed anomaly}
\begin{figure}
    \centering
    \includegraphics[width = 0.95\linewidth]{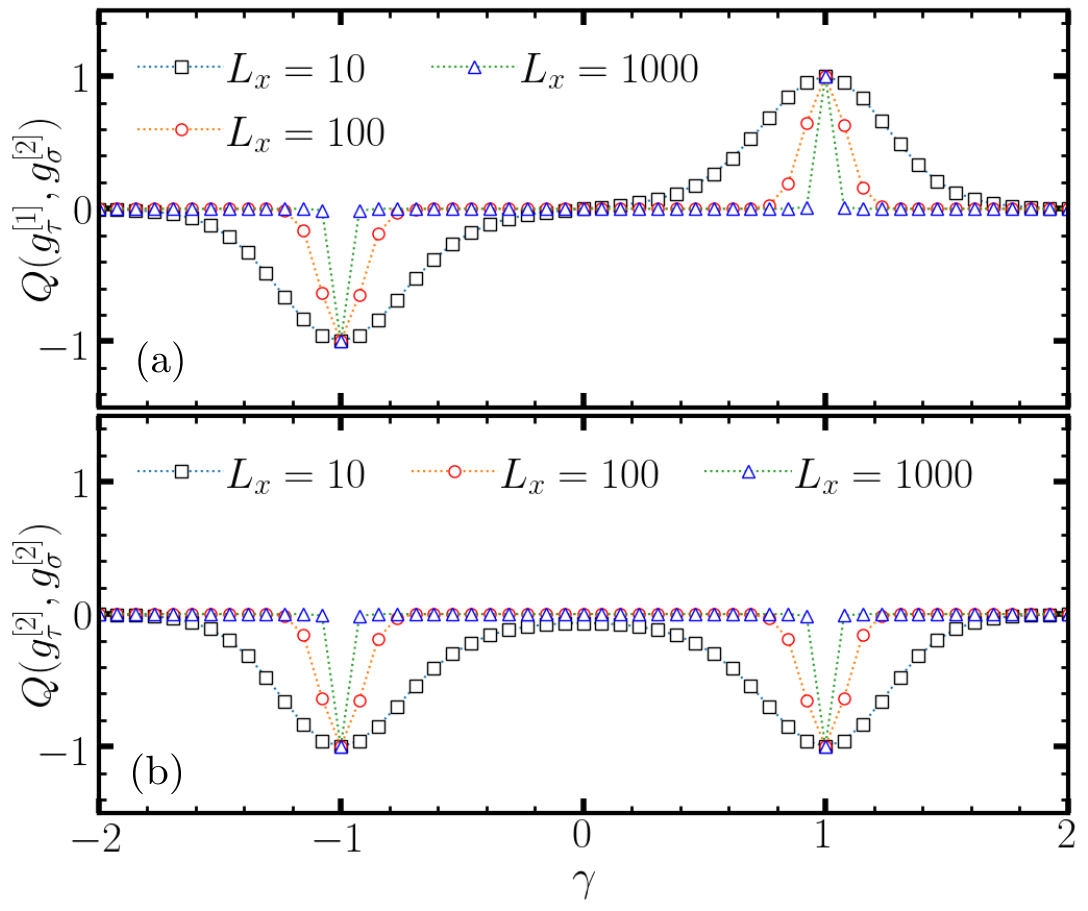}
    \caption{
    Mixed anomaly detection for the tunable tensor $T(\gamma)$. 
    Convergence of (a) $Q(g_\tau^{[1]},g_\sigma^{[2]})$ and (b) $Q(g_\tau^{[2]},g_\sigma^{[2]})$ with respect to system length $L_x$.
    }
    \label{fig cluster response}
\end{figure}
In this section, we apply our method to search for nontrivial topological phases based on a tunable tensor wave function. 
The explicit structure of the local tensor $T(\gamma)$ is represented graphically as
\begin{equation}
\begin{tikzpicture}
    \draw[thick]
    (6,0) -- (6.4,0) -- (7.1,0.7) -- (7.8,0) -- (7.1,-0.7) -- (6.4,0)
    (7.1,0.7) -- (7.1,1.1) (7.1,-0.7) -- (7.1,-1.1) (7.8,0) -- (8.2,0);
    \draw[very thick,white,fill=white] (6.65,0.35) circle[radius=0.2]
    (7.45,0.35) circle[radius=0.2]
    (6.65,-0.35) circle[radius=0.2]
    (7.45,-0.35) circle[radius=0.2];
    \draw (6.65,0.35) node {\small $H$}
    (6.65,-0.35) node {\small $\Gamma$}
    (7.45,0.35) node {\small $H$}
    (7.45,-0.35) node {\small $\Gamma$};
    \filldraw[blue] (7.1,0.7) circle[radius=0.15];

    \filldraw[red] (7.65,-0.15) -- (7.65,0.15) -- (7.95,0.15) -- (7.95,-0.15) -- cycle;
    \draw[very thick,white,fill=blue]
    (7.1,0.7) circle[radius=0.075];
    \draw[very thick,white,fill=red]
    (7.8,0) circle[radius=0.075];
    \draw (10.25,0) node {$\Gamma=
    \begin{pmatrix}
        1 & 1\\
        1 & \gamma
    \end{pmatrix}/\sqrt{2}.$}
    (8.5,-0.1) node {$,$};
\end{tikzpicture}
\end{equation} 
When $\gamma=-1$, $T$ represents the local tensor of 2D cluster state as given by Eq. (\ref{eq local tensor 2d cluster}), corresponding to the strong SSPT phase. 
When $\gamma=1$, the system disentangles between different rows, forming parallel 1D cluster chains belonging to the weak SSPT phase. 
The boundary symmetry operators of $T(\gamma=1)$ are
\begin{equation}
    W^{\mathrm{Right}}(g_\tau^{[y]}) = Z_{L_x,y},\quad W^{\mathrm{Right}}(g_\sigma^{[y]}) = X_{L_x,y}.
\end{equation}
The associated quantum anomaly is expressed graphically as
\begin{equation}
\begin{tikzpicture}
    \draw (-0.35,0) node {$g_{\tau}^{[y]}$} (-0.6,1) node {$g_{\tau}^{[y+1]}$} (1.6,1) node {$g_{\sigma}^{[y+1]}$} (1.4,0) node {$g_{\sigma}^{[y]}$};
    \filldraw[red] 
    (1,1) circle[radius=0.08] 
    (1,0) circle[radius=0.08];
    \filldraw[blue] 
    (0,0) circle[radius=0.08] 
    (0,1) circle[radius=0.08];
    \draw[very thick] (0.15,0) -- (0.85,0) (0.15,1) -- (0.85,1);
\end{tikzpicture}.
\end{equation}

To explore these two nontrivial SSPT phases, we vary $\gamma\in[-2,2]$ and calculate the anomaly indicators $Q$ associated with the transfer matrices
\begin{equation}
\begin{tikzpicture}
    \draw (1.9,0.2) node {$\mathbb{T}(g_\tau^{[1]},g_\sigma^{[2]})=$}
    (4.5,0.7) node {$X$} (3.8,1.1) node {$X$}; 
    \draw[thick] (4,0.8) -- (4,-0.8) (3.5,0.4) -- (4.6,0.4) (3.5,-0.4) -- (4.6,-0.4) (3.35,1.15) -- (3.35,-0.45) (2.85,0.75) -- (3.95,0.75) (2.85,-0.05) -- (3.95,-0.05);
    \draw[thick,fill=white] (3.75,0.15) rectangle ++(0.5,0.5) (3.75,-0.65) rectangle ++(0.5,0.5);
    \draw[thick] (4,0.8) arc[start angle=180, end angle=0, x radius=0.075, y radius=0.3]
    (4,-0.8) arc[start angle=-180, end angle=0, x radius=0.075, y radius=0.3] (3.35,1.15) arc[start angle=180, end angle=0, x radius=0.075, y radius=0.3]
    (3.35,-0.45) arc[start angle=-180, end angle=0, x radius=0.075, y radius=0.3];
    \draw[line width=0.1cm,white] (4,0.4) -- (3.35,0.75) (4,-0.4) -- (3.35,-0.05);
    \draw[ultra thick,purplish] (4,0.4) -- (3.35,0.75) (4,-0.4) -- (3.35,-0.05);
    \draw[thick,fill=white] 
    (3.1,0.5) rectangle ++(0.5,0.5) (3.1,-0.3) rectangle ++(0.5,0.5);
    \filldraw[white] (3.75,-0.265) circle[radius=0.1];
    \filldraw[blue] (3.75,-0.265) circle[radius=0.075];
    \filldraw[black] (4.4,0.4) circle[radius=0.075] (3.8,0.75) circle[radius=0.075];
    \draw[blue] (3.75,-0.85) node {$\tau^x$};
\end{tikzpicture}\label{eq mixed anomaly cluster}
\end{equation}
and
\begin{equation}
\begin{tikzpicture}
    \draw (5.9,0.2) node {$\mathbb{T}(g_\tau^{[2]},g_\sigma^{[2]})=$}
    (8.5,0.7) node {$X$} (7.8,1.1) node {$X$}; 
    \draw[thick] (8,0.8) -- (8,-0.8) (7.5,0.4) -- (8.6,0.4) (7.5,-0.4) -- (8.6,-0.4) (7.35,1.15) -- (7.35,-0.45) (6.85,0.75) -- (7.95,0.75) (6.85,-0.05) -- (7.95,-0.05);
    \draw[thick,fill=white] (7.75,0.15) rectangle ++(0.5,0.5) (7.75,-0.65) rectangle ++(0.5,0.5);
    \draw[thick] (8,0.8) arc[start angle=180, end angle=0, x radius=0.075, y radius=0.3]
    (8,-0.8) arc[start angle=-180, end angle=0, x radius=0.075, y radius=0.3] (7.35,1.15) arc[start angle=180, end angle=0, x radius=0.075, y radius=0.3]
    (7.35,-0.45) arc[start angle=-180, end angle=0, x radius=0.075, y radius=0.3];
    \draw[line width=0.1cm,white] (8,0.4) -- (7.35,0.75) (8,-0.4) -- (7.35,-0.05);
    \draw[ultra thick,purplish] (8,0.4) -- (7.35,0.75) (8,-0.4) -- (7.35,-0.05);
    \draw[thick,fill=white] 
    (7.1,0.5) rectangle ++(0.5,0.5) (7.1,-0.3) rectangle ++(0.5,0.5);
    \filldraw[white] (7.75,0.535) circle[radius=0.1];
    \filldraw[blue] (7.75,0.535) circle[radius=0.075];
    \filldraw[black] (8.4,0.4) circle[radius=0.075] (7.8,0.75) circle[radius=0.075];
    \draw[blue] (7.75,0.25) node {$\tau^x$};
\end{tikzpicture}.
\end{equation}
The simulation results are displayed in Fig. \ref{fig cluster response}, which are summarized as 
\begin{itemize}
    \item $\gamma=-1$, $Q(g_\tau^{[1]},g_\sigma^{[2]})=Q(g_\tau^{[2]},g_\sigma^{[2]})=-1$;
    \item $\gamma=1$, $Q(g_\tau^{[1]},g_\sigma^{[2]})=1$, $Q(g_\tau^{[2]},g_\sigma^{[2]})=-1$;
    \item $\gamma\neq \pm 1$, the subsystem symmetries are broken in the thermodynamic limit.
\end{itemize}
This indicates that the system exhibits $Z_2^\sigma\times Z_2^\tau$ subsystem symmetry at $\gamma=\pm 1$, where $\gamma= -1/1$ corresponds to the strong/weak SSPT phase. 

\begin{figure}
    \centering
    \includegraphics[width = 1\linewidth]{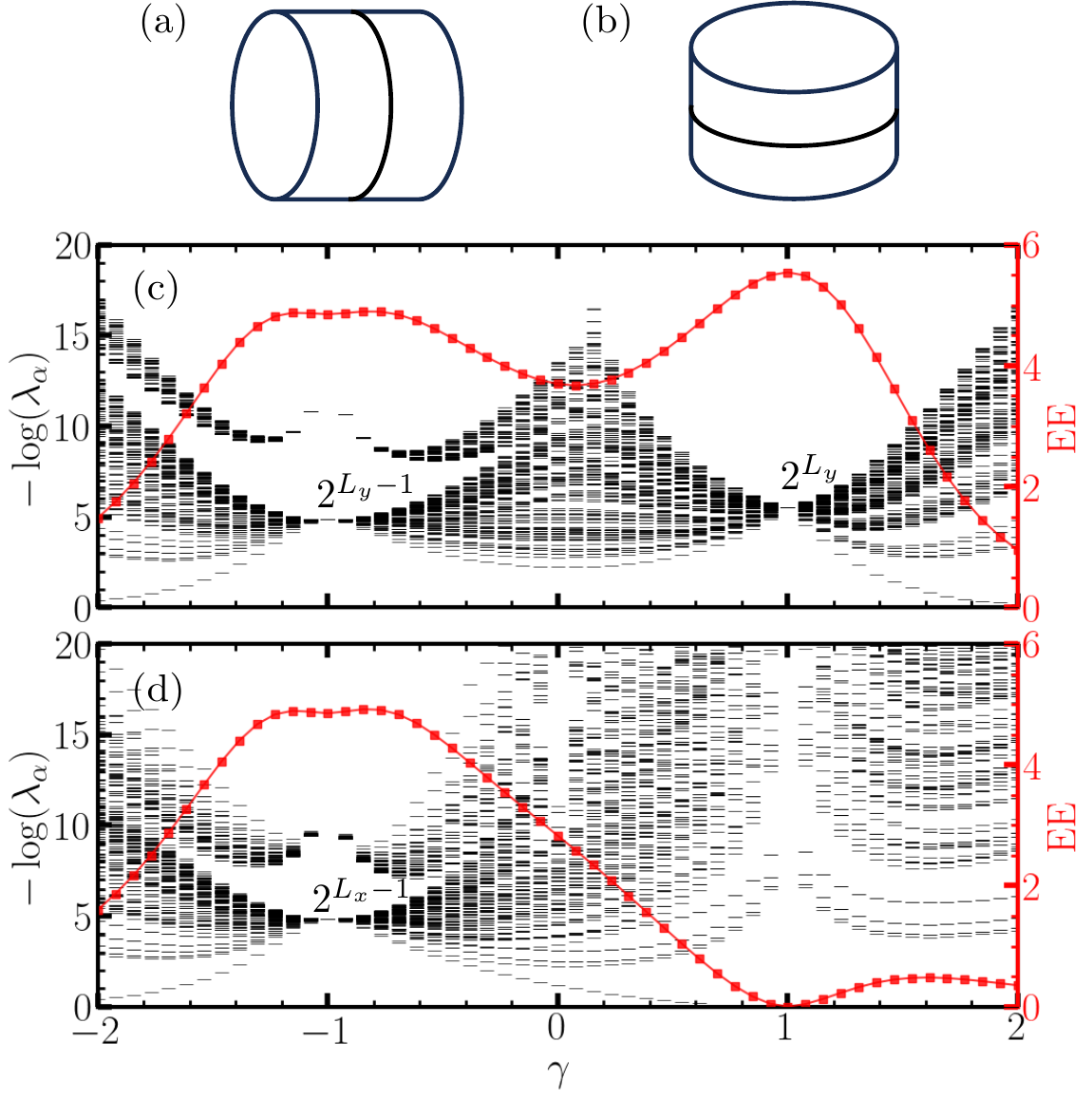}
    \caption{
    The entanglement spectrum (ES) and the entanglement entropy (EE) of the reduced density matrix $\rho_l(\gamma)$ on the cylinder along different directions with [(a) and (c)] $L_x=\infty,L_y=8$ and [(b) and (d)] $L_x=8,L_y=\infty$.
    }
    \label{fig es cluster}
\end{figure}

We further investigate the entanglement spectrum of $T(\gamma)$ and discuss the effective boundary theory of the SSPT phase. 
Since the PEPS built from $T(\gamma)$ is inhomogeneous in the vertical and horizontal directions, we characterize the entanglement spectra in two distinct cylinder geometries shown in Fig. \ref{fig es cluster}. 
The simulations in Fig. \ref{fig es cluster}(c) [Fig. \ref{fig es cluster}(d)] are based on the wave function on a cylinder with $L_y=8$ ($L_x=8$) shown in Fig. \ref{fig es cluster}(a) [Fig. \ref{fig es cluster}(b)]. 
The boundary 't Hooft anomaly of the 1D SPT is characterized by the vanishing entanglement gap \cite{PhysRevB.81.064439}, which is associated with the divergence of the string correlation length \cite{PhysRevLett.122.140506}. 
In the SSPT phase, the entanglement spectrum exhibits similar characteristics. 
As shown in Fig. \ref{fig es cluster}(c), the entanglement spectrum of the tunable tensor remains degenerate at $\gamma=\pm 1$, with the number of degeneracies exponentially increasing with the boundary scale. 
This implies that each subsystem hosts the long-range string order. 
For $\gamma=1$, the entanglement spectrum is fully degenerate, consistent with the energy spectrum of $L_y$ decoupled spin-$\frac{1}{2}$ degrees of freedom. 
This boundary degeneracy can be lifted by stacking 1D cluster chains and applying symmetric local unitary transformations. 
When $\gamma=-1$, the spectrum splits into two distinct branches, each branch being $2^{L_y-1}$-fold degenerate (with $\lambda_\alpha=0$ for the higher branch). 
Our simulation of the cylinder along the vertical direction is shown in Fig. \ref{fig es cluster}(d).
We observe that the entanglement gap along the vertical direction disappears at $\gamma=-1$, indicating that the string correlation length diverges only at this point. 

\section{Mixed anomaly of intrinsic SSPT phases}\label{sec z2}
In this section, we identify a nontrivial $Z_2$ SSPT phase by examining the entanglement spectrum and the anomaly indicator of the wave function. 
The absence of a weak $Z_2$ SSPT phase implies that the wave function exhibits mixed quantum anomalies between adjacent subsystems, indicative of an intrinsic SSPT phase.

\subsection{Intrinsic $Z_2$ SSPT wave function}
Considering only mixed anomalies between adjacent subsystems, there are $\mathcal{H}^2[Z_2^{[y]}\times Z_2^{[y+1]},U(1)]=\mathbb{Z}_2$ different types of boundary anomalies. 
Given that $\mathcal{H}^2[Z_2,U(1)]=\mathbb{Z}_1$, it is not feasible to construct a weak SSPT phase with decoupled 1D $Z_2$-SPT chains. 
This means that the unique boundary anomaly is encoded in the mixed anomaly $\phi(g^{[y]},g^{[y+1]})= -1$, which is visually represented as
\begin{equation}
\begin{tikzpicture}
    \draw[very thick] (0,0.15) -- (0,0.85);
    \filldraw[blue] (0,0) circle[radius=0.075] (0,1) circle[radius=0.075];
    \draw (-0.35,0) node {$g^{[y]}$} (-0.5,1) node {$g^{[y+1]}$};
\end{tikzpicture}.
\end{equation}
We use $g^{[y]}$ and $g^{[y+1]}$ to denote the group generators of $Z_2^{[y]}$ and $Z_2^{[y+1]}$ subsystem symmetries represented by $S^v(g^{[y]})=\prod_x X_{x,y}$ and $S^v(g^{[y+1]})=\prod_x X_{x,y+1}$ respectively. 
By solving the symmetry condition outlined in Eq. (\ref{eq tensor eq}), we reconstruct a bulk tensor wave function $|\psi_{Z_2}\rangle$ that belongs to the intrinsic $Z_2$ SSPT phase. 
The local bulk tensor of $|\psi_{Z_2}\rangle$ is
\begin{equation}
\begin{tikzpicture}
    \draw (-0.75,0) node {$T_{Z_2}=$}
    (1.5,0) node {$-$}
    (3.5,0) node {$+$}
    (-0.5,-1.5) node {$+$}
    (1.5,-1.5) node {$+$}
    (3.5,-1.5) node {$+$}
    (-0.5,-3) node {$+$}
    (1.5,-3) node {$-$}; 
    
    \draw[thick] (0,0) -- (1,0) 
    (0.5,0.5) -- (0.5,-0.5)
    (2.5,0) -- (3,0) 
    (2.5,0.5) -- (2.5,-0.5)
    (4,0) -- (5,0)
    (4.5,0.5) -- (4.5,0)
    (0.5,-1) -- (0.5,-1.5)
    (0.5,-1.5) -- (1,-1.5)
    (4,-1.5) -- (4.5,-1.5)
    (0.5,-3) -- (0.5,-3.5)
    (2,-3) -- (2.5,-3) -- (2.5,-3.5);

    \draw[very thick,dotted] (2,0) -- (2.5,0) (4.5,0) -- (4.5,-0.5) 
    (0,-1.5) -- (0.5,-1.5) -- (0.5,-2)
    (2.5,-1) -- (2.5,-2)
    (2,-1.5) -- (3,-1.5)
    (4.5,-1) -- (4.5,-2)
    (4.5,-1.5) -- (5,-1.5)
    (0.5,-2.5) -- (0.5,-3)
    (0,-3) -- (1,-3)
    (2.5,-2.5) -- (2.5,-3) -- (3,-3);
    
    \draw[very thick,blue,fill=white] 
    (2.5,-1.5) circle[radius=0.1]
    (4.5,-1.5) circle[radius=0.1]
    (0.5,-3) circle[radius=0.1]
    (2.5,-3) circle[radius=0.1];
    \filldraw[blue] (0.5,0) circle[radius=0.1]
    (2.5,0) circle[radius=0.1]
    (4.5,0) circle[radius=0.1]
    (0.5,-1.5) circle[radius=0.1];
\end{tikzpicture}.\label{eq Z2 wavefunction}
\end{equation}
Here, the solid and hollow blue balls denote the physical state $|\uparrow\rangle$ and $|\downarrow\rangle$, while solid and dotted lines represent the virtual vectors $|1\rangle$ and $|0\rangle$, respectively.

\subsection{Mixed anomaly detection}
The mixed quantum anomaly of the $Z_2$ subsystem symmetry is characterized by the topological invariant of the wave function. 
A tunable tensor $T_{Z_2}(\varphi)$ is constructed by modifying the $-1$ tensor elements of $T_{Z_2}$ as
\begin{equation}
\begin{tikzpicture}
    \draw[thick] (-0.5,0) -- (0,0) -- (0,-0.5) (2,0.5) -- (2,-0.5) (2,0) -- (2.5,0);
    \draw[very thick,dotted] (0.5,0) -- (0,0) -- (0,0.5) (1.5,0) -- (2,0);
    \draw[very thick,blue,fill=white] (0,0) circle[radius=0.1];
    \filldraw[blue] (2,0) circle[radius=0.1];
    \draw (1,-0.025) node {$=$}
    (3.4,0.05) node {$=e^{i\varphi}.$};
\end{tikzpicture}
\end{equation}
For $\varphi=(2m+1)\pi$ and $m\in\mathbb{Z}$, the tensor represents a $Z_2$ subsystem symmetric state, characterized by the mixed anomaly $\phi(g^{[y]},g^{[y+1]})=-1$. 
The boundary symmetry operators are given by
\begin{equation}
    W^{\mathrm{Right}}(g^{[y]})|_{\varphi=(2m+1)\pi} = X_{L_x,y}Z_{L_x,y+1},
\end{equation}
indicating that 
\begin{equation}
    \{W^{\mathrm{Right}}(g^{[y]}),W^{\mathrm{Right}}(g^{[y+1]})\}_{\varphi=(2m+1)\pi}=0.
\end{equation}
When $\varphi=2m\pi$, the tensor also preserves the $Z_2$ subsystem symmetry. 
However, the boundary symmetry operators form a linear representation of $G_h$ as
\begin{equation}
    W^{\mathrm{Right}}(g^{[y]})|_{\varphi=2m\pi} = X_{L_x,y}.
\end{equation}
\begin{figure}
    \centering
    \includegraphics[width = 0.95\linewidth]{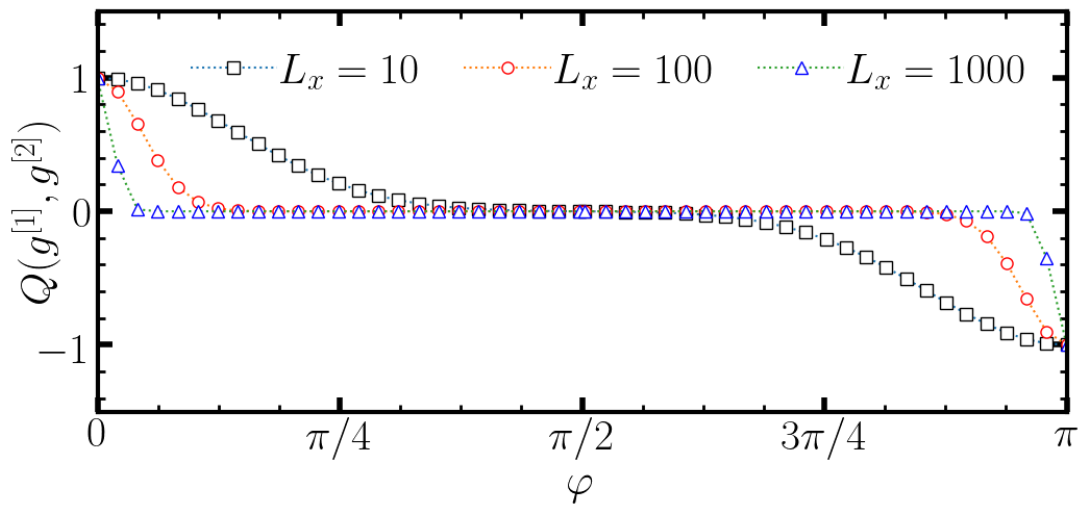}
    \caption{
    Mixed anomaly detection of the intrinsic $Z_2$ SSPT phase. Convergence of $Q(g^{[1]},g^{[2]})$ with respect to $L_x$.
    }
    \label{fig Z_2 response}
\end{figure}
To investigate the potential intrinsic SSPT phase within this tunable tensor $T_{Z_2}(\varphi)$, we compute the anomaly indicator $Q(g^{[1]},g^{[2]})$ associated with the transfer matrix
\begin{equation}
\begin{tikzpicture}
    \draw (1.75,0.2) node {$\mathbb{T}(g^{[1]},g^{[2]})=$}
    (4.5,0.7) node {$X$} (3.8,1.1) node {$X$}; 
    \draw[thick] (4,0.8) -- (4,-0.8) (3.5,0.4) -- (4.6,0.4) (3.5,-0.4) -- (4.6,-0.4) (3.35,1.15) -- (3.35,-0.45) (2.85,0.75) -- (3.95,0.75) (2.85,-0.05) -- (3.95,-0.05);
    \draw[thick,fill=white] (3.75,0.15) rectangle ++(0.5,0.5) (3.75,-0.65) rectangle ++(0.5,0.5);
    \draw[thick] (4,0.8) arc[start angle=180, end angle=0, x radius=0.075, y radius=0.3]
    (4,-0.8) arc[start angle=-180, end angle=0, x radius=0.075, y radius=0.3] (3.35,1.15) arc[start angle=180, end angle=0, x radius=0.075, y radius=0.3]
    (3.35,-0.45) arc[start angle=-180, end angle=0, x radius=0.075, y radius=0.3];
    \draw[line width=0.1cm,white] (4,0.4) -- (3.35,0.75) (4,-0.4) -- (3.35,-0.05);
    \draw[very thick,blue] (4,0.4) -- (3.35,0.75) (4,-0.4) -- (3.35,-0.05);
    \draw[thick,fill=white] 
    (3.1,0.5) rectangle ++(0.5,0.5) (3.1,-0.3) rectangle ++(0.5,0.5);
    \filldraw[white] (3.75,-0.265) circle[radius=0.1];
    \filldraw[blue] (3.75,-0.265) circle[radius=0.075];
    \filldraw[black] (4.4,0.4) circle[radius=0.075] (3.8,0.75) circle[radius=0.075];
    \draw[blue] (3.75,-0.85) node {$X$};
\end{tikzpicture}.
\end{equation}
Our simulation results shown in Fig. \ref{fig Z_2 response} are summarized by the equation
\begin{equation}
    \lim_{L_x\rightarrow\infty}Q(g^{[1]},g^{[2]}) = 
    \begin{cases}
        1, & \varphi = 2m\pi\\
        0, & \varphi\in \bigcup_m(2m\pi,(2m+1)\pi)\\
        -1, & \varphi = (2m+1)\pi
    \end{cases}.
\end{equation}
We observe $|Q(g^{[1]},g^{[2]})|=1$ for $\varphi=n\pi$ and $n\in\mathbb{Z}$, which signifies the restoration of $Z_2$ subsystem symmetry. 
In contrast, when $\varphi$ deviates from $n\pi$, we find $\lim_{L_x\rightarrow\infty}\langle\psi(\varphi)|S^h(g^{[y]})|\psi(\varphi)\rangle=0$, indicating the breaking of $Z_2$ subsystem symmetry.  

\begin{figure}
    \centering
    \includegraphics[width = 1\linewidth]{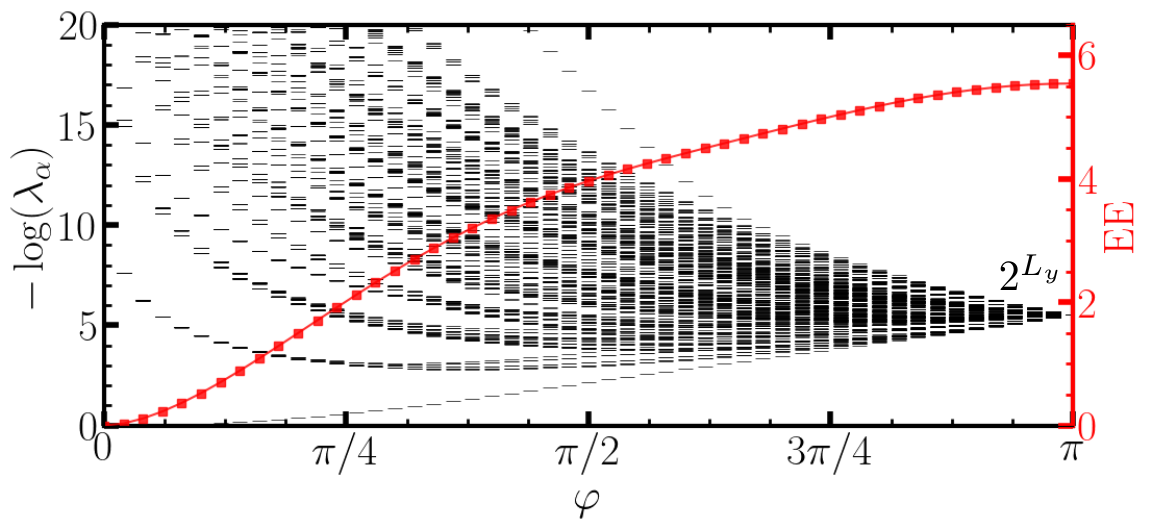}
    \caption{
    ES and EE of the reduced density matrix $\rho_l(\varphi)$ on a cylinder with $L_y=8$. 
    The entanglement gap vanishes at $\varphi=\pi$.
    }
    \label{fig Z_2 spectrum}
\end{figure}

In addition, we examine the mixed anomaly of subsystem symmetry through the entanglement spectrum and the effective edge theory. 
The entanglement spectrum of $\rho_l$ on a cylinder is shown in Fig. \ref{fig Z_2 spectrum}. 
At $\varphi=2m\pi$, the spectrum $\lambda_\alpha$ along the cylinder is given by $\{1,0,\cdots,0\}$. 
However, when $\varphi=(2m+1)\pi$, the spectrum exhibits full degeneracy as $\{2^{-L_y},\cdots,2^{-L_y}\}$, resembling the anomalous boundary of the nontrivial SSPT phase.
The closing of the entanglement gap at $\varphi=(2m+1)\pi$ suggests the existence of long-range string order for the $Z_2$ subsystem symmetry. 

Furthermore, we investigate the stability of the topological invariant and the degenerate entanglement spectrum in the intrinsic $Z_2$ SSPT phase.
By applying a subsystem symmetric perturbation 
\begin{equation}
    U_y^v(\eta)=\prod_x \mathrm{exp}({\eta\cdot X_{x,y}}),\quad \eta\in [0,1]
\end{equation}
to the 2D intrinsic $Z_2$ SSPT wave function, we find that $\phi(g^{[y]},g^{[y+1]})$ remains $-1$ and the entanglement spectrum remains fully degenerate throughout the nontrivial SSPT phase. 
Hence, our detection method effectively characterizes the wave function throughout the phase.

\section{Mixed-state anomaly of ASSPT phases}\label{sec disordered}
Recent studies have revealed the presence of mixed-state quantum anomalies associated with average symmetry \cite{deGroot2022symmetryprotected,PhysRevX.13.031016,ma2023topological,PhysRevLett.133.106503,lessa2024strongtoweakspontaneoussymmetrybreaking,zhang2024strongtoweakspontaneousbreaking1form,guo2024newframeworkquantumphases,PhysRevB.110.165160,PhysRevB.111.L201108,xu2024averageexactmixedanomaliescompatible}.
In this section, we expand our discussion to include average subsystem symmetry and provide an overview of recent developments.
We begin by introducing average symmetry and its characteristics before presenting methods for detecting mixed-state quantum anomalies of both average and exact symmetry in open quantum systems. 
Subsequently, we propose two scenarios for constructing ASSPT phases in open quantum systems: \textit{uniformly} breaking a subgroup of the subsystem symmetry, and \textit{alternately} breaking the subsystem symmetry. 
Finally, we demonstrate the effectiveness of our mixed-state anomaly detection technique in 2D uniform ASSPT phases.

\subsection{Anomaly indicator of average symmetries}
In this section, we focus on the SRE mixed state $\rho$, which can be deformed into a pure product state through a finite-depth quantum channel $\varepsilon$
\begin{equation}
    \rho = \varepsilon(|0\rangle\langle 0|).
\end{equation}
The SRE density matrix $\rho$ may display two types of symmetry: average symmetry and exact symmetry. 
We denote the average symmetry group by $\mathcal{G}$ and the exact symmetry group by $\mathcal{K}$. 
The condition for an average symmetry transformation $S(g)$ with $g\in \mathcal{G}$ is given by
\begin{equation}
    S(g)\rho S^\dagger(g) = \rho.
\end{equation}
The exact symmetry condition for $S(k)$ with $k\in \mathcal{K}$ is given by
\begin{equation}
    S(k)\rho = e^{i\theta}\rho,\quad \rho S^\dagger(k) = e^{-i\theta}\rho.
\end{equation}
The total symmetry group $\Tilde{G}$ of the density matrix is constructed via the group extension
\begin{equation}
1\rightarrow\mathcal{K}\rightarrow\mathcal{\Tilde{G}}\rightarrow\mathcal{G}\rightarrow 1,
\end{equation}
which is uniquely labeled by the $2$-cocycle $\nu_2\in\mathcal{H}^2[\mathcal{G},\mathcal{K}]$. 
We consider the case where $\nu_2$ is a trivial $2$-cocycle. 
The general elements of $\Tilde{G}$ are denoted as $\Tilde{g}=(g,k)\in \mathcal{\Tilde{G}}$. 
We introduce the following matrix product operator (MPO) representation for the density matrix $\rho$
\begin{equation}
\begin{tikzpicture}
    \draw (1.5,0) node {$\cdots$}
    (-0.75,0) node {$\rho=$};
    \draw[thick] (0,0) -- (1,0) (2,0) -- (4,0);
    \draw[ultra thick,purplish] (0.5,-0.6) -- (0.5,0.6) (2.5,-0.6) -- (2.5,0.6) (3.5,-0.6) -- (3.5,0.6);
    \draw[thick,fill=white] (0.25,-0.25) rectangle ++(0.5,0.5) (2.25,-0.25) rectangle ++(0.5,0.5) (3.25,-0.25) rectangle ++(0.5,0.5);
    \draw[thick] (0,0) arc[start angle=270, end angle=90, x radius=0.3, y radius=0.075]  (4,0) arc[start angle=-90, end angle=90, x radius=0.3, y radius=0.075];
\end{tikzpicture}.
\end{equation}
To detect the quantum anomaly of the total symmetry group $\mathcal{\Tilde{G}}$, we construct the twisted sector density matrix
\begin{equation}
\begin{tikzpicture}
    \draw (1.5,0) node {$\cdots$}
    (-0.9,0) node {$\rho_{\Tilde{g}}=$}
    (4.25,-0.35) node {$V(\Tilde{g})$};
    \draw[thick] (0,0) -- (1,0) (2,0) -- (4.5,0);
    \draw[ultra thick,purplish] (0.5,-0.6) -- (0.5,0.6) (2.5,-0.6) -- (2.5,0.6) (3.5,-0.6) -- (3.5,0.6);
    \draw[thick,fill=white] (0.25,-0.25) rectangle ++(0.5,0.5) (2.25,-0.25) rectangle ++(0.5,0.5) (3.25,-0.25) rectangle ++(0.5,0.5);
    \draw[thick] (0,0) arc[start angle=270, end angle=90, x radius=0.3, y radius=0.075]  (4.5,0) arc[start angle=-90, end angle=90, x radius=0.3, y radius=0.075];
    \filldraw[black] (4.25,0) circle[radius=0.075];
\end{tikzpicture}.
\end{equation}
The mixed-state anomaly between the exact symmetry $S(k)$ and the zero-dimensional symmetry defect $V(\Tilde{g})$ can be detected from the exact symmetry charge of the twisted sector density matrix
\begin{equation}
    Q^{\mathrm{exa}}(k,\Tilde{g}) = \frac{\mathrm{Tr}[S(k)\rho_{\Tilde{g}}]}{\mathrm{Tr}[\rho_{\Tilde{g}}]},\label{eq transfer 1}
\end{equation}
which is calculated with the transfer matrix 
\begin{equation}
\begin{tikzpicture}
    \draw (-1.4,0) node {$\mathbb{T}_{1}(k,\Tilde{g})=$}
    (0.75,0.25) node {$V(\Tilde{g})$};;
    \draw[blue] (-0.5,0.5) node {$U(k)$};
    \draw[thick] (-0.5,0) -- (0.75,0);
    \draw[ultra thick,purplish] (0,0.5) -- (0,-0.5) ;
    \draw[thick,fill=white] (-0.25,-0.25) rectangle ++(0.5,0.5);
    \draw[ultra thick,purplish] (0,0.5) arc[start angle=180, end angle=0, x radius=0.075, y radius=0.3] (0,-0.5) arc[start angle=-180, end angle=0, x radius=0.075, y radius=0.3];
    \filldraw[black] (0.5,0) circle[radius=0.075];
    \filldraw[blue] (0,0.5) circle[radius=0.075];
\end{tikzpicture}.
\end{equation}
The mixed-state anomaly associated with the average symmetry $g$ is evaluated using  
\begin{equation}
    Q^{\mathrm{ave}}(g,\Tilde{g}) = \frac{\mathrm{Tr}[S(g)\rho_{\Tilde{g}} S^\dagger(g) \rho_{\Tilde{g}}]}{\mathrm{Tr}[\rho_{\Tilde{g}}^2]},\label{eq transfer 2}
\end{equation}
which is calculated with the transfer matrix
\begin{equation}
\begin{tikzpicture}
    \draw (-2,0.5) node {$\mathbb{T}_{2}(g,\Tilde{g})=$}
    (0.75,0.25) node {$V(\Tilde{g})$}
    (0.75,1.25) node {$V(\Tilde{g})$};
    \draw[red] (-0.6,0.5) node {$U^\dagger(g)$} (-0.6,1.5) node {$U(g)$};
    \draw[thick] (-0.5,0) -- (0.75,0) (-0.5,1) -- (0.75,1);
    \draw[ultra thick,purplish] (0,-0.5) -- (0,1.5) ;
    \draw[thick,fill=white] (-0.25,-0.25) rectangle ++(0.5,0.5) (-0.25,0.75) rectangle ++(0.5,0.5);
    \draw[ultra thick,purplish] (0,1.5) arc[start angle=180, end angle=0, x radius=0.075, y radius=0.3] (0,-0.5) arc[start angle=-180, end angle=0, x radius=0.075, y radius=0.3];
    \filldraw[black] (0.5,0) circle[radius=0.075] (0.5,1) circle[radius=0.075];
    \filldraw[red] (0,0.5) circle[radius=0.075] (0,1.5) circle[radius=0.075];
\end{tikzpicture}.
\end{equation}
The relationship between the symmetry charge of $\rho_{\Tilde{g}}$ and the mixed-state anomaly of ASPT phase is discussed in Appendix \ref{app mixed-state anomaly detection}. 
The numerical calculations for $Q^{\mathrm{exa}}$ and $Q^{\mathrm{ave}}$ follow a methodology analogous to that presented in Appendix \ref{app anomaly detection numerical}. 

\subsection{Average subsystem symmetry}
This section shows two approaches for generating a nontrivial ASSPT phase. 
Using the extensibility of subsystem symmetry, one may break it either \textit{uniformly} or \textit{alternately} to realize an average subsystem symmetry. 

\subsubsection{Uniform average subsystem symmetry}
\begin{figure}
    \centering
    \begin{tikzpicture}
        \draw[thick] 
        (0.5,0) -- (6,0)
        (0.5,1.5) -- (6,1.5)
        (0.5,3) -- (6,3)
        (1,-0.5) -- (1,3.5)
        (2.5,-0.5) -- (2.5,3.5)
        (4,-0.5) -- (4,3.5)
        (5.5,-0.5) -- (5.5,3.5);
        \draw[very thick,blue] 
        (0.7,0.3) -- (1,0)
        (0.7,1.8) -- (1,1.5) 
        (0.7,3.3) -- (1,3) 
        (2.2,0.3) -- (2.5,0) 
        (2.2,1.8) -- (2.5,1.5) 
        (2.2,3.3) -- (2.5,3) 
        (3.7,0.3) -- (4,0) 
        (3.7,1.8) -- (4,1.5) 
        (3.7,3.3) -- (4,3) 
        (5.2,0.3) -- (5.5,0) 
        (5.2,1.8) -- (5.5,1.5) 
        (5.2,3.3) -- (5.5,3);
        \draw[very thick,red] 
        (0.7,-0.3) -- (1,0) 
        (0.7,1.2) -- (1,1.5) 
        (0.7,2.7) -- (1,3)
        (2.2,-0.3) -- (2.5,0) 
        (2.2,1.2) -- (2.5,1.5)
        (2.2,2.7) -- (2.5,3) 
        (3.7,-0.3) -- (4,0) 
        (3.7,1.2) -- (4,1.5) 
        (3.7,2.7) -- (4,3) 
        (5.2,-0.3) -- (5.5,0) 
        (5.2,1.2) -- (5.5,1.5) 
        (5.2,2.7) -- (5.5,3);
        \filldraw[blue,opacity=0.2] (0.6,1.55) rectangle ++(5.2,0.35);
        \filldraw[red,opacity=0.2] (0.6,1.45) rectangle ++(5.2,-0.35);
        \draw[blue] (6.4,1.75) node {$S^h(k^{[y]})$};
        \draw[red] (6.4,1.25) node {$S^h(g^{[y]})$};
    \end{tikzpicture}
    \caption{
    The $\mathcal{G}_s\times \mathcal{K}_s$ ASSPT phase with uniform disorder. 
    The red lines denote the local disorder that breaks $G_s$ to the average subsystem symmetry $\mathcal{G}_s$. 
    The exact subsystem symmetry operators $S^h(k^{[y]})$ and average subsystem symmetry operators $S^h(g^{[y]})$ are indicated using blue and red colors, respectively.
    }
    \label{2D disordered cluster state}
\end{figure}

In the first case, the subsystem symmetry $\Tilde{G}_s$ of a pure state SSPT wave function $|\psi_{\mathrm{SSPT}}\rangle$ is considered as a product group $G_s\times K_s$. 
The total horizontal subsystem symmetry of $|\psi_{\mathrm{SSPT}}\rangle$ is denoted by $G_h=\prod_y \Tilde{G}_s^{[y]}$. 
To prepare an ASSPT phase $\rho_{\mathrm{ASSPT}}$, local disorders are introduced uniformly in the SSPT state, leading to breaking $G_s$ into an average $\mathcal{G}_s$ subsystem symmetry shown in Fig. \ref{2D disordered cluster state}. 
The total symmetry group of $\rho_{\mathrm{ASSPT}}$ is
\begin{equation}
    \mathcal{G}_h = \prod_y \mathcal{\Tilde{G}}_s^{[y]} = \prod_y (\mathcal{G}_s^{[y]} \times \mathcal{K}_s^{[y]}).
\end{equation}
where $\mathcal{K}_s^{[y]}$ remains the exact subsystem symmetry within the $y$-th row.

To illustrate, we break down the subsystem symmetry of 2D cluster state from $Z_2^\sigma$ to the average $\mathcal{Z}_2^\sigma$ symmetry by introducing random disorders $\{h_{x,y}^I=\pm 1\}$ at $\sigma^z_{x,y}$ degrees of freedom. 
Each disordered Hamiltonian $H_I$ is given by
\begin{align}
\begin{aligned}
    H_I &= H_0 + H_{\mathrm{dis}}^I\\
    H_0 &= -\sum_{x,y}\tau_{x,y}^x\sigma^z_{x,y}\sigma^z_{x-1,y}\sigma^z_{x,y+1}\sigma^z_{x-1,y+1}\\
    H_{\mathrm{dis}}^I &= -\sum_{x,y} h_{x,y}^I\sigma^z_{x,y}
\end{aligned}\label{eq 2d dis H}
\end{align}
The unique ground state of $H_I$ is a fixed domain configuration determined by the local disorder $h^I$
\begin{equation}
    |\psi_I\rangle = \sum_{x,y}|\sigma_{x,y}^z=h_{x,y}^I\rangle|\tau_{x,y}^x=\sigma^z_{x,y}\sigma^z_{x-1,y}\sigma^z_{x,y+1}\sigma^z_{x-1,y+1}\rangle.\label{eq disordered wf}
\end{equation}
This wave function can also be realized through imaginary time evolution from a 2D cluster state
\begin{equation}
|\psi_I\rangle=\lim_{\beta\rightarrow\infty}O^I(\beta)|\psi_{\mathrm{cluster}}\rangle,\ O^I(\beta) = e^{-\beta H_{\mathrm{dis}}^I},
\end{equation}
where $\beta$ is the imaginary time. 
The mixed state resulting from an equal-weight superposition of all disorder patterns $|\psi_I\rangle$ is given by
\begin{equation}
    \rho_{\mathrm{cluster}}=\sum_I \frac{1}{2^N}|\psi_I\rangle\langle\psi_I|.\label{eq density matrix}
\end{equation}
We note that $\mathcal{Z}_2^\tau$ remains to be the exact subsystem symmetry $S^h(k_\tau^{[y]})\rho_{\mathrm{cluster}}=\rho_{\mathrm{cluster}} [S^h(k_\tau^{[y]})]^\dagger=\rho_{\mathrm{cluster}}$. 
$\mathcal{Z}_2^\sigma$ is broken down to the average symmetry as $S^h(g_\sigma^{[y]})\rho_{\mathrm{cluster}}[S^h(g_\sigma^{[y]})]^\dagger = \rho_{\mathrm{cluster}}$. 
We represent $\rho_{\mathrm{cluster}}$ using a projected entangled-pair density operator (PEPDO) $(A_\sigma$, $A_\tau)$ based on the domain wall decoration structure. 
The $\sigma^z$ domain is mapped to the virtual space via the local tensor
\begin{equation}
\begin{tikzpicture}
    \draw[ultra thick,red] (0.75,0.1) -- (0.75,-0.25) (0.75,-0.75) -- (0.75,-1.1);
    \draw (-0.5,-0.5) node {$A_\sigma=$}
    (3.5,-0.51) node {$=\delta_{i_\sigma k j_1 \cdots j_4}\cdot\delta_{\overline{i}_\sigma k \overline{j}_1 \cdots \overline{j}_4}$};
    \draw[thick] 
    (0.25,-0.25) -- (1.25,-0.25) (0.5,-0.5) -- (1,0) (0.25,-0.75) -- (1.25,-0.75) (0.5,-1) -- (1,-0.5);
    \draw[densely dotted, very thick] (0.75,-0.25) -- (0.75,-0.75);

    \filldraw[red] (0.65,-0.35) -- (0.65,-0.15) -- (0.85,-0.15) -- (0.85,-0.35) -- cycle
    (0.65,-0.85) -- (0.65,-0.65) -- (0.85,-0.65) -- (0.85,-0.85) -- cycle;
\end{tikzpicture},
\end{equation}
which is composed of two rank-$6$ tensors
\begin{equation}
\begin{tikzpicture}
    \draw[ultra thick,red] (0.75,0.15) -- (0.75,-0.25)
    (4.75,-0.25) -- (4.75,-0.65);
    \draw[thick] (0.25,-0.25) -- (1.25,-0.25) (0.5,-0.5) -- (1,0) (4.25,-0.25) -- (5.25,-0.25) (4.5,-0.5) -- (5,0);
    \draw[very thick, densely dotted] (0.75,-0.65) -- (0.75,-0.25) (4.75,-0.25) -- (4.75,0.15);

    \filldraw[red] (0.65,-0.35) -- (0.65,-0.15) -- (0.85,-0.15) -- (0.85,-0.35) -- cycle
    (4.65,-0.35) -- (4.65,-0.15) -- (4.85,-0.15) -- (4.85,-0.35) -- cycle;
    \draw (2.5,-0.25) node {$=\delta_{i_\sigma k j_1 \cdots j_4}$}
    (6.5,-0.25) node {$=\delta_{\overline{i}_\sigma k \overline{j}_1 \cdots \overline{j}_4}$};
\end{tikzpicture}.
\end{equation}
Here, the dotted leg is labeled by the Kraus index $k\in\{0,1\}$. 
The local operator $A_\tau$ is constructed from the decoration map Eq. (\ref{eq decoration rule})
\begin{equation}
\begin{tikzpicture}
    \draw[ultra thick,blue]  (1,0) -- (1,0.5) (1,-0.75) -- (1,-1.25);
    \draw (-0.75,-0.4) node {$A_\tau=$};
    \draw[thick] (0,0) -- (2,0) (1.5,0.5) -- (0.5,-0.5) (0,-0.75) -- (2,-0.75) (1.5,-0.25) -- (0.5,-1.25);
    \draw[thick,white,fill=white] 
    (0.5,0) circle[radius=0.15]
    (1.5,0) circle[radius=0.15]
    (1.25,0.25) circle[radius=0.175]
    (0.75,-0.25) circle[radius=0.175]
    (0.5,-0.75) circle[radius=0.15]
    (1.5,-0.75) circle[radius=0.15]
    (1.25,-0.5) circle[radius=0.175]
    (0.75,-1) circle[radius=0.175];
    \draw (0.5,0) node {\small $H$}
    (1.5,0) node {\small $H$}
    (1.25,0.25) node {\small $H$}
    (0.75,-0.25) node {\small $H$}
    (0.5,-0.75) node {\small $H$}
    (1.5,-0.75) node {\small $H$}
    (1.25,-0.5) node {\small $H$}
    (0.75,-1) node {\small $H$};
    \filldraw[blue] (1,0) circle[radius=0.1]
    (1,-0.75) circle[radius=0.1];
\end{tikzpicture}.
\end{equation}
Applying the deformation in Eq. (\ref{eq deformation}) to the PEPDO structure, we obtain the local tensor $A$
\begin{equation}
\begin{tikzpicture}
    \draw[ultra thick,purplish] (0.7,0.25) -- (0.7,-1.25);
    \draw[thick] (0,-0.5) -- (1.5,-0.5) (0.25,-1) -- (1.25,0);
    \draw[thick,fill=white]
    (0.3,-0.3) -- (0.8,-0.3) -- (0.8,-0.8) -- (0.3,-0.8) -- cycle (0.3,-0.3) -- (0.55,-0.15) -- (1.05,-0.15) -- (1.05,-0.65) -- (0.8,-0.8) (1.05,-0.15) -- (0.8,-0.3);
    \draw[line width=0.075cm,white] (0.9,-0.5) -- (1.5,-0.5) (0.25,-1) -- (0.6,-0.65);
    \draw[line width=0.1cm,white] (0.7,0.25) -- (0.7,-0.25);
    \draw[ultra thick,purplish] (0.7,0.25) -- (0.7,-0.25);
    \draw[thick] (0.9,-0.5) -- (1.5,-0.5) (0.25,-1) -- (0.6,-0.65);
    
    \draw[ultra thick,blue](4.5,0.5) -- (4.5,1) (4.5,-0.5) -- (4.5,-1.125) (4.5,-1.225) -- (4.5,-1.5); \draw[ultra thick,red] (5,-2) -- (5,-1) (5,0) -- (5,0.5);
    \draw (2,-0.5) node {$=$};
    \draw[purplish] (0.75,0.5) node {$(i_\tau,i_\sigma)$} (0.75,-1.5) node {$(\overline{i}_\tau,\overline{i}_\sigma)$};
    \draw[blue] (4.5,1.25) node {$i_\tau$} (4.5,-1.75) node {$\overline{i}_\tau$};
    \draw[red] (5,0.75) node {$i_\sigma$} (5,-2.25) node {$\overline{i}_\sigma$};
    \draw[thick] (2.5,0) -- (3,0) (5,0) -- (5.5,0) 
    (3.25,-0.75) -- (3.5,-0.5) (4.5,0.5) -- (4.75,0.75) (3,0) -- (4.5,0.5) -- (5,0) -- (3.5,-0.5) -- (3,0);
    \draw[densely dotted, very thick] (5,-1) -- (5,0);
    \filldraw[thick,white,fill=white] (3.75,0.25) circle[radius=0.15]
    (4.75,0.25) circle[radius=0.15]
    (4.25,-0.25) circle[radius=0.15]
    (3.25,-0.25) circle[radius=0.15];
    \draw (3.75,0.25) node {\small $H$}
    (4.75,0.25) node {\small $H$}
    (4.25,-0.25) node {\small $H$}
    (3.25,-0.25) node {\small $H$};

    \draw[thick] (2.5,-1) -- (3,-1) (5,-1) -- (5.5,-1) 
    (3.25,-1.75) -- (3.5,-1.5) (4.5,-0.5) -- (4.75,-0.25) (3,-1) -- (4.5,-0.5) -- (5,-1) -- (3.5,-1.5) -- (3,-1);

    \filldraw[thick,white,fill=white] (3.75,-0.75) circle[radius=0.15]
    (4.75,-0.75) circle[radius=0.15]
    (4.25,-1.25) circle[radius=0.15]
    (3.25,-1.25) circle[radius=0.15];
    \draw (3.75,-0.75) node {\small $H$}
    (4.75,-0.75) node {\small $H$}
    (4.25,-1.25) node {\small $H$}
    (3.25,-1.25) node {\small $H$};

    \filldraw[blue] (4.5,0.5) circle[radius=0.1] (4.5,-0.5) circle[radius=0.1];

    \filldraw[red] (4.9,-0.1) -- (4.9,0.1) -- (5.1,0.1) -- (5.1,-0.1) -- cycle
    (4.9,-0.9) -- (4.9,-1.1) -- (5.1,-1.1) -- (5.1,-0.9) -- cycle;
\end{tikzpicture}.
\end{equation}

Through the Choi-Jamiolkowski map, $\rho_{\mathrm{cluster}}$ is equivalent to a Choi state $|\rho_{\mathrm{cluster}}\rangle\rangle$ within the double Hilbert space $\mathcal{H}_l\otimes\mathcal{H}_r$.
The exact and average subsystem symmetries are transformed into the $Z_{2,d}^{\sigma,[y]}\times Z_{2,l}^{\tau,[y]}\times Z_{2,r}^{\tau,[y]}$ subsystem symmetry of the Choi state. 
We use subscripts $l$ and $r$ to denote symmetry operators acting on the spaces $\mathcal{H}_l$ and $\mathcal{H}_r$, respectively, and subscript $d$ for diagonal operators in $\mathcal{H}_l\otimes\mathcal{H}_r$. 
The tensor equation corresponding to the exact subsystem symmetry operator is given by
\begin{equation}
\begin{tikzpicture}
    \draw (-2.5,-0.4) node {$k_{\tau,l}^{[y]}$:};
    \draw[ultra thick,blue](0,0.5) -- (0,1) (0,-0.5) -- (0,-1.125) (0,-1.225) -- (0,-1.5); \draw[ultra thick,red] (0.5,-2) -- (0.5,-1) (0.5,0) -- (0.5,0.5);
    \draw[blue] (0,1.25) node {$\tau^x$} (0,-1.75) node {$I$};
    \draw[red] (0.5,0.75) node {$I$} (0.5,-2.25) node {$I$};
    \draw[thick] (-2,0) -- (-1.5,0) (0.5,0) -- (1,0) 
    (-1.25,-0.75) -- (-1,-0.5) (0,0.5) -- (0.25,0.75) (-1.5,0) -- (0,0.5) -- (0.5,0) -- (-1,-0.5) -- (-1.5,0);
    \draw[densely dotted, very thick] (0.5,-1) -- (0.5,0);
    \filldraw[thick,white,fill=white] (-0.75,0.25) circle[radius=0.15]
    (0.25,0.25) circle[radius=0.15]
    (-0.25,-0.25) circle[radius=0.15]
    (-1.25,-0.25) circle[radius=0.15];
    \draw (-0.75,0.25) node {\small $H$}
    (0.25,0.25) node {\small $H$}
    (-0.25,-0.25) node {\small $H$}
    (-1.25,-0.25) node {\small $H$};

    \draw[thick] (-2,-1) -- (-1.5,-1) (0.5,-1) -- (1,-1) 
    (-1.25,-1.75) -- (-1,-1.5) (0,-0.5) -- (0.25,-0.25) (-1.5,-1) -- (0,-0.5) -- (0.5,-1) -- (-1,-1.5) -- (-1.5,-1);

    \filldraw[thick,white,fill=white] (-0.75,-0.75) circle[radius=0.15]
    (0.25,-0.75) circle[radius=0.15]
    (-0.25,-1.25) circle[radius=0.15]
    (-1.25,-1.25) circle[radius=0.15];
    \draw (-0.75,-0.75) node {\small $H$}
    (0.25,-0.75) node {\small $H$}
    (-0.25,-1.25) node {\small $H$}
    (-1.25,-1.25) node {\small $H$};

    \filldraw[blue] (0,0.5) circle[radius=0.1] (0,-0.5) circle[radius=0.1];

    \filldraw[red] (0.4,-0.1) -- (0.4,0.1) -- (0.6,0.1) -- (0.6,-0.1) -- cycle
    (0.4,-0.9) -- (0.4,-1.1) -- (0.6,-1.1) -- (0.6,-0.9) -- cycle;

    \draw (4.95,1) node {$X$}
    (5.75,0) node {$Z$}
    (2.25,0) node {$Z$};
    \draw[ultra thick,blue](4.5,0.5) -- (4.5,1) (4.5,-0.5) -- (4.5,-1.125) (4.5,-1.225) -- (4.5,-1.5); \draw[ultra thick,red] (5,-2) -- (5,-1) (5,0) -- (5,0.5);
    \draw (1.5,-0.5) node {$=$};
    \draw[thick] (2.5,0) -- (3,0) (5,0) -- (5.5,0) 
    (3.25,-0.75) -- (3.5,-0.5) (4.5,0.5) -- (4.75,0.75) (3,0) -- (4.5,0.5) -- (5,0) -- (3.5,-0.5) -- (3,0);
    \draw[densely dotted, very thick] (5,-1) -- (5,0);
    \filldraw[thick,white,fill=white] (3.75,0.25) circle[radius=0.15]
    (4.75,0.25) circle[radius=0.15]
    (4.25,-0.25) circle[radius=0.15]
    (3.25,-0.25) circle[radius=0.15];
    \draw (3.75,0.25) node {\small $H$}
    (4.75,0.25) node {\small $H$}
    (4.25,-0.25) node {\small $H$}
    (3.25,-0.25) node {\small $H$};

    \draw[thick] (2.5,-1) -- (3,-1) (5,-1) -- (5.5,-1) 
    (3.25,-1.75) -- (3.5,-1.5) (4.5,-0.5) -- (4.75,-0.25) (3,-1) -- (4.5,-0.5) -- (5,-1) -- (3.5,-1.5) -- (3,-1);

    \filldraw[thick,white,fill=white] (3.75,-0.75) circle[radius=0.15]
    (4.75,-0.75) circle[radius=0.15]
    (4.25,-1.25) circle[radius=0.15]
    (3.25,-1.25) circle[radius=0.15];
    \draw (3.75,-0.75) node {\small $H$}
    (4.75,-0.75) node {\small $H$}
    (4.25,-1.25) node {\small $H$}
    (3.25,-1.25) node {\small $H$};

    \filldraw[blue] (4.5,0.5) circle[radius=0.1] (4.5,-0.5) circle[radius=0.1];

    \filldraw[red] (4.9,-0.1) -- (4.9,0.1) -- (5.1,0.1) -- (5.1,-0.1) -- cycle
    (4.9,-0.9) -- (4.9,-1.1) -- (5.1,-1.1) -- (5.1,-0.9) -- cycle;
\end{tikzpicture}.
\end{equation}
The virtual $X$ operator transforms the adjacent tensor by
\begin{equation}
\begin{tikzpicture}
    \draw[ultra thick,blue](0,0.5) -- (0,1) (0,-0.5) -- (0,-1.125) (0,-1.225) -- (0,-1.5); \draw[ultra thick,red] (0.5,-2) -- (0.5,-1) (0.5,0) -- (0.5,0.5);
    \draw (-1.35,-0.75) node {$X$};
    \draw[thick] (-2,0) -- (-1.5,0) (0.5,0) -- (1,0) 
    (-1.25,-0.75) -- (-1,-0.5) (0,0.5) -- (0.25,0.75) (-1.5,0) -- (0,0.5) -- (0.5,0) -- (-1,-0.5) -- (-1.5,0);
    \draw[densely dotted, very thick] (0.5,-1) -- (0.5,0);
    \filldraw[thick,white,fill=white] (-0.75,0.25) circle[radius=0.15]
    (0.25,0.25) circle[radius=0.15]
    (-0.25,-0.25) circle[radius=0.15]
    (-1.25,-0.25) circle[radius=0.15];
    \draw (-0.75,0.25) node {\small $H$}
    (0.25,0.25) node {\small $H$}
    (-0.25,-0.25) node {\small $H$}
    (-1.25,-0.25) node {\small $H$};

    \draw[thick] (-2,-1) -- (-1.5,-1) (0.5,-1) -- (1,-1) 
    (-1.25,-1.75) -- (-1,-1.5) (0,-0.5) -- (0.25,-0.25) (-1.5,-1) -- (0,-0.5) -- (0.5,-1) -- (-1,-1.5) -- (-1.5,-1);

    \filldraw[thick,white,fill=white] (-0.75,-0.75) circle[radius=0.15]
    (0.25,-0.75) circle[radius=0.15]
    (-0.25,-1.25) circle[radius=0.15]
    (-1.25,-1.25) circle[radius=0.15];
    \draw (-0.75,-0.75) node {\small $H$}
    (0.25,-0.75) node {\small $H$}
    (-0.25,-1.25) node {\small $H$}
    (-1.25,-1.25) node {\small $H$};

    \filldraw[blue] (0,0.5) circle[radius=0.1] (0,-0.5) circle[radius=0.1];

    \filldraw[red] (0.4,-0.1) -- (0.4,0.1) -- (0.6,0.1) -- (0.6,-0.1) -- cycle
    (0.4,-0.9) -- (0.4,-1.1) -- (0.6,-1.1) -- (0.6,-0.9) -- cycle;

    \draw
    (5.75,0) node {$Z$}
    (2.25,0) node {$Z$};
    \draw[ultra thick,blue](4.5,0.5) -- (4.5,1) (4.5,-0.5) -- (4.5,-1.125) (4.5,-1.225) -- (4.5,-1.5); \draw[ultra thick,red] (5,-2) -- (5,-1) (5,0) -- (5,0.5);
    \draw (1.5,-0.5) node {$=$};
    \draw[thick] (2.5,0) -- (3,0) (5,0) -- (5.5,0) 
    (3.25,-0.75) -- (3.5,-0.5) (4.5,0.5) -- (4.75,0.75) (3,0) -- (4.5,0.5) -- (5,0) -- (3.5,-0.5) -- (3,0);
    \draw[densely dotted, very thick] (5,-1) -- (5,0);
    \filldraw[thick,white,fill=white] (3.75,0.25) circle[radius=0.15]
    (4.75,0.25) circle[radius=0.15]
    (4.25,-0.25) circle[radius=0.15]
    (3.25,-0.25) circle[radius=0.15];
    \draw (3.75,0.25) node {\small $H$}
    (4.75,0.25) node {\small $H$}
    (4.25,-0.25) node {\small $H$}
    (3.25,-0.25) node {\small $H$};

    \draw[thick] (2.5,-1) -- (3,-1) (5,-1) -- (5.5,-1) 
    (3.25,-1.75) -- (3.5,-1.5) (4.5,-0.5) -- (4.75,-0.25) (3,-1) -- (4.5,-0.5) -- (5,-1) -- (3.5,-1.5) -- (3,-1);

    \filldraw[thick,white,fill=white] (3.75,-0.75) circle[radius=0.15]
    (4.75,-0.75) circle[radius=0.15]
    (4.25,-1.25) circle[radius=0.15]
    (3.25,-1.25) circle[radius=0.15];
    \draw (3.75,-0.75) node {\small $H$}
    (4.75,-0.75) node {\small $H$}
    (4.25,-1.25) node {\small $H$}
    (3.25,-1.25) node {\small $H$};

    \filldraw[blue] (4.5,0.5) circle[radius=0.1] (4.5,-0.5) circle[radius=0.1];

    \filldraw[red] (4.9,-0.1) -- (4.9,0.1) -- (5.1,0.1) -- (5.1,-0.1) -- cycle
    (4.9,-0.9) -- (4.9,-1.1) -- (5.1,-1.1) -- (5.1,-0.9) -- cycle;
\end{tikzpicture}.
\end{equation}
The boundary symmetry operators for exact symmetry are summarized as
\begin{align}
\begin{aligned}
    &W^{\mathrm{Right}}(k_{\tau,l}^{[y]}) = Z_{L_x,y}Z_{L_x,y+1}\otimes I,\\  &W^{\mathrm{Right}}(k_{\tau,r}^{[y]}) = I\otimes Z_{L_x,y}Z_{L_x,y+1}.
\end{aligned}\label{eq disordered cluster w}
\end{align}
The tensor equation for the average subsystem symmetry is
\begin{equation}
\begin{tikzpicture}
    \draw (-2.5,-0.4) node {$g_{\sigma,d}^{[y]}$:};
    \draw[ultra thick,blue](0,0.5) -- (0,1) (0,-0.5) -- (0,-1.125) (0,-1.225) -- (0,-1.5); \draw[ultra thick,red] (0.5,-2) -- (0.5,-1) (0.5,0) -- (0.5,0.5);
    \draw[blue] (0,1.25) node {$I$} (0,-1.75) node {$I$};
    \draw[red] (0.6,0.75) node {$\sigma^x$} (0.6,-2.25) node {$\sigma^x$};
    \draw[thick] (-2,0) -- (-1.5,0) (0.5,0) -- (1,0) 
    (-1.25,-0.75) -- (-1,-0.5) (0,0.5) -- (0.25,0.75) (-1.5,0) -- (0,0.5) -- (0.5,0) -- (-1,-0.5) -- (-1.5,0);
    \draw[densely dotted, very thick] (0.5,-1) -- (0.5,0);
    \filldraw[thick,white,fill=white] (-0.75,0.25) circle[radius=0.15]
    (0.25,0.25) circle[radius=0.15]
    (-0.25,-0.25) circle[radius=0.15]
    (-1.25,-0.25) circle[radius=0.15];
    \draw (-0.75,0.25) node {\small $H$}
    (0.25,0.25) node {\small $H$}
    (-0.25,-0.25) node {\small $H$}
    (-1.25,-0.25) node {\small $H$};

    \draw[thick] (-2,-1) -- (-1.5,-1) (0.5,-1) -- (1,-1) 
    (-1.25,-1.75) -- (-1,-1.5) (0,-0.5) -- (0.25,-0.25) (-1.5,-1) -- (0,-0.5) -- (0.5,-1) -- (-1,-1.5) -- (-1.5,-1);

    \filldraw[thick,white,fill=white] (-0.75,-0.75) circle[radius=0.15]
    (0.25,-0.75) circle[radius=0.15]
    (-0.25,-1.25) circle[radius=0.15]
    (-1.25,-1.25) circle[radius=0.15];
    \draw (-0.75,-0.75) node {\small $H$}
    (0.25,-0.75) node {\small $H$}
    (-0.25,-1.25) node {\small $H$}
    (-1.25,-1.25) node {\small $H$};

    \filldraw[blue] (0,0.5) circle[radius=0.1] (0,-0.5) circle[radius=0.1];

    \filldraw[red] (0.4,-0.1) -- (0.4,0.1) -- (0.6,0.1) -- (0.6,-0.1) -- cycle
    (0.4,-0.9) -- (0.4,-1.1) -- (0.6,-1.1) -- (0.6,-0.9) -- cycle;

    \draw (5.75,0) node {$X$}
    (2.25,0) node {$X$}
    (5.75,-1) node {$X$}
    (2.25,-1) node {$X$};
    \draw[ultra thick,blue](4.5,0.5) -- (4.5,1) (4.5,-0.5) -- (4.5,-1.125) (4.5,-1.225) -- (4.5,-1.5); \draw[ultra thick,red] (5,-2) -- (5,-1) (5,0) -- (5,0.5);
    \draw (1.5,-0.5) node {$=$};
    \draw[thick] (2.5,0) -- (3,0) (5,0) -- (5.5,0) 
    (3.25,-0.75) -- (3.5,-0.5) (4.5,0.5) -- (4.75,0.75) (3,0) -- (4.5,0.5) -- (5,0) -- (3.5,-0.5) -- (3,0);
    \draw[densely dotted, very thick] (5,-1) -- (5,0);
    \filldraw[thick,white,fill=white] (3.75,0.25) circle[radius=0.15]
    (4.75,0.25) circle[radius=0.15]
    (4.25,-0.25) circle[radius=0.15]
    (3.25,-0.25) circle[radius=0.15];
    \draw (3.75,0.25) node {\small $H$}
    (4.75,0.25) node {\small $H$}
    (4.25,-0.25) node {\small $H$}
    (3.25,-0.25) node {\small $H$};

    \draw[thick] (2.5,-1) -- (3,-1) (5,-1) -- (5.5,-1) 
    (3.25,-1.75) -- (3.5,-1.5) (4.5,-0.5) -- (4.75,-0.25) (3,-1) -- (4.5,-0.5) -- (5,-1) -- (3.5,-1.5) -- (3,-1);

    \filldraw[thick,white,fill=white] (3.75,-0.75) circle[radius=0.15]
    (4.75,-0.75) circle[radius=0.15]
    (4.25,-1.25) circle[radius=0.15]
    (3.25,-1.25) circle[radius=0.15];
    \draw (3.75,-0.75) node {\small $H$}
    (4.75,-0.75) node {\small $H$}
    (4.25,-1.25) node {\small $H$}
    (3.25,-1.25) node {\small $H$};

    \filldraw[blue] (4.5,0.5) circle[radius=0.1] (4.5,-0.5) circle[radius=0.1];

    \filldraw[red] (4.9,-0.1) -- (4.9,0.1) -- (5.1,0.1) -- (5.1,-0.1) -- cycle
    (4.9,-0.9) -- (4.9,-1.1) -- (5.1,-1.1) -- (5.1,-0.9) -- cycle;
    \draw[white,fill=white] (5,-0.3) circle[radius=0.15] (5,-0.7) circle[radius=0.15];
    \draw (5,-0.3) node {$X$} (5,-0.7) node {$X$};
\end{tikzpicture}.
\end{equation}
The corresponding boundary operator has the form
\begin{align}
\begin{aligned}
    W^{\mathrm{Right}}(g_{\sigma,d}^{[y]}) = X_{L_x,y}\otimes X_{L_x,y}.
\end{aligned}\label{eq disordered cluster w2}
\end{align}
Therefore, the mixed anomalies between $S^h(k^{[y]})$ and $S^h(g^{[y+1]})$ are
\begin{align}
\begin{aligned}
    &\{W^{\mathrm{Right}}(k_{\tau,l}^{[y]}),W^{\mathrm{Right}}(g_{\sigma,d}^{[y+1]})\} =0,\\
    &\{W^{\mathrm{Right}}(k_{\tau,r}^{[y]}),W^{\mathrm{Right}}(g_{\sigma,d}^{[y+1]})\} = 0,
\end{aligned}
\end{align}
which is encoded in the graph
\begin{equation}
\begin{tikzpicture}
    \draw (-1,-0.35) node {$k_{\tau,l}^{[y]}$}
    (0,-0.35) node {$g_{\sigma,d}^{[y]}$}
    (1,-0.35) node {$k_{\tau,r}^{[y]}$}
    (-1,1.35) node {$k_{\tau,l}^{[y+1]}$}
    (0,1.35) node {$g_{\sigma,d}^{[y+1]}$}
    (1,1.35) node {$k_{\tau,r}^{[y+1]}$};
    \filldraw[red] 
    (0,0) circle[radius=0.08] 
    (0,1) circle[radius=0.08];
    \filldraw[blue] 
    (-1,0) circle[radius=0.08]
    (1,0) circle[radius=0.08]
    (-1,1) circle[radius=0.08]
    (1,1) circle[radius=0.08];
    \draw[very thick] (0.15,0) -- (0.85,0) (-0.15,0) -- (-0.85,0) (0.15,1) -- (0.85,1) (-0.15,1) -- (-0.85,1) (-0.9,0.1) -- (-0.1,0.9) (0.9,0.1) -- (0.1,0.9);
\end{tikzpicture}.
\end{equation}

\subsubsection{Alternating average subsystem symmetry}
\begin{figure}
    \centering
    \begin{tikzpicture}
        \draw[thick] 
        (0.5,0) -- (6,0)
        (0.5,1.5) -- (6,1.5)
        (0.5,3) -- (6,3)
        (1,-0.5) -- (1,3.5)
        (2.5,-0.5) -- (2.5,3.5)
        (4,-0.5) -- (4,3.5)
        (5.5,-0.5) -- (5.5,3.5);
        \draw[red,very thick] 
        (0.7,0.3) -- (1,0)
        (0.7,3.3) -- (1,3) 
        (2.2,0.3) -- (2.5,0) 
        (2.2,3.3) -- (2.5,3) 
        (3.7,0.3) -- (4,0) 
        (3.7,3.3) -- (4,3) 
        (5.2,0.3) -- (5.5,0) 
        (5.2,3.3) -- (5.5,3);
        \draw[blue,very thick] 
        (0.7,1.8) -- (1,1.5) 
        (2.2,1.8) -- (2.5,1.5) 
        (3.7,1.8) -- (4,1.5) 
        (5.2,1.8) -- (5.5,1.5);
        \filldraw[red,opacity=0.2] (0.65,-0.1) rectangle ++(5.1,0.45)
        (0.65,2.9) rectangle ++(5.1,0.45);
        \filldraw[blue,opacity=0.2] (0.65,1.4) rectangle ++(5.1,0.45);
        \draw[blue] (6.8,1.5) node {$S^h(k^{[2y+1]})$};
        \draw[red] (6.8,3) node {$S^h(g^{[2y+2]})$} (6.7,0) node {$S^h(g^{[2y]})$};
    \end{tikzpicture}
    \caption{
    $\mathcal{G}_s$ ASSPT phase with alternating disorders. 
    The blue and red squares denote the exact subsystem symmetry operators $S^h(k^{[2y+1]})$ and the average subsystem symmetry operators $S^h(g^{[2y]})$, respectively. 
    }
    \label{fig stripe disorder}
\end{figure}

In the second case, we construct the ASSPT phase by breaking the subsystem symmetry in an alternating way shown in Fig. \ref{fig stripe disorder}. 
Starting from a $G_s$ SSPT wave function $|\psi_{\mathrm{SSPT}}\rangle$, we introduce impurity interactions alternately in even rows to break the horizontal subsystem symmetry group $G_s$ down to $\mathcal{G}_s$. 
As a result, the total symmetry group $G_h=\prod_y G_s^{[y]}$ of the SSPT wave function is reduced to
\begin{equation}
    \mathcal{G}_h=\prod_y (\mathcal{G}_s^{[2y]}\times\mathcal{K}_s^{[2y+1]})
\end{equation}
for the ASSPT density matrix $\rho_{\mathrm{ASSPT}}$. 
Here, $\mathcal{K}_s$ denotes the exact symmetry of the subsystem in odd rows.

We take $G_s=Z_2$ as an example.
Our initial state is an intrinsic $Z_2$ SSPT wave function $\rho_{\mathrm{SSPT}}=|\psi_{Z_2}\rangle\langle\psi_{Z_2}|$, with a tensor network representation given by Eq. (\ref{eq Z2 wavefunction}). 
By introducing random projections
\begin{equation}
O^I(\beta)=\prod_{x,y}e^{-\beta h^I_{x,2y}Z_{x,2y}},\ h^I_{x,2y}=\pm 1
\end{equation}
to even rows, we explicitly break the $Z_2^{[2y]}$ subsystem symmetry
\begin{equation}
    |\psi_{Z_2,I}\rangle = \lim_{\beta\rightarrow\infty} O^I(\beta)|\psi_{Z_2}\rangle.
\end{equation}
To restore the average subsystem $\mathcal{Z}_2^{[2y]}$ symmetry, we consider the statistical ensemble consisting of all possible disorder patterns $\{h^I_{x,2y}=\pm 1\}$
\begin{equation}
    \rho_{\mathrm{ASSPT}} = \sum_I \frac{1}{2^N} |\psi_{Z_2,I}\rangle\langle\psi_{Z_2,I}|,
\end{equation}
where $\frac{1}{2^N}$ represents the uniform statistical probability. 
We construct the PEPDO representation of the alternating $Z_2$ ASSPT density matrix $\rho_{\mathrm{ASSPT}}$ based on the PEPS representation of the $Z_2$ SSPT wave function.
The local tensors for even rows are fixed by projection operators
\begin{equation}
\begin{tikzpicture}
    \draw[ultra thick,red] (0.75,0.2) -- (0.75,-0.25) (0.75,-0.75) -- (0.75,-1.2);
    \draw (-0.5,-0.5) node {$A_{\mathrm{even}}=$}
    (2.2,-0.51) node {, where}
    (4.75,-0.5) node {$=T_{Z_2}^i\delta_{ik}$.};
    \draw (3.5,0.2) node {$i$} (3.5,-1.15) node {$k$};
    \draw[thick] 
    (0.25,-0.25) -- (1.25,-0.25) (0.5,-0.5) -- (1,0) (0.25,-0.75) -- (1.25,-0.75) (0.5,-1) -- (1,-0.5);
    \draw[densely dotted, very thick] (0.75,-0.25) -- (0.75,-0.75);

    \draw[ultra thick,red] (3.5,-0.1) -- (3.5,-0.5);
    \draw[densely dotted, very thick] (3.5,-0.5) -- (3.5,-0.9);
    \draw[thick] 
    (3,-0.5) -- (4,-0.5) (3.25,-0.75) -- (3.75,-0.25);
\end{tikzpicture}
\end{equation}
Here, $T$ is the local tensor defined in Eq. (\ref{eq Z2 wavefunction}).
We use subscripts $i$ and $k$ to denote physical and Kraus indices. 
The spin configurations within odd rows are decorated between two adjacent subsystems via the local tensor
\begin{equation}
\begin{tikzpicture}
    \draw[ultra thick,blue] (0.75,0.2) -- (0.75,-0.25) (0.75,-0.75) -- (0.75,-1.2);
    \draw (-0.5,-0.5) node {$A_{\mathrm{odd}}=$}
    (2.2,-0.51) node {, where}
    (4.6,-0.5) node {$=T_{Z_2}$.};
    \draw[thick] 
    (0.25,-0.25) -- (1.25,-0.25) (0.5,-0.5) -- (1,0) (0.25,-0.75) -- (1.25,-0.75) (0.5,-1) -- (1,-0.5);

    \draw[ultra thick,blue] (3.5,-0.1) -- (3.5,-0.5);
    \draw[thick] 
    (3,-0.5) -- (4,-0.5) (3.25,-0.75) -- (3.75,-0.25);
\end{tikzpicture}
\end{equation} 
The exact symmetry operators $S^h(k^{[2y+1]})$ and average symmetry operators $S^h(g^{[2y]})$ of the system are defined on the subsystems with odd and even indices
\begin{align}
\begin{aligned}
    S^h(k^{[2y+1]}) = \prod_x X_{x,2y+1},\quad
    S^h(g^{[2y]}) = \prod_x X_{x,2y}.
\end{aligned}
\end{align}
The local tensor equation for the average subsystem operator in the $2y$-th row is
\begin{equation}
\begin{tikzpicture}
    \draw (-0.5,-0.4) node {$g^{[2y]}_d$:};
    \draw[ultra thick,red] (0.75,0.35) -- (0.75,0) (0.75,-1) -- (0.75,-1.35)
    (2.75,0.45) -- (2.75,0) (2.75,-1) -- (2.75,-1.45);
    \draw (1.6,-0.51) node {$=$}
    (2.1,0) node {$X$}
    (3.4,0) node {$X$}
    (3.1,0.3) node {$X$}
    (2.1,-1) node {$X$}
    (3.4,-1) node {$X$}
    (3.1,-0.7) node {$X$};
    \draw[red]
    (0.75,0.55) node {$X$}
    (0.75,-1.55) node {$X$};
    \draw[thick] 
    (0.25,0) -- (1.25,0) (0.5,-0.25) -- (1,0.25) (0.25,-1) -- (1.25,-1) (0.5,-1.25) -- (1,-0.75)    
    (2.25,0) -- (3.25,0) (2.5,-0.25) -- (3,0.25) (2.25,-1) -- (3.25,-1) (2.5,-1.25) -- (3,-0.75);
    \draw[densely dotted, very thick] (0.75,0) -- (0.75,-1) (2.75,0) -- (2.75,-1);
    \draw[white,fill=white] (2.75,-0.3) circle[radius=0.15] (2.75,-0.7) circle[radius=0.15];
    \draw (2.75,-0.3) node {$X$} (2.75,-0.7) node {$X$};
\end{tikzpicture}.
\end{equation}
The $X$ operator acting within the vertical virtual space flips $A_{\mathrm{odd}}$ in the $(2y+1)$-th row
\begin{equation}
\begin{tikzpicture}
    \draw[ultra thick,blue] (0.75,0.2) -- (0.75,-0.25) (0.75,-0.75) -- (0.75,-1.2)
    (2.75,0.2) -- (2.75,-0.25) (2.75,-0.75) -- (2.75,-1.2);
    \draw (1.6,-0.51) node {$=$}
    (2.1,-0.25) node {$Z$}
    (3.4,-0.25) node {$Z$}
    (0.4,-0.55) node {$X$}
    (2.1,-0.75) node {$Z$}
    (3.4,-0.75) node {$Z$}
    (0.4,-1.05) node {$X$};
    \draw[thick] 
    (0.25,-0.25) -- (1.25,-0.25) (0.5,-0.5) -- (1,0) (0.25,-0.75) -- (1.25,-0.75) (0.5,-1) -- (1,-0.5)    
    (2.25,-0.25) -- (3.25,-0.25) (2.5,-0.5) -- (3,0) (2.25,-0.75) -- (3.25,-0.75) (2.5,-1) -- (3,-0.5);
\end{tikzpicture}.
\end{equation}
The tensor equations for exact subsystem symmetry transformations are
\begin{equation}
\begin{tikzpicture}
    \draw (-0.75,-0.4) node {$k_l^{[2y+1]}$:}
    (-0.75,-2.4) node {$k_r^{[2y+1]}$:};
    \foreach \y in {0, -2}{
    \draw[ultra thick,blue] (0.75,0.1+\y) -- (0.75,-0.25+\y) (0.75,-0.75+\y) -- (0.75,-1.1+\y)
    (2.75,0.2+\y) -- (2.75,-0.25+\y) (2.75,-0.75+\y) -- (2.75,-1.2+\y);
    \draw[thick] 
    (0.25,-0.25+\y) -- (1.25,-0.25+\y) (0.5,-0.5+\y) -- (1,0+\y) (0.25,-0.75+\y) -- (1.25,-0.75+\y) (0.5,-1+\y) -- (1,-0.5+\y)    
    (2.25,-0.25+\y) -- (3.25,-0.25+\y) (2.5,-0.5+\y) -- (3,0+\y) (2.25,-0.75+\y) -- (3.25,-0.75+\y) (2.5,-1+\y) -- (3,-0.5+\y);
    \draw (1.6,-0.51+\y) node {$=$};}
    \draw 
    (2.1,-0.25) node {$X$}
    (3.4,-0.25) node {$X$}
    (3.1,0.05) node {$X$}
    (2.1,-2.75) node {$X$}
    (3.4,-2.75) node {$X$}
    (3.1,-2.5) node {$X$}
    (3.75,-0.51) node {,}
    (3.75,-2.51) node {,};
    \draw[blue]
    (0.75,0.3) node {$X$}
    (0.75,-3.3) node {$X$};
\end{tikzpicture}
\end{equation}
which transforms $A_{\mathrm{even}}$ in the $(2y+2)$-th row by
\begin{equation}
\begin{tikzpicture}
    \foreach \x in {0, 4}{
    \draw[ultra thick,red] (\x+0.75,0.2) -- (\x+0.75,-0.25) (\x+0.75,-0.75) -- (\x+0.75,-1.2)
    (\x+2.75,0.2) -- (\x+2.75,-0.25) (\x+2.75,-0.75) -- (\x+2.75,-1.2);
    \draw (\x+1.6,-0.51) node {$=$};
    \draw[thick] 
    (\x+0.25,-0.25) -- (\x+1.25,-0.25) (\x+0.5,-0.5) -- (\x+1,0) (\x+0.25,-0.75) -- (\x+1.25,-0.75) (\x+0.5,-1) -- (\x+1,-0.5)    
    (\x+2.25,-0.25) -- (\x+3.25,-0.25) (\x+2.5,-0.5) -- (\x+3,0) (\x+2.25,-0.75) -- (\x+3.25,-0.75) (\x+2.5,-1) -- (\x+3,-0.5);
    \draw[densely dotted, very thick] (\x+0.75,-0.25) -- (\x+0.75,-0.75) (\x+2.75,-0.25) -- (\x+2.75,-0.75);
    }
    \draw (2.1,-0.25) node {$Z$}
    (3.4,-0.25) node {$Z$}
    (0.4,-0.55) node {$X$}
    (6.1,-0.75) node {$Z$}
    (7.4,-0.75) node {$Z$}
    (4.4,-1.05) node {$X$}
    (3.75,-0.51) node {,}
    (7.75,-0.51) node {.};
\end{tikzpicture}
\end{equation}
By deforming the density matrix into the double Hilbert space, we obtain a double state $|\rho_{\mathrm{ASSPT}}\rangle\rangle$ protected by the group $G_h=\prod_y(Z_{2,d}^{[2y]}\times Z_{2,l}^{[2y+1]}\times Z_{2,r}^{[2y+1]})$. 
According to the tensor equations above, the boundary operators acting on $|\rho_{\mathrm{ASSPT}}\rangle\rangle$ have the form
\begin{align}
\begin{aligned}
    &W^{\mathrm{Right}}(k_l^{[2y+1]}) = X_{L_x,2y+1}Z_{L_x,2y+2}\otimes I,\\ &W^{\mathrm{Right}}(k_r^{[2y+1]}) = I\otimes X_{L_x,2y+1}Z_{L_x,2y+2},\\ &W^{\mathrm{Right}}(g_d^{[2y]}) = X_{L_x,2y}Z_{L_x,2y+1}\otimes X_{L_x,2y}Z_{L_x,2y+1}
\end{aligned}
\end{align}
The mixed-state anomalies between $S^h(k^{[2y+1]})$ and $S^h(g^{[2y]}),S^h(g^{[2y+2]})$ are given by
\begin{align}
\begin{aligned}
    &\phi(k_l^{[2y+1]},g_d^{[2y]}) = \phi(k_r^{[2y+1]},g_d^{[2y]}) = -1,\\
    &\phi(k_l^{[2y+1]},g_d^{[2y+2]}) = \phi(k_r^{[2y+1]},g_d^{[2y+2]}) = -1,
\end{aligned}
\end{align}
which is represented by the graph
\begin{equation}
\begin{tikzpicture}
    \draw[very thick] (-0.075,0.15) -- (-0.425,0.85)
    (0.075,0.15) -- (0.425,0.85)
    (-0.075,1.85) -- (-0.425,1.15)
    (0.075,1.85) -- (0.425,1.15);
    \filldraw[red] (0,0) circle[radius=0.075]
    (0,2) circle[radius=0.075];
    \filldraw[blue] (-0.5,1) circle[radius=0.075] (0.5,1) circle[radius=0.075];
    \draw (-1.2,1) node {$k_l^{[2y+1]}$}
    (1.2,1) node {$k_r^{[2y+1]}$}
    (-0.5,0) node {$g_d^{[2y]}$}
    (-0.65,2) node {$g_d^{[2y+2]}$};
\end{tikzpicture}.
\end{equation}
Although the subsystem symmetry in the even rows is fully broken down to average symmetry, the $\mathcal{K}_s^{[2y+1]}$ subsystem symmetry remains exact. 
This leads to a mixed anomaly between the subsystem symmetries of odd and even rows. 
As a result, the system still exhibits nontrivial boundary states.

\subsection{Mixed-state anomaly detection}\label{mixed-state anomaly detection}
\begin{figure}
    \centering
    \includegraphics[width = 0.95\linewidth]{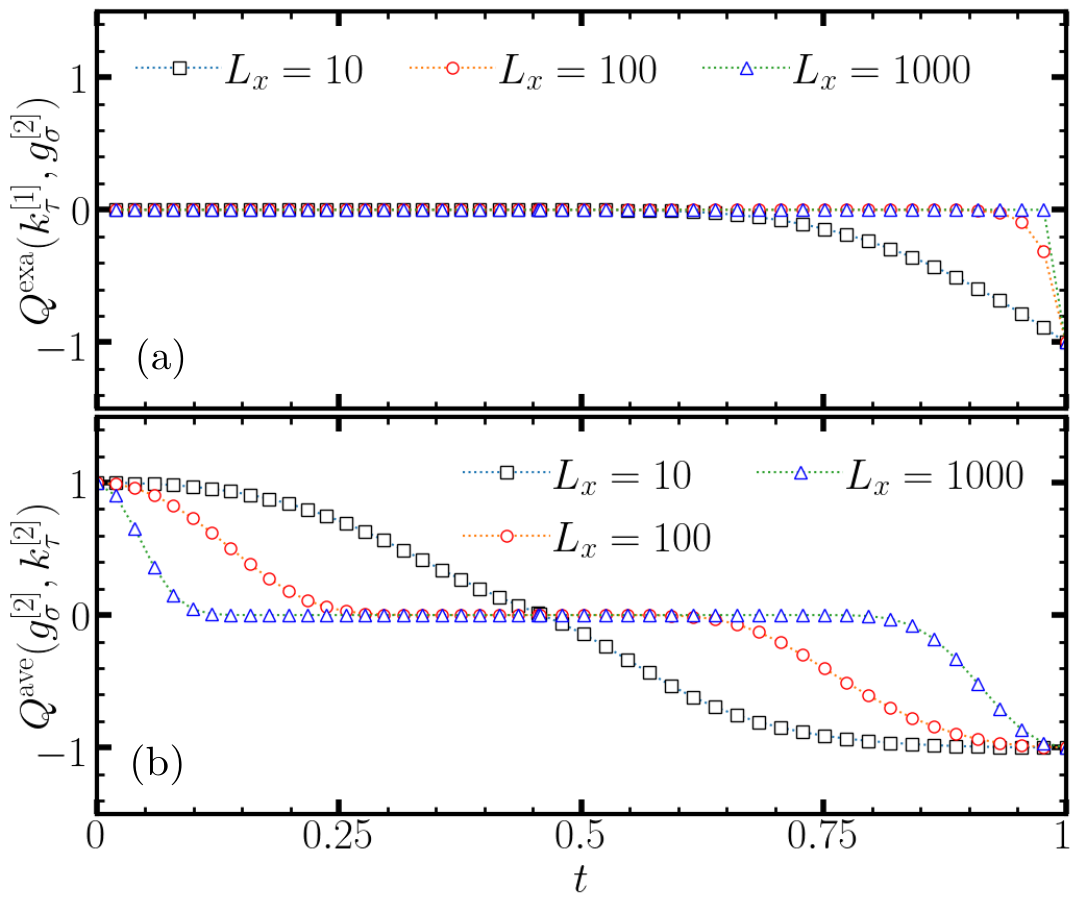}
    \caption{
    Mixed-state anomaly detection of the interpolation density matrix $\rho(t)$. 
    Convergence of (a) $Q^{\mathrm{exa}}$ and (b) $Q^{\mathrm{ave}}$ with respect to $L_x$.
    }
    \label{fig disordered cluster response}
\end{figure}
To apply our method to detect the mixed-state anomaly of the average subsystem symmetry, we consider the following density matrix interpolation
\begin{equation}
    \rho(t) = t\rho_{\mathrm{cluster}} + (1-t)\rho_{\mathrm{trivial}},\ t\in[0,1].
\end{equation}
When $t=1$, $\rho_{\mathrm{cluster}}$ denotes the fixed-point density matrix of the disordered cluster state, which satisfies both average $\mathcal{Z}_2^\sigma$ symmetry and exact $\mathcal{Z}_2^\tau$ symmetry. 
The boundary operator associated with this symmetry is given by Eqs. (\ref{eq disordered cluster w}) and (\ref{eq disordered cluster w2}). 
When $t=0$, all subsystem symmetries of the system are reduced to average symmetries.
The density matrix $\rho_{\mathrm{trivial}}$ lies in a trivial phase, with the specific form 
\begin{equation}
    \rho_{\mathrm{trivial}} = \Bigl(\sum_{\{\sigma\}} |\{\sigma\}\rangle\langle\{\sigma\}|\Bigr)\otimes\Bigl(\sum_{\{\tau\}}|\{\tau\}\rangle\langle\{\tau\}|\Bigr),
\end{equation}
where the summation over $\{\sigma\}$ and $\{\tau\}$ includes all possible spin configurations. 
We detect the mixed-state quantum anomaly $\phi^{\mathrm{exa}}(k_{\tau}^{[1]},g_{\sigma}^{[2]})$ via the transfer matrix
\begin{equation}
\begin{tikzpicture}
    \draw (2,0.4) node {$X\otimes X$};
    \draw[blue] (-0.2,0.7) node {$\tau^x$};
    \draw[thick] (1,0.6) -- (1.75,1.05);
    \draw[thick,fill=white]
    (0.75,0.85) -- (1.25,0.85) -- (1.25,0.35) -- (0.75,0.35) -- cycle (0.75,0.85) -- (1,1) -- (1.5,1) -- (1.5,0.5) -- (1.25,0.35) (1.5,1) -- (1.25,0.85);
    \draw[line width=0.075cm,white] (0.25,0.15) -- (1,0.6);
    \draw[thick] (0.25,0.15) -- (1,0.6);
    \draw[thick,fill=white] (-0.25,0.25) -- (0.25,0.25) -- (0.25,-0.25) -- (-0.25,-0.25) -- cycle (-0.25,0.25) -- (0,0.4) -- (0.5,0.4) -- (0.5,-0.1) -- (0.25,-0.25) (0.5,0.4) -- (0.25,0.25) ;
    \draw[line width=0.075cm,white] (0,0) -- (-0.4,-0.24) (0.375,0.075) -- (0.875,0.075)  (1.375,0.675) -- (1.875,0.675); 
    \draw[thick] (0,0) -- (-0.4,-0.24) (0.375,0.075) -- (0.875,0.075) (1.375,0.675) -- (1.875,0.675) (-0.25,0.075) -- (-0.65,0.075) (0.75,0.675) -- (0.35,0.675); 
    \draw[line width=0.1cm,white] (0.125,0.325) -- (0.125,0.725) (1.125,0.925) -- (1.125,1.325);
    \draw[ultra thick,purplish] (0.125,0.325) -- (0.125,0.725) (1.125,0.925) -- (1.125,1.325) (1.125,1.325) arc[start angle=180, end angle=0, x radius=0.075, y radius=0.3] (0.125,0.725) arc[start angle=180, end angle=0, x radius=0.075, y radius=0.3];
    \draw[ultra thick,purplish] (0.125,-0.25) -- (0.125,-0.55) (1.125,0.35) -- (1.125,0.05) (1.125,0.05) arc[start angle=-180, end angle=0, x radius=0.075, y radius=0.3] (0.125,-0.55) arc[start angle=-180, end angle=0, x radius=0.075, y radius=0.3];
    \filldraw[blue] (0.125,0.65) circle[radius=0.075];
    \filldraw[black] (1.7,0.675) circle[radius=0.075];
    \draw[thick,rotate around={-55:(-0.4,-0.24)}] (-0.4,-0.24) arc[start angle=-180, end angle=0, x radius=0.075, y radius=0.3];
    \draw[thick,rotate around={-55:(1.75,1.05)}] (1.75,1.05) arc[start angle=180, end angle=0, x radius=0.075, y radius=0.3];
\end{tikzpicture}.
\end{equation}
Our numerical results in Fig. \ref{fig disordered cluster response} show that 
\begin{equation}    
\lim_{L_x\rightarrow\infty}Q^{\mathrm{exa}}(k_\tau^{[1]},g_\sigma^{[2]}) = 
    \begin{cases}
        0, & 0 \leq t < 1\\
        -1, & t = 1
    \end{cases}.
\end{equation}
Therefore, the exact subsystem symmetry of the interpolation density matrix is summarized as
\begin{equation}
    \lim_{L_x\rightarrow\infty}\mathrm{Tr}[S^h(g_\sigma^{[2]})\rho(0)]=
    \begin{cases}
        0,\  t\in[0,1)\\
        1,\  t=1
    \end{cases}.
\end{equation}
The mixed-state anomaly $\phi^{\mathrm{ave}}(g_{\sigma}^{[2]},k_{\tau}^{[2]})$ is extracted from the spectrum of 
\begin{equation}
\begin{tikzpicture}
    \draw (2.2,0.4) node {$Z\otimes I$} (2.2,1.4) node {$Z\otimes I$};
    \draw[red] (0.8,2.2) node {$\sigma^x$}
    (1.5,1.25) node {$\sigma^x$};
    \draw[thick] (1,0.6) -- (1.75,1.05);
    \draw[thick,fill=white]
    (0.75,0.85) -- (1.25,0.85) -- (1.25,0.35) -- (0.75,0.35) -- cycle (0.75,0.85) -- (1,1) -- (1.5,1) -- (1.5,0.5) -- (1.25,0.35) (1.5,1) -- (1.25,0.85);
    \draw[line width=0.075cm,white] (0.25,0.15) -- (1,0.6);
    \draw[thick] (0.25,0.15) -- (1,0.6);
    \draw[thick,fill=white] (-0.25,0.25) -- (0.25,0.25) -- (0.25,-0.25) -- (-0.25,-0.25) -- cycle (-0.25,0.25) -- (0,0.4) -- (0.5,0.4) -- (0.5,-0.1) -- (0.25,-0.25) (0.5,0.4) -- (0.25,0.25) ;

    \draw[line width=0.075cm,white] (0,0) -- (-0.4,-0.24) (0.375,0.075) -- (0.875,0.075)  (1.375,0.675) -- (1.875,0.675); 
    \draw[thick] (0,0) -- (-0.4,-0.24) (0.375,0.075) -- (0.875,0.075) (1.375,0.675) -- (1.875,0.675) (-0.25,0.075) -- (-0.65,0.075) (0.75,0.675) -- (0.35,0.675); 
    \draw[line width=0.1cm,white] (0.125,0.325) -- (0.125,0.825) (1.125,0.925) -- (1.125,1.325);
    \draw[ultra thick,purplish] (0.125,0.325) -- (0.125,0.825) (1.125,0.925) -- (1.125,1.325) ;
    \draw[ultra thick,purplish] (0.125,-0.25) -- (0.125,-0.55) (1.125,0.35) -- (1.125,0.05) (0.125,-0.55) arc[start angle=-180, end angle=0, x radius=0.075, y radius=0.3] (1.125,0.05) arc[start angle=-180, end angle=0, x radius=0.075, y radius=0.3];
    \draw[thick,rotate around={-55:(1.75,2.05)}] (1.75,2.05) arc[start angle=180, end angle=0, x radius=0.075, y radius=0.3];
    \draw[thick,rotate around={-55:(1.75,1.05)}] (1.75,1.05) arc[start angle=180, end angle=0, x radius=0.075, y radius=0.3];

    \draw[thick] (1,1.6) -- (1.75,2.05);
    \draw[thick,fill=white]
    (0.75,1.85) -- (1.25,1.85) -- (1.25,1.35) -- (0.75,1.35) -- cycle (0.75,1.85) -- (1,2) -- (1.5,2) -- (1.5,1.5) -- (1.25,1.35) (1.5,2) -- (1.25,1.85);
    \draw[line width=0.075cm,white] (0.25,1.15) -- (1,1.6);
    \draw[thick] (0.25,1.15) -- (1,1.6);
    \draw[thick,fill=white] (-0.25,1.25) -- (0.25,1.25) -- (0.25,0.75) -- (-0.25,0.75) -- cycle (-0.25,1.25) -- (0,1.4) -- (0.5,1.4) -- (0.5,0.9) -- (0.25,0.75) (0.5,1.4) -- (0.25,1.25) ;
    \draw[line width=0.075cm,white] (0,1) -- (-0.4,0.76) (0.375,1.075) -- (0.875,1.075)  (1.375,1.675) -- (1.875,1.675); 
    \draw[thick] (0,1) -- (-0.4,0.76) (0.375,1.075) -- (0.875,1.075) (1.375,1.675) -- (1.875,1.675) (-0.25,1.075) -- (-0.65,1.075) (0.75,1.675) -- (0.35,1.675); 
    \draw[line width=0.1cm,white] (0.125,1.325) -- (0.125,1.825) (1.125,1.925) -- (1.125,2.325);
    \draw[ultra thick,purplish] (0.125,1.325) -- (0.125,1.725) (1.125,1.925) -- (1.125,2.225) (0.125,1.725) arc[start angle=180, end angle=0, x radius=0.075, y radius=0.3] (1.125,2.225) arc[start angle=180, end angle=0, x radius=0.075, y radius=0.3];
    \filldraw[red] (1.125,1.15) circle[radius=0.075] (1.125,2.2) circle[radius=0.075];
    \filldraw[black] (1.7,0.675) circle[radius=0.075] (1.7,1.675) circle[radius=0.075];

    \draw[thick,rotate around={-55:(-0.4,-0.24)}] (-0.4,-0.24) arc[start angle=-180, end angle=0, x radius=0.075, y radius=0.3];
    \draw[thick,rotate around={-55:(-0.4,0.75)}] (-0.4,0.75) arc[start angle=-180, end angle=0, x radius=0.075, y radius=0.3];
\end{tikzpicture}.
\end{equation}
The associated average subsystem symmetry charge is given by
\begin{equation}    
\lim_{L_x\rightarrow\infty}Q^{\mathrm{ave}}(g_\sigma^{[2]},k_\tau^{[2]}) = 
    \begin{cases}
        1, & t = 0\\
        0, & 0 < t < 1 \\
        -1, & t = 1
    \end{cases},
\end{equation}
indicating that $\rho(t)$ restores the average subsystem symmetry at $t=0$ and $t=1$. 
When $t=0$, the exact symmetry also breaks down to average symmetry. 
According to the proof in Ref. \cite{PhysRevX.13.031016}, there is no nontrivial topological phase protected solely by average symmetry. 
This means that the system resides in a trivial phase. 
When $t=1$, the system carries the boundary anomalies labeled by $\phi(g_\sigma^{[2]},k_\tau^{[2]})=\phi(k_\tau^{[1]},g_\sigma^{[2]})=-1$, indicating that the density matrix is exactly a nontrivial ASSPT phase.

\begin{figure}
    \centering
    \includegraphics[width = 1\linewidth]{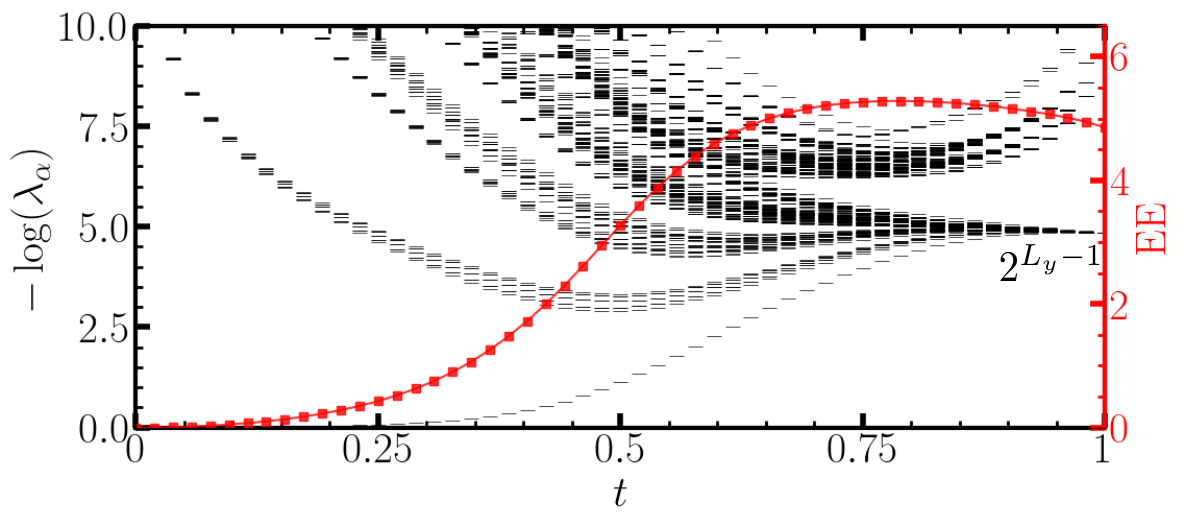}
    \caption{
    The ES and the EE of $\rho_l(t)$ for the cylinder with $L_y=8$. 
    The entanglement gap closes at $t=1$.
    }
    \label{fig ASSPT spectrum}
\end{figure}
In Fig. \ref{fig ASSPT spectrum}, we also compute the entanglement spectrum of the density matrix $\rho(t)$ on a cylinder with a circumference of $L_y=8$. 
The entanglement gap closes at $t=1$, indicating a nonzero average value of the string order parameter in the density matrix, which indicates the presence of a nontrivial mixed-state quantum anomaly. 
The entanglement spectrum exhibits a $2^{L_y-1}$-fold degeneracy at $t=1$, consistent with the anomalous edge theory of the 2D cluster state.

\section{Conclusion and Discussion}
\begin{figure}
    \centering
    \begin{tikzpicture}
        \draw[thick] (0,-3.3) -- (0,4) (-4.5,0) -- (4,0) (4,4) -- (4,-3.4) -- (-3.5,-3.4) -- (-3.5,4) -- (4,4) (-4.5,3) -- (4,3) (-3.5,-3.4) -- (-4.5,-3.4) -- (-4.5,3);
        \draw (-1.7,3.5) node {SSPT phases} (2,3.5) node {ASSPT phases};
        \draw (-1.7,2.6) node {2D cluster state};
        \draw (-4,1.5) node[rotate=90] {$G_s=Z_2^\tau\times Z_2^\sigma$}
        (-4,-1.7) node[rotate=90] {$G_s=Z_2$};
        \draw (-2.55,1-0.25) node {$g_{\tau}^{[y]}$} (-2.75,2-0.25) node {$g_{\tau}^{[y+1]}$} (1.6-2.2,2-0.25) node {$g_{\sigma}^{[y+1]}$} (1.4-2.2,1-0.25) node {$g_{\sigma}^{[y]}$};
        \filldraw[red] 
        (-1.2,2-0.25) circle[radius=0.08] 
        (-1.2,1-0.25) circle[radius=0.08];
        \filldraw[blue] 
        (-2.2,1-0.25) circle[radius=0.08] 
        (-2.2,2-0.25) circle[radius=0.08];
        \draw[very thick] (0.15-2.2,1-0.25) -- (0.85-2.2,1-0.25) (0.15-2.2,2-0.25) -- (0.85-2.2,2-0.25);
        \draw[very thick] (0.1-2.2,1.1-0.25) -- (0.9-2.2,1.9-0.25);

        \draw (-1.7,-0.4) node {2D intrinsic SSPT};
        \draw (-2.2,-1) node {$g^{[y+2]}$} (-2.2,-2) node {$g^{[y+1]}$} (-2.05,-3) node {$g^{[y]}$}
;
        \draw[very thick] (-1.7,0.15-2) -- (-1.7,0.85-2) (-1.7,0.15-3) -- (-1.7,0.85-3);
        \filldraw[blue] (-1.7,-3) circle[radius=0.075] (-1.7,-2) circle[radius=0.075] (-1.7,-1) circle[radius=0.075];

        \draw (2,2.7) node {2D disorder cluster state};
        \draw (1,-0.35+0.75) node {$k_{\tau,l}^{[y]}$}
        (2,-0.35+0.75) node {$g_{\sigma,d}^{[y]}$}
        (3,-0.35+0.75) node {$k_{\tau,r}^{[y]}$}
        (1,1.35+0.75) node {$k_{\tau,l}^{[y+1]}$}
        (2,1.35+0.75) node {$g_{\sigma,d}^{[y+1]}$}
        (3,1.35+0.75) node {$k_{\tau,r}^{[y+1]}$};
        \filldraw[red] 
        (2,0.75) circle[radius=0.08] 
        (2,1.75) circle[radius=0.08];
        \filldraw[blue] 
        (1,0.75) circle[radius=0.08]
        (3,0.75) circle[radius=0.08]
        (1,1.75) circle[radius=0.08]
        (3,1.75) circle[radius=0.08];
        \draw[very thick] (2.15,0.75) -- (2.85,0.75) (2-0.15,0.75) -- (2-0.85,0.75) (2.15,1.75) -- (2.85,1.75) (2-0.15,1.75) -- (2-0.85,1.75) (2-0.9,1.1-0.25) -- (2-0.1,1.9-0.25) (2.9,1.1-0.25) -- (2.1,1.9-0.25);

        \draw (2,-0.4) node {2D alternating ASSPT};
        \draw (0.85,-2) node {$k_l^{[2y+1]}$}
        (3.2,-2) node {$k_r^{[2y+1]}$}
        (1.5,-3) node {$g_d^{[2y]}$}
        (2-0.65,-1) node {$g_d^{[2y+2]}$};
        \draw[very thick] (2-0.075,0.15-3) -- (2-0.425,0.85-3)
        (2.075,0.15-3) -- (2.425,0.85-3)
        (2-0.075,1.85-3) -- (2-0.425,1.15-3)
        (2.075,1.85-3) -- (2.425,1.15-3);
        \filldraw[red] (2,-3) circle[radius=0.075]
        (2,-1) circle[radius=0.075];
        \filldraw[blue] (2-0.5,-2) circle[radius=0.075] (2.5,-2) circle[radius=0.075];
    \end{tikzpicture}
    \caption{
    Summary of boundary anomalies in various topological phases. 
    Blue and red dots represent different group generators, while the solid black line indicates anti-commutativity between these generators.
    }
    \label{tab anomaly summary}
\end{figure}
In summary, this study showcases the power of tensor network representations in studying 2D strong SSPT phases. 
By extending the concept of anomaly indicator to subsystem symmetries, we systematically investigate the boundary quantum anomalies of various SSPT wave functions. 
Our study reveals new insights into strong and weak SSPT phases, including the discovery of an intrinsic SSPT phase and the verification of nontrivial edge theory through numerical simulations. 
Furthermore, our method has been successfully extended to investigate mixed-state density matrices with average subsystem symmetries, demonstrating the persistence of boundary anomalies in systems with uniform and alternating disorders. 
This extension also establishes a connection between quantum anomalies in pure and mixed states, as illustrated by Fig. \ref{tab anomaly summary}. 
Importantly, we show that the transition from the left to the right side of this figure can be achieved by duplicating the generators of $K_s$, which extends the $K_s$ subsystem symmetry to $K_{l}\times K_{r}$ within the Choi representation of the mixed-state density matrix.
Since the average subsystem symmetry acts diagonally in the doubled Hilbert space, the mixed quantum anomalies for $K_l\times G_d$ and $K_r\times G_d$ are identical. 

Our findings have significant implications for studying topological quantum matters protected by foliated subsystem symmetries. 
Utilizing tensor network formalism enables us to investigate higher-order fractonic topological phases \cite{PhysRevB.103.245128,PhysRevB.108.045133,PhysRevB.108.155123,sun2024holographicviewmixedstatesymmetryprotected}, which are a new class of topological insulators. 
The tunable tensor wave function also facilitates the exploration of quantum phase transitions in these systems, providing valuable insights into their behaviors. 
Moreover, our tensor representation can be generalized to 3D systems, allowing us to study the duality between fracton models and planar SSPT phases \cite{10.21468/SciPostPhys.6.4.041,doi:10.1142/S0217751X20300033,SHIRLEY2019167922,PhysRevX.8.031051}. 
This has the potential to reveal new aspects of topological physics and provide a deeper understanding of these complex systems.

\begin{acknowledgements}
We thank Frank Pollmann and Yuchen Guo for helpful discussions. 
This work is supported by the National Natural Science Foundation of China (NSFC) (Grant No. 12475022 and No. 12174214) and the Innovation Program for Quantum Science and Technology (Grant No. 2021ZD0302100).
\end{acknowledgements}

\bibliography{references}
\clearpage
\onecolumngrid
\appendix

\section{Topological response theory}\label{app inflow}
In this appendix, we discuss the relationship between the topological response theory and the anomaly indicator. 
Turning on the background gauge field along temporal and spatial dimensions of a $(1+1)$D $G$-SPT phase, we obtain the torus partition function with symmetry defects at zero temperature
\begin{equation}
\begin{tikzpicture}
    \draw (-0.5,1) node {$Z\biggl[$}
    (7.25,1) node {$\biggr]=\mathrm{Tr}[e^{-\beta H(f)}S(g)] = \langle\psi_{f}|S(g)|\psi_{f}\rangle=\omega(g,f)/\omega(f,g),\quad \beta\rightarrow\infty$};
    \draw (1.25,1.75) node {\textcolor{blue}{$f$}} (1.75,0.75) node {\textcolor{red}{$g$}};
    \draw[very thick] (0,0) -- (2,0) -- (2,2) -- (0,2) -- cycle;
    \draw[blue,dashed,ultra thick] (1,0) -- (1,2);
    \draw[red,ultra thick] (0,1) -- (2,1);
    \draw[very thick,->] (-0.35,-0.35) -- (0.15,-0.35);
    \draw[very thick,->] (-0.35,-0.35) -- (-0.35,0.15);
    \draw (0.35,-0.35) node {$x$}
    (-0.35,0.35) node {$t$};
\end{tikzpicture}\label{eq topo response global}
\end{equation}
where the solid red and dashed blue lines are the temporal and spatial symmetry defects, respectively.
We define $H(f)$ as the Hamiltonian with $f$-defect insertion and $|\psi_f\rangle$ as its ground state, which coincides with the twisted sector state from $V(f)$ spatial defect insertion. 
The zero-temperature partition function with symmetry defects can thus be computed via the symmetry charges $\langle\psi_{f}|S(g)|\psi_{f}\rangle$ of the twisted sector state \cite{10.21468/SciPostPhys.15.2.051,kapustin2014anomaliesdiscretesymmetriesvarious}, yielding exactly the anomaly indicator
\begin{equation}
    Q(g,f) = \frac{\langle\psi_{f}|S(g)|\psi_{f}\rangle}{\langle\psi_{f}|\psi_{f}\rangle}
\end{equation}
under the normalization condition $\langle\psi_{f}|\psi_{f}\rangle=1$.

This framework allows systematic extension to analyze the boundary 't Hooft anomaly of subsystem symmetry group $G_h=\prod_y G_s^{[y]}$.
The topological response theory of $G_h$ is obtained by turning on the background gauge fields of $g^{[y_1]}\in G_s^{[y_1]}$ and $f^{[y_2]}\in G_s^{[y_2]}$. 
By setting $y_1=y+1$ and $y_2=y$, the explicit form of the partition function with subsystem symmetry defects manifest as
\begin{equation}
\begin{tikzpicture}
\draw (-0.5,2.5) node {$Z\biggl[$}
(12.2,2.5) node {$\biggr]=\langle\Psi_{f^{[y]}}|S^h(g^{[y+1]})|\Psi_{f^{[y]}}\rangle=\omega(g^{[y+1]},f^{[y]})/\omega(f^{[y]},g^{[y+1]})$};
\draw[draw=blue,thick] (2.5,1.25) -- (7,0.75) -- (7,4.75) -- (2.5,5.25) -- cycle;
\draw[red,ultra thick] (2.5,3.3) -- (7,2.8);
\draw (3.25,4.125) node[rotate=30] {$\cdots$};
\draw[draw=blue,thick] (1.5,0.75) -- (6,0.25) -- (6,4.25) -- (1.5,4.75) -- cycle;

\draw[draw=blue,thick] (0,0) -- (4.5,-0.5) -- (4.5,3.5) -- (0,4) -- cycle;

\draw[ultra thick] (4,-0.75) -- (4,3.25) -- (7.5,5) -- (7.5,1) -- cycle;
\filldraw[white,opacity=0.75] (6,0.25) -- (6,4.25) -- (7,4.75) -- (7,0.75) -- cycle;
\draw[pattern=north east lines,pattern color=blue,draw=blue,dashed,ultra thick] (6,0.25) -- (6,4.25) -- (7,4.75) -- (7,0.75) -- cycle;
\draw (3,2.8) node {\textcolor{red}{$g^{[y+1]}$}};
\filldraw[white] (6.1,1) -- (6.1,1.5) -- (6.9,2) -- (6.9,1.5) -- cycle; 
\draw (6.5,1.5) node[rotate=30] {\textcolor{blue}{$f^{[y]}$}};
\draw (2.25,5.5) node {$[y+1]$}
(1.25,5) node {$[y]$};
\draw (-1.2,0.85) node {$t$}
(-0.15,-0.425) node {$x$}
(-0.2,0.21) node {$y$};
\draw[very thick,->] (-1.2,-0.25) -- (-0.3,-0.35);
\draw[very thick,->] (-1.2,-0.25) -- (-1.2,0.65);
\draw[very thick,->] (-1.2,-0.25) -- (-0.4,0.15);
\end{tikzpicture}.\label{eq topo response subsystem}
\end{equation}
Eq. (\ref{eq topo response subsystem}) provides a higher-dimensional generalization of Eq. (\ref{eq topo response global}), extending topological response theory to linear foliated subsystem symmetries $G_h$ in $(2+1)$D. 
Although spatial symmetry defects of $G_h$ generally extend across the entire $y-t$ plane, two distinct scenarios emerge
\begin{itemize}
    \item Weak SSPT phases without mixed subsystem symmetry anomalies exhibit spatial symmetry defect $f^{[y]}$ localized within its $y$-th subsystem;
    \item Strong SSPT phases manifest mixed anomalies that extend the spatial symmetry defect $f^{[y]}$ beyond a single subsystem. 
\end{itemize}
Our analysis specifically focuses on cases where the spatial symmetry defect of $f^{[y]}\in G_s^{[y]}$ spans only adjacent subsystems at $y$ and $y+1$.
This constraint localizes spatial defect $f^{[y]}$ within the blue-shaded subregion $[y\sim y+1]$ of the $y-t$ plane, while the temporal symmetry defect $g^{[y+1]}$ is explicitly marked by the red line. 
This partition function with symmetry defects is calculated via evaluating subsystem symmetry charge $\langle\Psi_{f^{[y]}}|S^h(g^{[y+1]})|\Psi_{f^{[y]}}\rangle$ of the $(2+1)$D twisted sector state with $W^\mathrm{Right}(f^{[y]})=V_{y}(f^{[y]})V_{y+1}(f^{[y]})$ flux insertion. 
The symmetry action $S^h(g^{[y+1]})$ on $|\Psi_{f^{[y]}}\rangle$ takes the form
\begin{equation}
\begin{tikzpicture}[>=stealth]
    \draw (-1,-1) node {$S^h(g^{[y+1]})|\Psi_{f^{[y]}}\rangle=$}
    (3.5,0.4) node {$\cdots$} (3.5,-0.4) node {$\cdots$}
    (3.5,-2.6) node {$\cdots$}
    (1.5,-1.85) node[rotate=90] {$\cdots$} (2.5,-1.85) node[rotate=90] {$\cdots$}
    (4.5,-1.85) node[rotate=90] {$\cdots$} (6.5,0.45) node {$[y+2]$} (6.5,-0.35) node {$[y+1]$} (6.3,-1.15) node {$[y]$};
    \draw[thick] (1,0.4) -- (3,0.4) (4,0.4) -- (5.5,0.4) (1,-0.4) -- (3,-0.4) (4,-0.4) -- (5.5,-0.4) (1,-1.2) -- (3,-1.2) (4,-1.2) -- (5.5,-1.2) (1,-2.6) -- (3,-2.6) (4,-2.6) -- (5.5,-2.6) (1.5,0.8) -- (1.5,-1.6) (1.5,-2.2) -- (1.5,-3) (2.5,0.8) -- (2.5,-1.6) (2.5,-2.2) -- (2.5,-3) (4.5,0.8) -- (4.5,-1.6) (4.5,-2.2) -- (4.5,-3) (1,0.4) arc[start angle=270, end angle=90, x radius=0.25, y radius=0.075] (1,-0.4) arc[start angle=270, end angle=90, x radius=0.25, y radius=0.075] (1,-1.2) arc[start angle=270, end angle=90, x radius=0.25, y radius=0.075] (1,-2.6) arc[start angle=270, end angle=90, x radius=0.25, y radius=0.075] (5.5,0.4) arc[start angle=-90, end angle=90, x radius=0.25, y radius=0.075] (5.5,-0.4) arc[start angle=-90, end angle=90, x radius=0.25, y radius=0.075] (5.5,-1.2) arc[start angle=-90, end angle=90, x radius=0.25, y radius=0.075] (5.5,-2.6) arc[start angle=-90, end angle=90, x radius=0.25, y radius=0.075];
    \draw[thick,fill=white] (1.25,0.15) rectangle ++(0.5,0.5) (1.25,-0.65) rectangle ++(0.5,0.5) (1.25,-1.45) rectangle ++(0.5,0.5) (2.25,0.15) rectangle ++(0.5,0.5) (2.25,-0.65) rectangle ++(0.5,0.5) (2.25,-1.45) rectangle ++(0.5,0.5) (4.25,0.15) rectangle ++(0.5,0.5) (4.25,-0.65) rectangle ++(0.5,0.5) (4.25,-1.45) rectangle ++(0.5,0.5) (1.25,-2.85) rectangle ++(0.5,0.5) (2.25,-2.85) rectangle ++(0.5,0.5) (4.25,-2.85) rectangle ++(0.5,0.5);
    \draw[line width=0.1cm,white] (1.5,0.4) -- (1.2,0.7) (1.5,-0.4) -- (1.2,-0.1)
    (1.5,-1.2) -- (1.2,-0.9)
    (1.5,-2.6) -- (1.2,-2.3)(2.5,0.4) -- (2.2,0.7) (2.5,-0.4) -- (2.2,-0.1)
    (2.5,-1.2) -- (2.2,-0.9) 
    (2.5,-2.6) -- (2.2,-2.3)(4.5,0.4) -- (4.2,0.7) (4.5,-0.4) -- (4.2,-0.1)
    (4.5,-1.2) -- (4.2,-0.9)
    (4.5,-2.6) -- (4.2,-2.3);
    \draw[ultra thick,purplish] (1.5,0.4) -- (1.2,0.7) (1.5,-0.4) -- (1.2,-0.1)
    (1.5,-1.2) -- (1.2,-0.9)
    (1.5,-2.6) -- (1.2,-2.3)(2.5,0.4) -- (2.2,0.7) (2.5,-0.4) -- (2.2,-0.1)
    (2.5,-1.2) -- (2.2,-0.9) 
    (2.5,-2.6) -- (2.2,-2.3)(4.5,0.4) -- (4.2,0.7) (4.5,-0.4) -- (4.2,-0.1)
    (4.5,-1.2) -- (4.2,-0.9)
    (4.5,-2.6) -- (4.2,-2.3);
    \draw[thick] (1.5,0.8) arc[start angle=180, end angle=0, x radius=0.075, y radius=0.3]
    (1.5,-3) arc[start angle=-180, end angle=0, x radius=0.075, y radius=0.3] (2.5,0.8) arc[start angle=180, end angle=0, x radius=0.075, y radius=0.3]
    (2.5,-3) arc[start angle=-180, end angle=0, x radius=0.075, y radius=0.3] (4.5,0.8) arc[start angle=180, end angle=0, x radius=0.075, y radius=0.3]
    (4.5,-3) arc[start angle=-180, end angle=0, x radius=0.075, y radius=0.3];
    \filldraw[blue] (5.25,-0.4) circle[radius=0.06] (5.25,-1.2) circle[radius=0.06];
    \draw[blue] (5.6,0.1) node {$V_{y+1}(f^{[y]})$}
    (5.45,-0.8) node {$V_{y}(f^{[y]})$};
    \foreach \x in {0.75, 2.25, 3.75}{
    \draw[red] (\x,0.15) node {$U(g^{[y+1]})$};
    }
\end{tikzpicture}.
\end{equation}
By diagonalizing the flux-inserted SSPT Hamiltonian $H(f^{[y]})$, we obtain the ground state wave function $|\Psi_{f^{[y]}}\rangle$, which exactly coincides with the twisted sector state defined above.

We now establish the connection between the anomaly indicator in Eq. (\ref{eq topo response}) and the topological response theory in Eq. (\ref{eq topo response subsystem}). 
The mixed quantum anomaly of $G_s^{[y]}\times G_s^{[y+1]}$ can also be detected through the topological response $\langle\Psi_{f^{[y+1]}}|S^h(g^{[y]})|\Psi_{f^{[y+1]}}\rangle$ of the $(2+1)$D linear SSPT phase with $W^\mathrm{Right}(f^{[y+1]})=V_{y+1}(f^{[y+1]})V_{y+2}(f^{[y+1]})$ flux insertion. 
The corresponding symmetry action on this twisted sector state is given by
\begin{equation}
\begin{tikzpicture}[>=stealth]
    \draw (-1,-1) node {$S^h(g^{[y]})|\Psi_{f^{[y+1]}}\rangle=$}
    (3.5,0.4) node {$\cdots$} (3.5,-0.4) node {$\cdots$}
    (3.5,-2.6) node {$\cdots$}
    (1.5,-1.85) node[rotate=90] {$\cdots$} (2.5,-1.85) node[rotate=90] {$\cdots$}
    (4.5,-1.85) node[rotate=90] {$\cdots$} (6.5,0.45) node {$[y+2]$} (6.5,-0.35) node {$[y+1]$} (6.3,-1.15) node {$[y]$};
    \draw[thick] (1,0.4) -- (3,0.4) (4,0.4) -- (5.5,0.4) (1,-0.4) -- (3,-0.4) (4,-0.4) -- (5.5,-0.4) (1,-1.2) -- (3,-1.2) (4,-1.2) -- (5.5,-1.2) (1,-2.6) -- (3,-2.6) (4,-2.6) -- (5.5,-2.6) (1.5,0.8) -- (1.5,-1.6) (1.5,-2.2) -- (1.5,-3) (2.5,0.8) -- (2.5,-1.6) (2.5,-2.2) -- (2.5,-3) (4.5,0.8) -- (4.5,-1.6) (4.5,-2.2) -- (4.5,-3) (1,0.4) arc[start angle=270, end angle=90, x radius=0.25, y radius=0.075] (1,-0.4) arc[start angle=270, end angle=90, x radius=0.25, y radius=0.075] (1,-1.2) arc[start angle=270, end angle=90, x radius=0.25, y radius=0.075] (1,-2.6) arc[start angle=270, end angle=90, x radius=0.25, y radius=0.075] (5.5,0.4) arc[start angle=-90, end angle=90, x radius=0.25, y radius=0.075] (5.5,-0.4) arc[start angle=-90, end angle=90, x radius=0.25, y radius=0.075] (5.5,-1.2) arc[start angle=-90, end angle=90, x radius=0.25, y radius=0.075] (5.5,-2.6) arc[start angle=-90, end angle=90, x radius=0.25, y radius=0.075];
    \draw[thick,fill=white] (1.25,0.15) rectangle ++(0.5,0.5) (1.25,-0.65) rectangle ++(0.5,0.5) (1.25,-1.45) rectangle ++(0.5,0.5) (2.25,0.15) rectangle ++(0.5,0.5) (2.25,-0.65) rectangle ++(0.5,0.5) (2.25,-1.45) rectangle ++(0.5,0.5) (4.25,0.15) rectangle ++(0.5,0.5) (4.25,-0.65) rectangle ++(0.5,0.5) (4.25,-1.45) rectangle ++(0.5,0.5) (1.25,-2.85) rectangle ++(0.5,0.5) (2.25,-2.85) rectangle ++(0.5,0.5) (4.25,-2.85) rectangle ++(0.5,0.5);
    \draw[line width=0.1cm,white] (1.5,0.4) -- (1.2,0.7) (1.5,-0.4) -- (1.2,-0.1)
    (1.5,-1.2) -- (1.2,-0.9)
    (1.5,-2.6) -- (1.2,-2.3)(2.5,0.4) -- (2.2,0.7) (2.5,-0.4) -- (2.2,-0.1)
    (2.5,-1.2) -- (2.2,-0.9) 
    (2.5,-2.6) -- (2.2,-2.3)(4.5,0.4) -- (4.2,0.7) (4.5,-0.4) -- (4.2,-0.1)
    (4.5,-1.2) -- (4.2,-0.9)
    (4.5,-2.6) -- (4.2,-2.3);
    \draw[ultra thick,purplish] (1.5,0.4) -- (1.2,0.7) (1.5,-0.4) -- (1.2,-0.1)
    (1.5,-1.2) -- (1.2,-0.9)
    (1.5,-2.6) -- (1.2,-2.3)(2.5,0.4) -- (2.2,0.7) (2.5,-0.4) -- (2.2,-0.1)
    (2.5,-1.2) -- (2.2,-0.9) 
    (2.5,-2.6) -- (2.2,-2.3)(4.5,0.4) -- (4.2,0.7) (4.5,-0.4) -- (4.2,-0.1)
    (4.5,-1.2) -- (4.2,-0.9)
    (4.5,-2.6) -- (4.2,-2.3);
    \draw[thick] (1.5,0.8) arc[start angle=180, end angle=0, x radius=0.075, y radius=0.3]
    (1.5,-3) arc[start angle=-180, end angle=0, x radius=0.075, y radius=0.3] (2.5,0.8) arc[start angle=180, end angle=0, x radius=0.075, y radius=0.3]
    (2.5,-3) arc[start angle=-180, end angle=0, x radius=0.075, y radius=0.3] (4.5,0.8) arc[start angle=180, end angle=0, x radius=0.075, y radius=0.3]
    (4.5,-3) arc[start angle=-180, end angle=0, x radius=0.075, y radius=0.3];
    \filldraw[blue] (5.25,0.4) circle[radius=0.06] (5.25,-0.4) circle[radius=0.06];
    \draw[blue] (5.7,0.9) node {$V_{y+2}(f^{[y+1]})$}
    (5.7,0) node {$V_{y+1}(f^{[y+1]})$};
    \foreach \x in {0.75, 2, 3.75}{
    \draw[red] (\x,-0.65) node {$U(g^{[y]})$};
    }
\end{tikzpicture}.\label{eq two rows insertion}
\end{equation}
Since $V_{y+2}(f^{[y+1]})$ always commutes with the boundary operator of $U(g^{[y]})$, inserting $V_{y+1}(f^{[y+1]})$ flux in the overlapping region suffices to fully extract the boundary anomalies. 
The corresponding twisted sector state under $S^h(g^{[y]})$ transformation is
\begin{equation}
\begin{tikzpicture}[>=stealth]
    \draw (-1,-1) node {$S^h(g^{[y]})|\psi_{f^{[y+1]}}\rangle=$}
    (3.5,0.4) node {$\cdots$} (3.5,-0.4) node {$\cdots$}
    (3.5,-2.6) node {$\cdots$}
    (1.5,-1.85) node[rotate=90] {$\cdots$} (2.5,-1.85) node[rotate=90] {$\cdots$}
    (4.5,-1.85) node[rotate=90] {$\cdots$} (6.5,0.45) node {$[y+2]$} (6.5,-0.35) node {$[y+1]$} (6.3,-1.15) node {$[y]$};
    \draw[thick] (1,0.4) -- (3,0.4) (4,0.4) -- (5.5,0.4) (1,-0.4) -- (3,-0.4) (4,-0.4) -- (5.5,-0.4) (1,-1.2) -- (3,-1.2) (4,-1.2) -- (5.5,-1.2) (1,-2.6) -- (3,-2.6) (4,-2.6) -- (5.5,-2.6) (1.5,0.8) -- (1.5,-1.6) (1.5,-2.2) -- (1.5,-3) (2.5,0.8) -- (2.5,-1.6) (2.5,-2.2) -- (2.5,-3) (4.5,0.8) -- (4.5,-1.6) (4.5,-2.2) -- (4.5,-3) (1,0.4) arc[start angle=270, end angle=90, x radius=0.25, y radius=0.075] (1,-0.4) arc[start angle=270, end angle=90, x radius=0.25, y radius=0.075] (1,-1.2) arc[start angle=270, end angle=90, x radius=0.25, y radius=0.075] (1,-2.6) arc[start angle=270, end angle=90, x radius=0.25, y radius=0.075] (5.5,0.4) arc[start angle=-90, end angle=90, x radius=0.25, y radius=0.075] (5.5,-0.4) arc[start angle=-90, end angle=90, x radius=0.25, y radius=0.075] (5.5,-1.2) arc[start angle=-90, end angle=90, x radius=0.25, y radius=0.075] (5.5,-2.6) arc[start angle=-90, end angle=90, x radius=0.25, y radius=0.075];
    \draw[thick,fill=white] (1.25,0.15) rectangle ++(0.5,0.5) (1.25,-0.65) rectangle ++(0.5,0.5) (1.25,-1.45) rectangle ++(0.5,0.5) (2.25,0.15) rectangle ++(0.5,0.5) (2.25,-0.65) rectangle ++(0.5,0.5) (2.25,-1.45) rectangle ++(0.5,0.5) (4.25,0.15) rectangle ++(0.5,0.5) (4.25,-0.65) rectangle ++(0.5,0.5) (4.25,-1.45) rectangle ++(0.5,0.5) (1.25,-2.85) rectangle ++(0.5,0.5) (2.25,-2.85) rectangle ++(0.5,0.5) (4.25,-2.85) rectangle ++(0.5,0.5);
    \draw[line width=0.1cm,white] (1.5,0.4) -- (1.2,0.7) (1.5,-0.4) -- (1.2,-0.1)
    (1.5,-1.2) -- (1.2,-0.9)
    (1.5,-2.6) -- (1.2,-2.3)(2.5,0.4) -- (2.2,0.7) (2.5,-0.4) -- (2.2,-0.1)
    (2.5,-1.2) -- (2.2,-0.9) 
    (2.5,-2.6) -- (2.2,-2.3)(4.5,0.4) -- (4.2,0.7) (4.5,-0.4) -- (4.2,-0.1)
    (4.5,-1.2) -- (4.2,-0.9)
    (4.5,-2.6) -- (4.2,-2.3);
    \draw[ultra thick,purplish] (1.5,0.4) -- (1.2,0.7) (1.5,-0.4) -- (1.2,-0.1)
    (1.5,-1.2) -- (1.2,-0.9)
    (1.5,-2.6) -- (1.2,-2.3)(2.5,0.4) -- (2.2,0.7) (2.5,-0.4) -- (2.2,-0.1)
    (2.5,-1.2) -- (2.2,-0.9) 
    (2.5,-2.6) -- (2.2,-2.3)(4.5,0.4) -- (4.2,0.7) (4.5,-0.4) -- (4.2,-0.1)
    (4.5,-1.2) -- (4.2,-0.9)
    (4.5,-2.6) -- (4.2,-2.3);
    \draw[thick] (1.5,0.8) arc[start angle=180, end angle=0, x radius=0.075, y radius=0.3]
    (1.5,-3) arc[start angle=-180, end angle=0, x radius=0.075, y radius=0.3] (2.5,0.8) arc[start angle=180, end angle=0, x radius=0.075, y radius=0.3]
    (2.5,-3) arc[start angle=-180, end angle=0, x radius=0.075, y radius=0.3] (4.5,0.8) arc[start angle=180, end angle=0, x radius=0.075, y radius=0.3]
    (4.5,-3) arc[start angle=-180, end angle=0, x radius=0.075, y radius=0.3];
    \filldraw[blue] (5.25,-0.4) circle[radius=0.06];
    \draw[blue]
    (5.7,0) node {$V_{y+1}(f^{[y+1]})$};
    \foreach \x in {0.75, 2, 3.75}{
    \draw[red] (\x,-0.65) node {$U(g^{[y]})$};
    }
\end{tikzpicture}.
\end{equation}
Here we use $|\psi_{f^{[y+1]}}\rangle$ to distinguish the bulk wave function with $V_{y+1}(f^{[y+1]})$ flux insertion from that with $W^{\mathrm{Right}}(f^{[y+1]})$ insertion in Eq. (\ref{eq two rows insertion}).
The subsystem symmetry charge of $|\psi_{f^{[y+1]}}\rangle$ provides the anomaly indicator
\begin{equation}
    Q(g^{[y]},f^{[y+1]})=\frac{\langle\psi_{f^{[y+1]}}|S^h(g^{[y]})|\psi_{f^{[y+1]}}\rangle}{\langle\psi_{f^{[y+1]}}|\psi_{f^{[y+1]}}\rangle}.\label{eq def Q}
\end{equation}
According to our discussion in Appendix \ref{app mixed anomaly}, we prove in Eq. (\ref{eq single bond phi}) that the mixed anomaly
\begin{equation}
    \phi(g^{[y]},f^{[y+1]})=\omega(g^{[y]},f^{[y+1]})/\omega(f^{[y+1]},g^{[y]})\label{eq def phi}
\end{equation}
arises from the non-commutativity of projective representations
\begin{equation}
    V_{y+1}(g^{[y]})V_{y+1}(f^{[y+1]}) = \phi(g^{[y]},f^{[y+1]}) V_{y+1}(f^{[y+1]})V_{y+1}(g^{[y]})
\end{equation}
within the $(y+1)$-th virtual space.
Therefore, the same boundary anomaly can be extracted by computing either the topological response or the anomaly indicator $Q$
\begin{equation}
    \langle\psi_{f^{[y+1]}}|S^h(g^{[y]})|\psi_{f^{[y+1]}}\rangle=\langle\Psi_{f^{[y+1]}}|S^h(g^{[y]})|\Psi_{f^{[y+1]}}\rangle=\phi(g^{[y]},f^{[y+1]})\label{eq reduce}
\end{equation}
in symmetric states with $L_y>2$ that satisfy the normalization condition $\langle\psi_{f^{[y+1]}}|\psi_{f^{[y+1]}}\rangle=\langle\Psi_{f^{[y+1]}}|\Psi_{f^{[y+1]}}\rangle=1$. A comparison of
\begin{align}
\begin{aligned}
    &\langle\Psi_{f^{[y]}}|S^h(g^{[y+1]})|\Psi_{f^{[y]}}\rangle = \omega(g^{[y+1]},f^{[y]})/\omega(f^{[y]},g^{[y+1]})\\
    &\langle\Psi_{f^{[y+1]}}|S^h(g^{[y]})|\Psi_{f^{[y+1]}}\rangle = \omega(g^{[y]},f^{[y+1]})/\omega(f^{[y+1]},g^{[y]})
\end{aligned}\label{eq comparison}
\end{align}
reveals that the zero-temperature partition function in Eq. (\ref{eq topo response subsystem}) exactly corresponds to the normalized anomaly indicator $Q$
\begin{equation}
    \langle\Psi_{f^{[y]}}|S^h(g^{[y+1]})|\Psi_{f^{[y]}}\rangle\stackrel{\text{(\ref{eq comparison})}}{=}\langle\Psi_{g^{[y+1]}}|S^h(f^{[y]})|\Psi_{g^{[y+1]}}\rangle^{-1}\stackrel{\text{(\ref{eq reduce})}}{=}\langle\psi_{g^{[y+1]}}|S^h(f^{[y]})|\psi_{g^{[y+1]}}\rangle^{-1}\stackrel{\text{(\ref{eq def Q})}}{=}Q^{-1}(f^{[y]},g^{[y+1]}).\label{eq response equations}
\end{equation}
Therefore, we have demonstrated the correspondence between the topological response of subsystem symmetry and the anomaly indicator. 
Since the topological response theory is general and model-independent, the quantity $Q$ does not rely on specific tensor network constructions. Furthermore, our anomaly indicator computation requires only symmetry defect insertion in the overlapping region, making it feasible for systems with $L_y=2$ where $Q$ reduces to Eq. (\ref{eq topo response}). 
This enables efficient numerical evaluation of the anomaly indicator using $L_y=2$ wave function in our calculations (Figs. \ref{fig cluster response}, \ref{fig Z_2 response}, \ref{fig disordered cluster response}).

To clarify our statement, we take the 2D cluster state as an example to verify the equivalence between the anomaly indicator and the topological response theory. 
The local tensor of 2D cluster state is given by Eq. (\ref{eq local tensor 2d cluster}) in the main text. 
Through contraction of these local tensors, we obtain the cluster state wave function on a $L_x\times L_y$ torus. 
Introducing a symmetry defect
\begin{equation}
    W^{\mathrm{Right}}(g_\tau^{[y]}) = Z_{L_x,y}Z_{L_x,y+1}.
\end{equation}
at $x=L_x$ of the wave function and performing normalization yields the twisted sector state $|\Psi_{g_\tau^{[y]}}\rangle$. 
Then we may detect the boundary anomaly $\phi(g_\tau^{[y]},g_\sigma^{[y+1]})$ by computing the topological response of $|\Psi_{g_\tau^{[y]}}\rangle$. 
The subsystem transformation on the wave function is evaluated by
\begin{equation}
\begin{tikzpicture}[>=stealth]
    \draw (-1,-1) node {$S^h(g_\sigma^{[y+1]})|\Psi_{g_\tau^{[y]}}\rangle=$}
    (3.5,0.4) node {$\cdots$} (3.5,-0.4) node {$\cdots$}
    (3.5,-2.6) node {$\cdots$}
    (1.5,-1.85) node[rotate=90] {$\cdots$} (2.5,-1.85) node[rotate=90] {$\cdots$}
    (4.5,-1.85) node[rotate=90] {$\cdots$} (6.5,0.45) node {$[y+2]$} (6.5,-0.35) node {$[y+1]$} (6.3,-1.15) node {$[y]$};
    \draw[thick] (1,0.4) -- (3,0.4) (4,0.4) -- (5.5,0.4) (1,-0.4) -- (3,-0.4) (4,-0.4) -- (5.5,-0.4) (1,-1.2) -- (3,-1.2) (4,-1.2) -- (5.5,-1.2) (1,-2.6) -- (3,-2.6) (4,-2.6) -- (5.5,-2.6) (1.5,0.8) -- (1.5,-1.6) (1.5,-2.2) -- (1.5,-3) (2.5,0.8) -- (2.5,-1.6) (2.5,-2.2) -- (2.5,-3) (4.5,0.8) -- (4.5,-1.6) (4.5,-2.2) -- (4.5,-3) (1,0.4) arc[start angle=270, end angle=90, x radius=0.25, y radius=0.075] (1,-0.4) arc[start angle=270, end angle=90, x radius=0.25, y radius=0.075] (1,-1.2) arc[start angle=270, end angle=90, x radius=0.25, y radius=0.075] (1,-2.6) arc[start angle=270, end angle=90, x radius=0.25, y radius=0.075] (5.5,0.4) arc[start angle=-90, end angle=90, x radius=0.25, y radius=0.075] (5.5,-0.4) arc[start angle=-90, end angle=90, x radius=0.25, y radius=0.075] (5.5,-1.2) arc[start angle=-90, end angle=90, x radius=0.25, y radius=0.075] (5.5,-2.6) arc[start angle=-90, end angle=90, x radius=0.25, y radius=0.075];
    \draw[thick,fill=white] (1.25,0.15) rectangle ++(0.5,0.5) (1.25,-0.65) rectangle ++(0.5,0.5) (1.25,-1.45) rectangle ++(0.5,0.5) (2.25,0.15) rectangle ++(0.5,0.5) (2.25,-0.65) rectangle ++(0.5,0.5) (2.25,-1.45) rectangle ++(0.5,0.5) (4.25,0.15) rectangle ++(0.5,0.5) (4.25,-0.65) rectangle ++(0.5,0.5) (4.25,-1.45) rectangle ++(0.5,0.5) (1.25,-2.85) rectangle ++(0.5,0.5) (2.25,-2.85) rectangle ++(0.5,0.5) (4.25,-2.85) rectangle ++(0.5,0.5);
    \draw[line width=0.1cm,white] (1.5,0.4) -- (1.2,0.7) (1.5,-0.4) -- (1.2,-0.1)
    (1.5,-1.2) -- (1.2,-0.9)
    (1.5,-2.6) -- (1.2,-2.3)(2.5,0.4) -- (2.2,0.7) (2.5,-0.4) -- (2.2,-0.1)
    (2.5,-1.2) -- (2.2,-0.9) 
    (2.5,-2.6) -- (2.2,-2.3)(4.5,0.4) -- (4.2,0.7) (4.5,-0.4) -- (4.2,-0.1)
    (4.5,-1.2) -- (4.2,-0.9)
    (4.5,-2.6) -- (4.2,-2.3);
    \draw[ultra thick,purplish] (1.5,0.4) -- (1.2,0.7) (1.5,-0.4) -- (1.2,-0.1)
    (1.5,-1.2) -- (1.2,-0.9)
    (1.5,-2.6) -- (1.2,-2.3)(2.5,0.4) -- (2.2,0.7) (2.5,-0.4) -- (2.2,-0.1)
    (2.5,-1.2) -- (2.2,-0.9) 
    (2.5,-2.6) -- (2.2,-2.3)(4.5,0.4) -- (4.2,0.7) (4.5,-0.4) -- (4.2,-0.1)
    (4.5,-1.2) -- (4.2,-0.9)
    (4.5,-2.6) -- (4.2,-2.3);
    \draw[thick] (1.5,0.8) arc[start angle=180, end angle=0, x radius=0.075, y radius=0.3]
    (1.5,-3) arc[start angle=-180, end angle=0, x radius=0.075, y radius=0.3] (2.5,0.8) arc[start angle=180, end angle=0, x radius=0.075, y radius=0.3]
    (2.5,-3) arc[start angle=-180, end angle=0, x radius=0.075, y radius=0.3] (4.5,0.8) arc[start angle=180, end angle=0, x radius=0.075, y radius=0.3]
    (4.5,-3) arc[start angle=-180, end angle=0, x radius=0.075, y radius=0.3];
    \filldraw[blue] (5.25,-0.4) circle[radius=0.06] (5.25,-1.2) circle[radius=0.06];
    \draw[blue] (5.25,-0.1) node {$Z$}
    (5.25,-0.9) node {$Z$};
    \foreach \x in {1.1, 2.2, 4.15}{
    \draw[red] (\x,0.15) node {$\sigma^x$};
    }
\end{tikzpicture}.
\end{equation}
Considering the commutativity between symmetry defects
\begin{equation}
\begin{tikzpicture}
    \draw (2.85,0) node {$=-1\cdot$}
    (6.35,0) node {$=-1\cdot$};
    \draw[blue] (1,0.4) node {$Z$} (4,0.4) node {$Z$} (7.5,0.4) node {$Z$};
    \draw[red] (0.5,-0.4) node {$X$} (1.5,-0.4) node {$X$} (4.5,-0.4) node {$X$} (5,-0.4) node {$X$};
    \draw[thick] (0,0) -- (2.,0) (3.5,0) -- (5.5,0) (7,0) -- (8,0);
    \filldraw[red] (0.5,0) circle[radius=0.075] (1.5,0) circle[radius=0.075] (4.5,0) circle[radius=0.075] (5,0) circle[radius=0.075];
    \filldraw[blue] (1,0) circle[radius=0.075] (4,0) circle[radius=0.075] (7.5,0) circle[radius=0.075];
\end{tikzpicture}\label{eq reordering XZX}
\end{equation}
and the local tensor equation of $g_\sigma^{[y+1]}$, we have
\begin{equation}
\begin{tikzpicture}[>=stealth]
    \draw (-1.25,-1) node {$S^h(g_\sigma^{[y+1]})|\Psi_{g_\tau^{[y]}}\rangle\stackrel{\text{(\ref{eq cluster tensor eq 1})}}{=}$}
    (6.5,-1) node {$\stackrel{\text{(\ref{eq reordering XZX})}}{=}-1\cdot$};
    \foreach \x in {0,6.8}{
    \draw (\x+3.5,0.4) node {$\cdots$} (\x+3.5,-0.4) node {$\cdots$}
    (\x+3.5,-2.6) node {$\cdots$}
    (\x+1.5,-1.85) node[rotate=90] {$\cdots$} (\x+2.5,-1.85) node[rotate=90] {$\cdots$}
    (\x+4.5,-1.85) node[rotate=90] {$\cdots$};
    \draw[thick] (\x+0.75,0.4) -- (\x+3,0.4) (\x+4,0.4) -- (\x+5.5,0.4) (\x+0.75,-0.4) -- (\x+3,-0.4) (\x+4,-0.4) -- (\x+5.5,-0.4) (\x+0.75,-1.2) -- (\x+3,-1.2) (\x+4,-1.2) -- (\x+5.5,-1.2) (\x+0.75,-2.6) -- (\x+3,-2.6) (\x+4,-2.6) -- (\x+5.5,-2.6) (\x+1.5,0.8) -- (\x+1.5,-1.6) (\x+1.5,-2.2) -- (\x+1.5,-3) (\x+2.5,0.8) -- (\x+2.5,-1.6) (\x+2.5,-2.2) -- (\x+2.5,-3) (\x+4.5,0.8) -- (\x+4.5,-1.6) (\x+4.5,-2.2) -- (\x+4.5,-3) 
    (\x+0.75,0.4) arc[start angle=270, end angle=90, x radius=0.25, y radius=0.075] 
    (\x+0.75,-0.4) arc[start angle=270, end angle=90, x radius=0.25, y radius=0.075] 
    (\x+0.75,-1.2) arc[start angle=270, end angle=90, x radius=0.25, y radius=0.075] 
    (\x+0.75,-2.6) arc[start angle=270, end angle=90, x radius=0.25, y radius=0.075] 
    (\x+5.5,0.4) arc[start angle=-90, end angle=90, x radius=0.25, y radius=0.075] 
    (\x+5.5,-0.4) arc[start angle=-90, end angle=90, x radius=0.25, y radius=0.075] 
    (\x+5.5,-1.2) arc[start angle=-90, end angle=90, x radius=0.25, y radius=0.075] 
    (\x+5.5,-2.6) arc[start angle=-90, end angle=90, x radius=0.25, y radius=0.075];
    \draw[thick,fill=white] (\x+1.25,0.15) rectangle ++(0.5,0.5) (\x+1.25,-0.65) rectangle ++(0.5,0.5) 
    (\x+1.25,-1.45) rectangle ++(0.5,0.5) 
    (\x+2.25,0.15) rectangle ++(0.5,0.5) 
    (\x+2.25,-0.65) rectangle ++(0.5,0.5) 
    (\x+2.25,-1.45) rectangle ++(0.5,0.5) 
    (\x+4.25,0.15) rectangle ++(0.5,0.5) 
    (\x+4.25,-0.65) rectangle ++(0.5,0.5) 
    (\x+4.25,-1.45) rectangle ++(0.5,0.5) 
    (\x+1.25,-2.85) rectangle ++(0.5,0.5) 
    (\x+2.25,-2.85) rectangle ++(0.5,0.5) 
    (\x+4.25,-2.85) rectangle ++(0.5,0.5);
    \draw[line width=0.1cm,white] (\x+1.5,0.4) -- (\x+1.2,0.7) (\x+1.5,-0.4) -- (\x+1.2,-0.1)
    (\x+1.5,-1.2) -- (\x+1.2,-0.9)
    (\x+1.5,-2.6) -- (\x+1.2,-2.3) (\x+2.5,0.4) -- (\x+2.2,0.7) (\x+2.5,-0.4) -- (\x+2.2,-0.1)
    (\x+2.5,-1.2) -- (\x+2.2,-0.9) 
    (\x+2.5,-2.6) -- (\x+2.2,-2.3) (\x+4.5,0.4) -- (\x+4.2,0.7) (\x+4.5,-0.4) -- (\x+4.2,-0.1)
    (\x+4.5,-1.2) -- (\x+4.2,-0.9)
    (\x+4.5,-2.6) -- (\x+4.2,-2.3);
    \draw[ultra thick,purplish] (\x+1.5,0.4) -- (\x+1.2,0.7) (\x+1.5,-0.4) -- (\x+1.2,-0.1)
    (\x+1.5,-1.2) -- (\x+1.2,-0.9)
    (\x+1.5,-2.6) -- (\x+1.2,-2.3) (\x+2.5,0.4) -- (\x+2.2,0.7) (\x+2.5,-0.4) -- (\x+2.2,-0.1)
    (\x+2.5,-1.2) -- (\x+2.2,-0.9) 
    (\x+2.5,-2.6) -- (\x+2.2,-2.3) (\x+4.5,0.4) -- (\x+4.2,0.7) (\x+4.5,-0.4) -- (\x+4.2,-0.1)
    (\x+4.5,-1.2) -- (\x+4.2,-0.9)
    (\x+4.5,-2.6) -- (\x+4.2,-2.3);
    \draw[thick] (\x+1.5,0.8) arc[start angle=180, end angle=0, x radius=0.075, y radius=0.3]
    (\x+1.5,-3) arc[start angle=-180, end angle=0, x radius=0.075, y radius=0.3] 
    (\x+2.5,0.8) arc[start angle=180, end angle=0, x radius=0.075, y radius=0.3]
    (\x+2.5,-3) arc[start angle=-180, end angle=0, x radius=0.075, y radius=0.3] 
    (\x+4.5,0.8) arc[start angle=180, end angle=0, x radius=0.075, y radius=0.3]
    (\x+4.5,-3) arc[start angle=-180, end angle=0, x radius=0.075, y radius=0.3];
    \filldraw[blue] (\x+5.25,-0.4) circle[radius=0.06] (\x+5.25,-1.2) circle[radius=0.06];
    \draw[blue] (\x+5.25,-0.1) node {$Z$}
    (\x+5.25,-0.9) node {$Z$};}
    \foreach \x in {1,5}{
    \filldraw[red] (\x,-0.4) circle[radius=0.06];
    \draw[red] (\x,-0.7) node {$X$};
    }
\end{tikzpicture}\label{eq cluster state sigma charge}
\end{equation}
The boundary anomaly can be extracted from the topological response theory as
\begin{equation}
    \phi(g_\tau^{[y]},g_\sigma^{[y+1]}) \stackrel{\text{(\ref{eq def phi})}}{\equiv} \frac{\omega(g_\tau^{[y]},g_\sigma^{[y+1]})}{\omega(g_\sigma^{[y+1]},g_\tau^{[y]})} \stackrel{\text{(\ref{eq topo response subsystem})}}{=} \biggl[\langle\Psi_{g_\tau^{[y]}}|S^h(g_\sigma^{[y+1]})|\Psi_{g_\tau^{[y]}}\rangle\biggr]^{-1}\stackrel{\text{(\ref{eq cluster state sigma charge})}}{=} \biggl[\langle\Psi_{g_\tau^{[y]}}|\cdot\biggl(-|\Psi_{g_\tau^{[y]}}\rangle\biggr)\biggr]^{-1}=-1.\label{phi cluster}
\end{equation}
The same quantity can be extracted from the anomaly indicator $Q(g_\tau^{[y]},g_\sigma^{[y+1]})$ defined as Eq. (\ref{eq def Q}). 
By introducing the spatial symmetry defect
\begin{equation}
    V_{y+1}(g_\sigma^{[y+1]})=X_{L_x,y+1}.
\end{equation}
and applying normalization to the 2D cluster state, we obtain the wave function $|\psi_{g_\sigma^{[y+1]}}\rangle$. 
The subsystem symmetry transformation on $|\psi_{g_\sigma^{[y+1]}}\rangle$ is evaluated via
\begin{equation}
\begin{tikzpicture}[>=stealth]
    \draw (-1,-1) node {$S^h(g_\tau^{[y]})|\psi_{g_\sigma^{[y+1]}}\rangle=$}
    (3.5,0.4) node {$\cdots$} (3.5,-0.4) node {$\cdots$}
    (3.5,-2.6) node {$\cdots$}
    (1.5,-1.85) node[rotate=90] {$\cdots$} (2.5,-1.85) node[rotate=90] {$\cdots$}
    (4.5,-1.85) node[rotate=90] {$\cdots$} (6.5,0.45) node {$[y+2]$} (6.5,-0.35) node {$[y+1]$} (6.3,-1.15) node {$[y]$};
    \draw[thick] (1,0.4) -- (3,0.4) (4,0.4) -- (5.5,0.4) (1,-0.4) -- (3,-0.4) (4,-0.4) -- (5.5,-0.4) (1,-1.2) -- (3,-1.2) (4,-1.2) -- (5.5,-1.2) (1,-2.6) -- (3,-2.6) (4,-2.6) -- (5.5,-2.6) (1.5,0.8) -- (1.5,-1.6) (1.5,-2.2) -- (1.5,-3) (2.5,0.8) -- (2.5,-1.6) (2.5,-2.2) -- (2.5,-3) (4.5,0.8) -- (4.5,-1.6) (4.5,-2.2) -- (4.5,-3) (1,0.4) arc[start angle=270, end angle=90, x radius=0.25, y radius=0.075] (1,-0.4) arc[start angle=270, end angle=90, x radius=0.25, y radius=0.075] (1,-1.2) arc[start angle=270, end angle=90, x radius=0.25, y radius=0.075] (1,-2.6) arc[start angle=270, end angle=90, x radius=0.25, y radius=0.075] (5.5,0.4) arc[start angle=-90, end angle=90, x radius=0.25, y radius=0.075] (5.5,-0.4) arc[start angle=-90, end angle=90, x radius=0.25, y radius=0.075] (5.5,-1.2) arc[start angle=-90, end angle=90, x radius=0.25, y radius=0.075] (5.5,-2.6) arc[start angle=-90, end angle=90, x radius=0.25, y radius=0.075];
    \draw[thick,fill=white] (1.25,0.15) rectangle ++(0.5,0.5) (1.25,-0.65) rectangle ++(0.5,0.5) (1.25,-1.45) rectangle ++(0.5,0.5) (2.25,0.15) rectangle ++(0.5,0.5) (2.25,-0.65) rectangle ++(0.5,0.5) (2.25,-1.45) rectangle ++(0.5,0.5) (4.25,0.15) rectangle ++(0.5,0.5) (4.25,-0.65) rectangle ++(0.5,0.5) (4.25,-1.45) rectangle ++(0.5,0.5) (1.25,-2.85) rectangle ++(0.5,0.5) (2.25,-2.85) rectangle ++(0.5,0.5) (4.25,-2.85) rectangle ++(0.5,0.5);
    \draw[line width=0.1cm,white] (1.5,0.4) -- (1.2,0.7) (1.5,-0.4) -- (1.2,-0.1)
    (1.5,-1.2) -- (1.2,-0.9)
    (1.5,-2.6) -- (1.2,-2.3)(2.5,0.4) -- (2.2,0.7) (2.5,-0.4) -- (2.2,-0.1)
    (2.5,-1.2) -- (2.2,-0.9) 
    (2.5,-2.6) -- (2.2,-2.3)(4.5,0.4) -- (4.2,0.7) (4.5,-0.4) -- (4.2,-0.1)
    (4.5,-1.2) -- (4.2,-0.9)
    (4.5,-2.6) -- (4.2,-2.3);
    \draw[ultra thick,purplish] (1.5,0.4) -- (1.2,0.7) (1.5,-0.4) -- (1.2,-0.1)
    (1.5,-1.2) -- (1.2,-0.9)
    (1.5,-2.6) -- (1.2,-2.3)(2.5,0.4) -- (2.2,0.7) (2.5,-0.4) -- (2.2,-0.1)
    (2.5,-1.2) -- (2.2,-0.9) 
    (2.5,-2.6) -- (2.2,-2.3)(4.5,0.4) -- (4.2,0.7) (4.5,-0.4) -- (4.2,-0.1)
    (4.5,-1.2) -- (4.2,-0.9)
    (4.5,-2.6) -- (4.2,-2.3);
    \draw[thick] (1.5,0.8) arc[start angle=180, end angle=0, x radius=0.075, y radius=0.3]
    (1.5,-3) arc[start angle=-180, end angle=0, x radius=0.075, y radius=0.3] (2.5,0.8) arc[start angle=180, end angle=0, x radius=0.075, y radius=0.3]
    (2.5,-3) arc[start angle=-180, end angle=0, x radius=0.075, y radius=0.3] (4.5,0.8) arc[start angle=180, end angle=0, x radius=0.075, y radius=0.3]
    (4.5,-3) arc[start angle=-180, end angle=0, x radius=0.075, y radius=0.3];
    \filldraw[red] (5.25,-0.4) circle[radius=0.06];
    \draw[red]
    (5.2,-0.65) node {$X$};
    \foreach \x in {1.1, 2.1, 4.1}{
    \draw[blue] (\x,-0.75) node {$\tau^x$};
    }
\end{tikzpicture}.
\end{equation}
Combing the commutativity between $V_{y+1}(g_\sigma^{[y+1]})$ and $V_{y+1}(g_\tau^{[y]})$
\begin{equation}
\begin{tikzpicture}
    \draw (2.85,0) node {$=-1\cdot$}
    (6.35,0) node {$=-1\cdot$};
    \draw[red] (1,-0.4) node {$X$} (4,-0.4) node {$X$} (7.5,-0.4) node {$X$};
    \draw[blue] (0.5,0.4) node {$Z$} (1.5,0.4) node {$Z$} (4.5,0.4) node {$Z$} (5,0.4) node {$Z$};
    \draw[thick] (0,0) -- (2.,0) (3.5,0) -- (5.5,0) (7,0) -- (8,0);
    \filldraw[blue] (0.5,0) circle[radius=0.075] (1.5,0) circle[radius=0.075] (4.5,0) circle[radius=0.075] (5,0) circle[radius=0.075];
    \filldraw[red] (1,0) circle[radius=0.075] (4,0) circle[radius=0.075] (7.5,0) circle[radius=0.075];
\end{tikzpicture}\label{eq reordering ZXZ}
\end{equation}
and the local tensor equation of $g_\tau^{[y]}$, we have
\begin{equation}
\begin{tikzpicture}[>=stealth]
    \draw (-1.25,-1) node {$S^h(g_\tau^{[y]})|\psi_{g_\sigma^{[y+1]}}\rangle\stackrel{\text{(\ref{eq cluster tensor eq 1})(\ref{eq cluster tensor eq 2})}}{=}$}
    (6.5,-1) node {$\stackrel{\text{(\ref{eq reordering ZXZ})}}{=}-1\cdot$};
    \foreach \x in {0,6.8}{
    \draw (\x+3.5,0.4) node {$\cdots$} (\x+3.5,-0.4) node {$\cdots$}
    (\x+3.5,-2.6) node {$\cdots$}
    (\x+1.5,-1.85) node[rotate=90] {$\cdots$} (\x+2.5,-1.85) node[rotate=90] {$\cdots$}
    (\x+4.5,-1.85) node[rotate=90] {$\cdots$};
    \draw[thick] (\x+0.75,0.4) -- (\x+3,0.4) (\x+4,0.4) -- (\x+5.5,0.4) (\x+0.75,-0.4) -- (\x+3,-0.4) (\x+4,-0.4) -- (\x+5.5,-0.4) (\x+0.75,-1.2) -- (\x+3,-1.2) (\x+4,-1.2) -- (\x+5.5,-1.2) (\x+0.75,-2.6) -- (\x+3,-2.6) (\x+4,-2.6) -- (\x+5.5,-2.6) (\x+1.5,0.8) -- (\x+1.5,-1.6) (\x+1.5,-2.2) -- (\x+1.5,-3) (\x+2.5,0.8) -- (\x+2.5,-1.6) (\x+2.5,-2.2) -- (\x+2.5,-3) (\x+4.5,0.8) -- (\x+4.5,-1.6) (\x+4.5,-2.2) -- (\x+4.5,-3) 
    (\x+0.75,0.4) arc[start angle=270, end angle=90, x radius=0.25, y radius=0.075] 
    (\x+0.75,-0.4) arc[start angle=270, end angle=90, x radius=0.25, y radius=0.075] 
    (\x+0.75,-1.2) arc[start angle=270, end angle=90, x radius=0.25, y radius=0.075] 
    (\x+0.75,-2.6) arc[start angle=270, end angle=90, x radius=0.25, y radius=0.075] 
    (\x+5.5,0.4) arc[start angle=-90, end angle=90, x radius=0.25, y radius=0.075] 
    (\x+5.5,-0.4) arc[start angle=-90, end angle=90, x radius=0.25, y radius=0.075] 
    (\x+5.5,-1.2) arc[start angle=-90, end angle=90, x radius=0.25, y radius=0.075] 
    (\x+5.5,-2.6) arc[start angle=-90, end angle=90, x radius=0.25, y radius=0.075];
    \draw[thick,fill=white] (\x+1.25,0.15) rectangle ++(0.5,0.5) (\x+1.25,-0.65) rectangle ++(0.5,0.5) 
    (\x+1.25,-1.45) rectangle ++(0.5,0.5) 
    (\x+2.25,0.15) rectangle ++(0.5,0.5) 
    (\x+2.25,-0.65) rectangle ++(0.5,0.5) 
    (\x+2.25,-1.45) rectangle ++(0.5,0.5) 
    (\x+4.25,0.15) rectangle ++(0.5,0.5) 
    (\x+4.25,-0.65) rectangle ++(0.5,0.5) 
    (\x+4.25,-1.45) rectangle ++(0.5,0.5) 
    (\x+1.25,-2.85) rectangle ++(0.5,0.5) 
    (\x+2.25,-2.85) rectangle ++(0.5,0.5) 
    (\x+4.25,-2.85) rectangle ++(0.5,0.5);
    \draw[line width=0.1cm,white] (\x+1.5,0.4) -- (\x+1.2,0.7) (\x+1.5,-0.4) -- (\x+1.2,-0.1)
    (\x+1.5,-1.2) -- (\x+1.2,-0.9)
    (\x+1.5,-2.6) -- (\x+1.2,-2.3) (\x+2.5,0.4) -- (\x+2.2,0.7) (\x+2.5,-0.4) -- (\x+2.2,-0.1)
    (\x+2.5,-1.2) -- (\x+2.2,-0.9) 
    (\x+2.5,-2.6) -- (\x+2.2,-2.3) (\x+4.5,0.4) -- (\x+4.2,0.7) (\x+4.5,-0.4) -- (\x+4.2,-0.1)
    (\x+4.5,-1.2) -- (\x+4.2,-0.9)
    (\x+4.5,-2.6) -- (\x+4.2,-2.3);
    \draw[ultra thick,purplish] (\x+1.5,0.4) -- (\x+1.2,0.7) (\x+1.5,-0.4) -- (\x+1.2,-0.1)
    (\x+1.5,-1.2) -- (\x+1.2,-0.9)
    (\x+1.5,-2.6) -- (\x+1.2,-2.3) (\x+2.5,0.4) -- (\x+2.2,0.7) (\x+2.5,-0.4) -- (\x+2.2,-0.1)
    (\x+2.5,-1.2) -- (\x+2.2,-0.9) 
    (\x+2.5,-2.6) -- (\x+2.2,-2.3) (\x+4.5,0.4) -- (\x+4.2,0.7) (\x+4.5,-0.4) -- (\x+4.2,-0.1)
    (\x+4.5,-1.2) -- (\x+4.2,-0.9)
    (\x+4.5,-2.6) -- (\x+4.2,-2.3);
    \draw[thick] (\x+1.5,0.8) arc[start angle=180, end angle=0, x radius=0.075, y radius=0.3]
    (\x+1.5,-3) arc[start angle=-180, end angle=0, x radius=0.075, y radius=0.3] 
    (\x+2.5,0.8) arc[start angle=180, end angle=0, x radius=0.075, y radius=0.3]
    (\x+2.5,-3) arc[start angle=-180, end angle=0, x radius=0.075, y radius=0.3] 
    (\x+4.5,0.8) arc[start angle=180, end angle=0, x radius=0.075, y radius=0.3]
    (\x+4.5,-3) arc[start angle=-180, end angle=0, x radius=0.075, y radius=0.3];
    \filldraw[red] (\x+5.25,-0.4) circle[radius=0.06];
    \draw[red] (\x+5.25,-0.65) node {$X$};}
    \foreach \x in {1,5}{
    \filldraw[blue] (\x,-0.4) circle[radius=0.06];
    \filldraw[blue] (\x,-1.2) circle[radius=0.06];
    \draw[blue] (\x,-0.1) node {$Z$};
    \draw[blue] (\x,-0.9) node {$Z$};
    }
\end{tikzpicture}\label{eq cluster state tau charge}
\end{equation}
Therefore, we obtain the anomaly indicator of the twisted sector state with normalization condition $\langle\psi_{g_\sigma^{[y+1]}}|\psi_{g_\sigma^{[y+1]}}\rangle=1$ as
\begin{equation}
    Q(g_\tau^{[y]},g_\sigma^{[y+1]}) \stackrel{\text{(\ref{eq def Q})}}{=} \langle\psi_{g_\sigma^{[y+1]}}|S^h(g_\tau^{[y]})|\psi_{g_\sigma^{[y+1]}}\rangle \stackrel{\text{(\ref{eq cluster state tau charge})}}{=} -\langle\psi_{g_\sigma^{[y+1]}}|\psi_{g_\sigma^{[y+1]}}\rangle =-1.\label{Q cluster}
\end{equation}
A direct comparison between Eq. (\ref{Q cluster}) and Eq. (\ref{phi cluster}) reveals that the anomaly indicator and topological response theory provide identical boundary quantum anomalies
\begin{equation}
    Q(g_\tau^{[y]},g_\sigma^{[y+1]}) =  \biggl[\langle\Psi_{g_\tau^{[y]}}|S^h(g_\sigma^{[y+1]})|\Psi_{g_\tau^{[y]}}\rangle\biggr]^{-1} = \phi(g_\tau^{[y]},g_\sigma^{[y+1]}).
\end{equation}
When the wave function deviates from the 2D cluster state, the tensor equations fail and the boundary anomalies become analytically intractable. 
Nevertheless, we numerically compute $Q$ to characterize the boundary quantum anomaly. Following the discussion in Sec. \ref{sec mixed anomaly}, we apply the transfer matrix in Eq. (\ref{eq mixed anomaly cluster}) to calculate $Q(g_\tau^{[y]},g_\sigma^{[y+1]})$ for $L_y=2$ system, obtaining the results shown in Fig. \ref{fig cluster response}. 
Notably, the $\gamma=-1$ point in the figure exhibits perfect agreement with our analytical prediction.

\section{Extract mixed anomaly from local boundary Hilbert space}\label{app mixed anomaly}
In this appendix, we examine the commutation relations of boundary subsystem symmetry operators $W^{\mathrm{Right}}(g^{[y]})$, which are obtained by mapping the bulk symmetry operator $S^h(g^{[y]})$ to the boundary space
\begin{equation}
    S^h(g^{[y]})|\psi\rangle = W^{\mathrm{Left}}(g^{[y]}) W^{\mathrm{Right}}(g^{[y]})|\psi\rangle.
\end{equation}
As outlined in the main text, we focus on scenarios where $W^{\mathrm{Right}}(g^{[y]})$ consists of two local boundary operators
\begin{equation}
    W^{\mathrm{Right}}(g^{[y]}) = V_y(g^{[y]})V_{y+1}(g^{[y]}).\label{eq decompose}
\end{equation}
By analyzing the commutation between $W^{\mathrm{Right}}(g^{[y_1]})$ and $W^{\mathrm{Right}}(f^{[y_2]})$, we identify the boundary quantum anomalies of the group $G_h$ in the boundary Hilbert space. 
According to Eq. (\ref{eq decompose}), we show three typical cases:
\begin{itemize}
    \item When $|y_1-y_2|>1$, the operators satisfy $[W^{\mathrm{Right}}(g^{[y_1]}),W^{\mathrm{Right}}(f^{[y_2]})]=0$ because of their spatial separation, as shown in Eq. (\ref{eq decompose}). 
    \item When $|y_1-y_2|=0$, $\phi(g^{[y]},f^{[y]})$ represents the boundary anomaly within the $y$-th row of the subsystem.
    \item When $|y_1-y_2|=1$, $\phi(g^{[y]},f^{[y+1]})$ denotes the \textit{mixed quantum anomaly} between two adjacent subsystems at the $y$-th and $(y+1)$-th rows.
\end{itemize}
We will characterize the above three cases using tensor equations on a $L_y=4$ system.

(1) For $|y_1-y_2|>1$, we take $y_1=1$ and $y_2=3$ as an example. 
The resulting tensor equation is
\begin{equation}
\begin{tikzpicture}
    \draw[thick] (0,2.4) -- (0,-0.8) (-0.5,0.4) -- (0.5,0.4) (-0.5,-0.4) -- (0.5,-0.4) (-0.5,1.2) -- (0.5,1.2) (-0.5,2) -- (0.5,2);
    \draw[thick,fill=white] (-0.25,0.15) rectangle ++(0.5,0.5) (-0.25,-0.65) rectangle ++(0.5,0.5) (-0.25,0.95) rectangle ++(0.5,0.5) (-0.25,1.75) rectangle ++(0.5,0.5);
    \draw[thick] (0,2.4) arc[start angle=180, end angle=0, x radius=0.05, y radius=0.2]
    (0,-0.8) arc[start angle=-180, end angle=0, x radius=0.05, y radius=0.2];
    \draw[line width=0.1cm,white] (0,0.4) -- (-0.3,0.7) (0,-0.4) -- (-0.3,-0.1) (0,-0.4) -- (-0.3,-0.1) (0,2) -- (-0.3,2.3) (0,1.2) -- (-0.3,1.5);
    \draw[ultra thick,purplish] (0,0.4) -- (-0.3,0.7) (0,-0.4) -- (-0.3,-0.1) (0,2) -- (-0.3,2.3) (0,1.2) -- (-0.3,1.5);
    \draw (0.8,0.8) node {$=$};
    \draw[purplish] (-1.2,0.1) node {$U_{x,y_1}(g^{[y_1]})$} (-1.2,1.7) node {$U_{x,y_2}(f^{[y_2]})$} (-0.45,2.35) node {$I$} (-0.45,0.75) node {$I$};
    \draw (2,0) node {$W^{\mathrm{Left}}(g^{[y_1]})$} (4.25,0.4) node {$V^{-1}_{y_1+1}(g^{[y_1]})$} (4.3,-0.4) node {$V^{-1}_{y_1}(g^{[y_1]})$};
    \draw (9.75,0) node {$W^{\mathrm{Right}}(g^{[y_1]})$} (7.25,0.4) node {$V_{y_1+1}(g^{[y_1]})$} (7.2,-0.4) node {$V_{y_1}(g^{[y_1]})$};
    \node[rotate=270] at (3.1,0) {$\underbrace{\hspace{1.4cm}}$};
    \node[rotate=-270] at (8.5,0) {$\underbrace{\hspace{1.4cm}}$};
    \draw (2,1.6) node {$W^{\mathrm{Left}}(f^{[y_2]})$} (4.25,2) node {$V^{-1}_{y_2+1}(f^{[y_2]})$} (4.3,1.2) node {$V^{-1}_{y_2}(f^{[y_2]})$};
    \draw (9.75,1.6) node {$W^{\mathrm{Right}}(f^{[y_2]})$} (7.25,2) node {$V_{y_2+1}(f^{[y_2]})$} (7.2,1.2) node {$V_{y_2}(f^{[y_2]})$};
    \node[rotate=270] at (3.1,1.6) {$\underbrace{\hspace{1.4cm}}$};
    \node[rotate=-270] at (8.5,1.6) {$\underbrace{\hspace{1.4cm}}$};
    \draw[thick] (5.75,2.4) -- (5.75,-0.8) (5.25,0.4) -- (6.25,0.4) (5.25,-0.4) -- (6.25,-0.4) (5.25,2) -- (6.25,2) (5.25,1.2) -- (6.25,1.2) ;
    \draw[thick,fill=white] (5.5,0.15) rectangle ++(0.5,0.5) (5.5,-0.65) rectangle ++(0.5,0.5) (5.5,0.95) rectangle ++(0.5,0.5) (5.5,1.75) rectangle ++(0.5,0.5);
    \draw[thick] (5.75,2.4) arc[start angle=180, end angle=0, x radius=0.05, y radius=0.2]
    (5.75,-0.8) arc[start angle=-180, end angle=0, x radius=0.05, y radius=0.2];
    \draw[line width=0.1cm,white] (5.75,0.4) -- (5.45,0.7) (5.75,-0.4) -- (5.45,-0.1) (5.75,2) -- (5.45,2.3) (5.75,1.2) -- (5.45,1.5);
    \draw[ultra thick,purplish] (5.75,0.4) -- (5.45,0.7) (5.75,-0.4) -- (5.45,-0.1) (5.75,2) -- (5.45,2.3) (5.75,1.2) -- (5.45,1.5);
\end{tikzpicture}.
\end{equation}
Since $W^{\mathrm{Right}}(g^{[y_1]})$ and $W^{\mathrm{Right}}(g^{[y_2]})$ act in different subspaces, they are always commutative. Therefore, we have
\begin{equation}
    W^{\mathrm{Right}}(g^{[y_1]})W^{\mathrm{Right}}(g^{[y_2]}) = W^{\mathrm{Right}}(g^{[y_2]})W^{\mathrm{Right}}(g^{[y_1]}),\quad |y_1-y_2|>1.
\end{equation}

(2) When $|y_1-y_2|=0$, the operators $S^h(g^{[y]})$ and $S^h(f^{[y]})$ either act simultaneously or sequentially on the tensor. 
Sequential symmetry action $S^h(g^{[y]})S^h(f^{[y]})$ yields
\begin{equation}
\begin{tikzpicture}
    \draw[thick] (0,2.4) -- (0,-0.8) (-0.5,0.4) -- (0.5,0.4) (-0.5,-0.4) -- (0.5,-0.4) (-0.5,1.2) -- (0.5,1.2) (-0.5,2) -- (0.5,2);
    \draw[thick,fill=white] (-0.25,0.15) rectangle ++(0.5,0.5) (-0.25,-0.65) rectangle ++(0.5,0.5) (-0.25,0.95) rectangle ++(0.5,0.5) (-0.25,1.75) rectangle ++(0.5,0.5);
    \draw[thick] (0,2.4) arc[start angle=180, end angle=0, x radius=0.05, y radius=0.2]
    (0,-0.8) arc[start angle=-180, end angle=0, x radius=0.05, y radius=0.2];
    \draw[line width=0.1cm,white] (0,0.4) -- (-0.3,0.7) (0,-0.4) -- (-0.3,-0.1) (0,-0.4) -- (-0.3,-0.1) (0,2) -- (-0.3,2.3) (0,1.2) -- (-0.3,1.5);
    \draw[ultra thick,purplish] (0,0.4) -- (-0.3,0.7) (0,-0.4) -- (-0.3,-0.1) (0,2) -- (-0.3,2.3) (0,1.2) -- (-0.3,1.5);
    
    \draw[purplish] (-1.8,0.1) node {$U_{x,y}(g^{[y]})U_{x,y}(f^{[y]})$} (-0.45,1.7) node {$I$} (-0.45,2.35) node {$I$} (-0.45,0.75) node {$I$};

    \draw[thick] (0,-1.6) -- (0,-4.8) (-0.5,-3.6) -- (0.5,-3.6) (-0.5,-4.4) -- (0.5,-4.4) (-0.5,-2.8) -- (0.5,-2.8) (-0.5,-2) -- (0.5,-2);
    \draw[thick,fill=white] (-0.25,-3.85) rectangle ++(0.5,0.5) (-0.25,-4.65) rectangle ++(0.5,0.5) (-0.25,-3.05) rectangle ++(0.5,0.5) (-0.25,-2.25) rectangle ++(0.5,0.5);
    \draw[thick] (0,-1.6) arc[start angle=180, end angle=0, x radius=0.05, y radius=0.2]
    (0,-4.8) arc[start angle=-180, end angle=0, x radius=0.05, y radius=0.2];
    \draw[line width=0.1cm,white] (0,-3.6) -- (-0.3,-3.3) (0,-4.4) -- (-0.3,-4.1) (0,-4.4) -- (-0.3,-4.1) (0,-2) -- (-0.3,-1.7) (0,-2.8) -- (-0.3,-2.5);
    \draw[ultra thick,purplish] (0,-3.6) -- (-0.3,-3.3) (0,-4.4) -- (-0.3,-4.1) (0,-2) -- (-0.3,-1.7) (0,-2.8) -- (-0.3,-2.5);
    \draw (-8,-3) node {$=$};
    \draw (-6.2,-4) node {$W^{\mathrm{Left}}(f^{[y]})W^{\mathrm{Left}}(g^{[y]})$} (-2.2,-3.6) node {$V^{-1}_{y+1}(f^{[y]})V^{-1}_{y+1}(g^{[y]})$} (-2.15,-4.4) node {$V^{-1}_{y}(f^{[y]})V^{-1}_{y}(g^{[y]})$};
    
    \draw (6.2,-4) node {$W^{\mathrm{Right}}(g^{[y]})W^{\mathrm{Right}}(f^{[y]})$} (2.2,-3.6) node {$V_{y+1}(g^{[y]})V_{y+1}(f^{[y]})$} (1.9,-4.4) node {$V_{y}(g^{[y]})V_{y}(f^{[y]})$};  \node[rotate=270] at (-4,-4) {$\underbrace{\hspace{1.4cm}}$};
    \node[rotate=-270] at (4,-4) {$\underbrace{\hspace{1.4cm}}$};
\end{tikzpicture}.\label{eq sequential same}
\end{equation}
The simultaneous symmetry transformation $S^h(g^{[y]}f^{[y]})$ produces
\begin{equation}
\begin{tikzpicture}
    \draw[thick] (0,2.4) -- (0,-0.8) (-0.5,0.4) -- (0.5,0.4) (-0.5,-0.4) -- (0.5,-0.4) (-0.5,1.2) -- (0.5,1.2) (-0.5,2) -- (0.5,2);
    \draw[thick,fill=white] (-0.25,0.15) rectangle ++(0.5,0.5) (-0.25,-0.65) rectangle ++(0.5,0.5) (-0.25,0.95) rectangle ++(0.5,0.5) (-0.25,1.75) rectangle ++(0.5,0.5);
    \draw[thick] (0,2.4) arc[start angle=180, end angle=0, x radius=0.05, y radius=0.2]
    (0,-0.8) arc[start angle=-180, end angle=0, x radius=0.05, y radius=0.2];
    \draw[line width=0.1cm,white] (0,0.4) -- (-0.3,0.7) (0,-0.4) -- (-0.3,-0.1) (0,-0.4) -- (-0.3,-0.1) (0,2) -- (-0.3,2.3) (0,1.2) -- (-0.3,1.5);
    \draw[ultra thick,purplish] (0,0.4) -- (-0.3,0.7) (0,-0.4) -- (-0.3,-0.1) (0,2) -- (-0.3,2.3) (0,1.2) -- (-0.3,1.5);
    
    \draw[purplish] (-1.5,0.1) node {$U_{x,y}(g^{[y]}f^{[y]})$} (-0.45,1.7) node {$I$} (-0.45,2.35) node {$I$} (-0.45,0.75) node {$I$};

    \draw[thick] (6.6,2.4) -- (6.6,-0.8) (6.1,0.4) -- (7.1,0.4) (6.1,-0.4) -- (7.1,-0.4) (6.1,1.2) -- (7.1,1.2) (6.1,2) -- (7.1,2);
    \draw[thick,fill=white] (6.35,0.15) rectangle ++(0.5,0.5) (6.35,-0.65) rectangle ++(0.5,0.5) (6.35,0.95) rectangle ++(0.5,0.5) (6.35,1.75) rectangle ++(0.5,0.5);
    \draw[thick] (6.6,2.4) arc[start angle=180, end angle=0, x radius=0.05, y radius=0.2]
    (6.6,-0.8) arc[start angle=-180, end angle=0, x radius=0.05, y radius=0.2];
    \draw[line width=0.1cm,white] (6.6,0.4) -- (6.3,0.7) (6.6,-0.4) -- (6.3,-0.1) (6.6,-0.4) -- (6.3,-0.1) (6.6,2) -- (6.3,2.3) (6.6,1.2) -- (6.3,1.5);
    \draw[ultra thick,purplish] (6.6,0.4) -- (6.3,0.7) (6.6,-0.4) -- (6.3,-0.1) (6.6,2) -- (6.3,2.3) (6.6,1.2) -- (6.3,1.5);
    \draw (1,0.8) node {$=$};
    \draw (2.5,0) node {$W^{\mathrm{Left}}(g^{[y]}f^{[y]})$} (5,0.4) node {$V^{-1}_{y+1}(g^{[y]}f^{[y]})$} (5,-0.4) node {$V^{-1}_{y}(g^{[y]}f^{[y]})$};
    
    \draw (11,0) node {$W^{\mathrm{Right}}(g^{[y]}f^{[y]})$} (8.3,0.4) node {$V_{y+1}(g^{[y]}f^{[y]})$} (8.15,-0.4) node {$V_{y}(g^{[y]}f^{[y]})$};  \node[rotate=270] at (3.8,0) {$\underbrace{\hspace{1.4cm}}$};
    \node[rotate=-270] at (9.5,0) {$\underbrace{\hspace{1.4cm}}$};
\end{tikzpicture}.\label{eq simutaneous same}
\end{equation}
Combining Eq. (\ref{eq sequential same}) and Eq. (\ref{eq simutaneous same}) results in the following tensor equation
\begin{equation}
\begin{tikzpicture}
    \draw[thick] (0,2.4) -- (0,-0.8) (-0.5,0.4) -- (0.5,0.4) (-0.5,-0.4) -- (0.5,-0.4) (-0.5,1.2) -- (0.5,1.2) (-0.5,2) -- (0.5,2);
    \draw[thick,fill=white] (-0.25,0.15) rectangle ++(0.5,0.5) (-0.25,-0.65) rectangle ++(0.5,0.5) (-0.25,0.95) rectangle ++(0.5,0.5) (-0.25,1.75) rectangle ++(0.5,0.5);
    \draw[thick] (0,2.4) arc[start angle=180, end angle=0, x radius=0.05, y radius=0.2]
    (0,-0.8) arc[start angle=-180, end angle=0, x radius=0.05, y radius=0.2];
    \draw[line width=0.1cm,white] (0,0.4) -- (-0.3,0.7) (0,-0.4) -- (-0.3,-0.1) (0,-0.4) -- (-0.3,-0.1) (0,2) -- (-0.3,2.3) (0,1.2) -- (-0.3,1.5);
    \draw[ultra thick,purplish] (0,0.4) -- (-0.3,0.7) (0,-0.4) -- (-0.3,-0.1) (0,2) -- (-0.3,2.3) (0,1.2) -- (-0.3,1.5);
    \draw (-6.2,0) node {$W^{\mathrm{Left}}(f^{[y]})W^{\mathrm{Left}}(g^{[y]})$} (-2.2,0.4) node {$V_{y+1}^{-1}(f^{[y]})V^{-1}_{y+1}(g^{[y]})$} (-2.1,-0.4) node {$V_{y}^{-1}(f^{[y]})V^{-1}_{y}(g^{[y]})$};
    
    \draw (6.3,0) node {$W^{\mathrm{Right}}(g^{[y]})W^{\mathrm{Right}}(f^{[y]})$} (2.2,0.4) node {$V_{y+1}(g^{[y]})V_{y+1}(f^{[y]})$} (1.8,-0.4) node {$V_{y}(g^{[y]})V_{y}(f^{[y]})$};  
    \node[rotate=270] at (-4,0) {$\underbrace{\hspace{1.4cm}}$};
    \node[rotate=-270] at (4,0) {$\underbrace{\hspace{1.4cm}}$};

    \draw[thick] (0,-1.6) -- (0,-4.8) (-0.5,-3.6) -- (0.5,-3.6) (-0.5,-4.4) -- (0.5,-4.4) (-0.5,-2.8) -- (0.5,-2.8) (-0.5,-2) -- (0.5,-2);
    \draw[thick,fill=white] (-0.25,-3.85) rectangle ++(0.5,0.5) (-0.25,-4.65) rectangle ++(0.5,0.5) (-0.25,-3.05) rectangle ++(0.5,0.5) (-0.25,-2.25) rectangle ++(0.5,0.5);
    \draw[thick] (0,-1.6) arc[start angle=180, end angle=0, x radius=0.05, y radius=0.2]
    (0,-4.8) arc[start angle=-180, end angle=0, x radius=0.05, y radius=0.2];
    \draw[line width=0.1cm,white] (0,-3.6) -- (-0.3,-3.3) (0,-4.4) -- (-0.3,-4.1) (0,-4.4) -- (-0.3,-4.1) (0,-2) -- (-0.3,-1.7) (0,-2.8) -- (-0.3,-2.5);
    \draw[ultra thick,purplish] (0,-3.6) -- (-0.3,-3.3) (0,-4.4) -- (-0.3,-4.1) (0,-2) -- (-0.3,-1.7) (0,-2.8) -- (-0.3,-2.5);
    \draw (-6,-3) node {$=$};
    \draw (-4.3,-4) node {$W^{\mathrm{Left}}(g^{[y]}f^{[y]})$} (-1.7,-3.6) node {$V^{-1}_{y+1}(g^{[y]}f^{[y]})$} (-1.6,-4.4) node {$V^{-1}_{y}(g^{[y]}f^{[y]})$};
    
    \draw (4.3,-4) node {$W^{\mathrm{Right}}(g^{[y]}f^{[y]})$} (1.6,-3.6) node {$V_{y+1}(g^{[y]}f^{[y]})$} (1.45,-4.4) node {$V_{y}(g^{[y]}f^{[y]})$};  \node[rotate=270] at (-2.9,-4) {$\underbrace{\hspace{1.4cm}}$};
    \node[rotate=-270] at (2.8,-4) {$\underbrace{\hspace{1.4cm}}$};
\end{tikzpicture}
\end{equation}
The formula above may differ by a $U(1)$ factor of $\omega^{*}(g^{[y]},f^{[y]})$ and $\omega(g^{[y]},f^{[y]})$ on the left and right boundaries, leading to the boundary symmetry condition
\begin{equation}
    W^{\mathrm{Right}}(g^{[y]})W^{\mathrm{Right}}(f^{[y]}) = \omega(g^{[y]},f^{[y]}) W^{\mathrm{Right}}(g^{[y]}f^{[y]})
\end{equation}
The corresponding topological invariant $\phi(g^{[y]},f^{[y]})=\omega(g^{[y]},f^{[y]})/\omega(f^{[y]},g^{[y]})$ characterizes the quantum anomaly associated with the subsystem symmetry transformation $S^h(g^{[y]})$ within the $y$-th row.

(3) When $|y_1-y_2|=1$, we examine the symmetry transformations $S^h(g^{[y]})$ and $S^h(f^{[y+1]})$ of adjacent subsystems. 
Sequential application of $S^h(g^{[y]})S^h(f^{[y+1]})$ results in the tensor equation
\begin{equation}
\begin{tikzpicture}
    \draw[thick] (0,2.4) -- (0,-0.8) (-0.5,0.4) -- (0.5,0.4) (-0.5,-0.4) -- (0.5,-0.4) (-0.5,1.2) -- (0.5,1.2) (-0.5,2) -- (0.5,2);
    \draw[thick,fill=white] (-0.25,0.15) rectangle ++(0.5,0.5) (-0.25,-0.65) rectangle ++(0.5,0.5) (-0.25,0.95) rectangle ++(0.5,0.5) (-0.25,1.75) rectangle ++(0.5,0.5);
    \draw[thick] (0,2.4) arc[start angle=180, end angle=0, x radius=0.05, y radius=0.2]
    (0,-0.8) arc[start angle=-180, end angle=0, x radius=0.05, y radius=0.2];
    \draw[line width=0.1cm,white] (0,0.4) -- (-0.3,0.7) (0,-0.4) -- (-0.3,-0.1) (0,-0.4) -- (-0.3,-0.1) (0,2) -- (-0.3,2.3) (0,1.2) -- (-0.3,1.5);
    \draw[ultra thick,purplish] (0,0.4) -- (-0.3,0.7) (0,-0.4) -- (-0.3,-0.1) (0,2) -- (-0.3,2.3) (0,1.2) -- (-0.3,1.5);
    
    \draw[purplish] (-1.2,0.1) node {$U_{x,y}(g^{[y]})$} (-0.45,1.7) node {$I$} (-0.45,2.35) node {$I$} (-1.5,0.75) node {$U_{x,y+1}(f^{[y+1]})$};

    \draw[thick] (0,-1.6) -- (0,-4.8) (-0.5,-3.6) -- (0.5,-3.6) (-0.5,-4.4) -- (0.5,-4.4) (-0.5,-2.8) -- (0.5,-2.8) (-0.5,-2) -- (0.5,-2);
    \draw[thick,fill=white] (-0.25,-3.85) rectangle ++(0.5,0.5) (-0.25,-4.65) rectangle ++(0.5,0.5) (-0.25,-3.05) rectangle ++(0.5,0.5) (-0.25,-2.25) rectangle ++(0.5,0.5);
    \draw[thick] (0,-1.6) arc[start angle=180, end angle=0, x radius=0.05, y radius=0.2]
    (0,-4.8) arc[start angle=-180, end angle=0, x radius=0.05, y radius=0.2];
    \draw[line width=0.1cm,white] (0,-3.6) -- (-0.3,-3.3) (0,-4.4) -- (-0.3,-4.1) (0,-4.4) -- (-0.3,-4.1) (0,-2) -- (-0.3,-1.7) (0,-2.8) -- (-0.3,-2.5);
    \draw[ultra thick,purplish] (0,-3.6) -- (-0.3,-3.3) (0,-4.4) -- (-0.3,-4.1) (0,-2) -- (-0.3,-1.7) (0,-2.8) -- (-0.3,-2.5);
    \draw (-9,-3) node {$=$};
    \draw (-6.5,-3.6) node {$W^{\mathrm{Left}}(f^{[y+1]})W^{\mathrm{Left}}(g^{[y]})$} (-2.35,-3.6) node {$V^{-1}_{y+1}(f^{[y+1]})V^{-1}_{y+1}(g^{[y]})$} (-1.3,-4.4) node {$V^{-1}_{y}(g^{[y]})$}
    (-1.55,-2.8) node {$V^{-1}_{y+2}(f^{[y+1]})$};
    
    \draw (6.6,-3.6) node {$W^{\mathrm{Right}}(g^{[y]})W^{\mathrm{Right}}(f^{[y+1]})$} (2.3,-3.6) node {$V_{y+1}(g^{[y]})V_{y+1}(f^{[y+1]})$} (1.2,-4.4) node {$V_{y}(g^{[y]})$} (1.55,-2.8) node {$V_{y+2}(f^{[y+1]})$};  \node[rotate=270] at (-4.3,-3.6) {$\underbrace{\hspace{2cm}}$};
    \node[rotate=-270] at (4.3,-3.6) {$\underbrace{\hspace{2cm}}$};
\end{tikzpicture},\label{eq sequential mixed}
\end{equation}
whereas simultaneous transformation $S^h(g^{[y]}f^{[y+1]})$ gives
\begin{equation}
\begin{tikzpicture}
    \draw[thick] (0,2.4) -- (0,-0.8) (-0.5,0.4) -- (0.5,0.4) (-0.5,-0.4) -- (0.5,-0.4) (-0.5,1.2) -- (0.5,1.2) (-0.5,2) -- (0.5,2);
    \draw[thick,fill=white] (-0.25,0.15) rectangle ++(0.5,0.5) (-0.25,-0.65) rectangle ++(0.5,0.5) (-0.25,0.95) rectangle ++(0.5,0.5) (-0.25,1.75) rectangle ++(0.5,0.5);
    \draw[thick] (0,2.4) arc[start angle=180, end angle=0, x radius=0.05, y radius=0.2]
    (0,-0.8) arc[start angle=-180, end angle=0, x radius=0.05, y radius=0.2];
    \draw[line width=0.1cm,white] (0,0.4) -- (-0.3,0.7) (0,-0.4) -- (-0.3,-0.1) (0,-0.4) -- (-0.3,-0.1) (0,2) -- (-0.3,2.3) (0,1.2) -- (-0.3,1.5);
    \draw[ultra thick,purplish] (0,0.4) -- (-0.3,0.7) (0,-0.4) -- (-0.3,-0.1) (0,2) -- (-0.3,2.3) (0,1.2) -- (-0.3,1.5);
    
    \draw[purplish] (-1.2,0.1) node {$U_{x,y}(g^{[y]})$} (-0.45,1.7) node {$I$} (-0.45,2.35) node {$I$} (-1.5,0.75) node {$U_{x,y+1}(f^{[y+1]})$};

    \draw[thick] (7.1,2.4) -- (7.1,-0.8) (6.6,0.4) -- (7.6,0.4) (6.6,-0.4) -- (7.6,-0.4) (6.6,1.2) -- (7.6,1.2) (6.6,2) -- (7.6,2);
    \draw[thick,fill=white] (6.85,0.15) rectangle ++(0.5,0.5) (6.85,-0.65) rectangle ++(0.5,0.5) (6.85,0.95) rectangle ++(0.5,0.5) (6.85,1.75) rectangle ++(0.5,0.5);
    \draw[thick] (7.1,2.4) arc[start angle=180, end angle=0, x radius=0.05, y radius=0.2]
    (7.1,-0.8) arc[start angle=-180, end angle=0, x radius=0.05, y radius=0.2];
    \draw[line width=0.1cm,white] (7.1,0.4) -- (6.8,0.7) (7.1,-0.4) -- (6.8,-0.1) (7.1,-0.4) -- (6.8,-0.1) (7.1,2) -- (6.8,2.3) (7.1,1.2) -- (6.8,1.5);
    \draw[ultra thick,purplish] (7.1,0.4) -- (6.8,0.7) (7.1,-0.4) -- (6.8,-0.1) (7.1,2) -- (6.8,2.3) (7.1,1.2) -- (6.8,1.5);
    \draw (1,0.8) node {$=$};
    \draw (2.5,0.4) node {$W^{\mathrm{Left}}(g^{[y]}f^{[y+1]})$} (5.35,0.4) node {$V^{-1}_{y+1}(g^{[y]}f^{[y+1]})$} (5.8,-0.4) node {$V^{-1}_{y}(g^{[y]})$} (5.55,1.2) node {$V^{-1}_{y+2}(f^{[y+1]})$};
    
    \draw (11.85,0.4) node {$W^{\mathrm{Right}}(g^{[y]}f^{[y+1]})$} (8.9,0.4) node {$V_{y+1}(g^{[y]}f^{[y+1]})$} (8.3,-0.4) node {$V_{y}(g^{[y]})$} (8.65,1.2) node {$V_{y+2}(f^{[y+1]})$};  \node[rotate=270] at (4,0.4) {$\underbrace{\hspace{2cm}}$};
    \node[rotate=-270] at (10.25,0.4) {$\underbrace{\hspace{2cm}}$};
\end{tikzpicture}.\label{eq simutaneous mixed}
\end{equation}
Combining Eq. (\ref{eq sequential mixed}) and Eq. (\ref{eq simutaneous mixed}) produces
\begin{equation}
\begin{tikzpicture}
    \draw[thick] (0,2.4) -- (0,-0.8) (-0.5,0.4) -- (0.5,0.4) (-0.5,-0.4) -- (0.5,-0.4) (-0.5,1.2) -- (0.5,1.2) (-0.5,2) -- (0.5,2);
    \draw[thick,fill=white] (-0.25,0.15) rectangle ++(0.5,0.5) (-0.25,-0.65) rectangle ++(0.5,0.5) (-0.25,0.95) rectangle ++(0.5,0.5) (-0.25,1.75) rectangle ++(0.5,0.5);
    \draw[thick] (0,2.4) arc[start angle=180, end angle=0, x radius=0.05, y radius=0.2]
    (0,-0.8) arc[start angle=-180, end angle=0, x radius=0.05, y radius=0.2];
    \draw[line width=0.1cm,white] (0,0.4) -- (-0.3,0.7) (0,-0.4) -- (-0.3,-0.1) (0,-0.4) -- (-0.3,-0.1) (0,2) -- (-0.3,2.3) (0,1.2) -- (-0.3,1.5);
    \draw[ultra thick,purplish] (0,0.4) -- (-0.3,0.7) (0,-0.4) -- (-0.3,-0.1) (0,2) -- (-0.3,2.3) (0,1.2) -- (-0.3,1.5);
    \draw (-6.2,0.4) node {$W^{\mathrm{Left}}(f^{[y+1]})W^{\mathrm{Left}}(g^{[y]})$} (-2.25,0.4) node {$V_{y+1}^{-1}(f^{[y+1]})V^{-1}_{y+1}(g^{[y]})$} (-1.3,-0.4) node {$V^{-1}_{y}(g^{[y]})$}
    (-1.5,1.2) node {$V_{y+2}^{-1}(f^{[y+1]})$};
    
    \draw (6.3,0.4) node {$W^{\mathrm{Right}}(g^{[y]})W^{\mathrm{Right}}(f^{[y+1]})$} (2.2,0.4) node {$V_{y+1}(g^{[y]})V_{y+1}(f^{[y+1]})$} (1.1,-0.4) node {$V_{y}(g^{[y]})$} (1.5,1.2) node {$V_{y+2}(f^{[y+1]})$};  
    \node[rotate=270] at (-4.1,0.4) {$\underbrace{\hspace{2cm}}$};
    \node[rotate=-270] at (4,0.4) {$\underbrace{\hspace{2cm}}$};
    \draw[red,thick,dashed] (0.5,0.8) -- (3.9,0.8) -- (3.9,0) -- (0.5,0) -- cycle;

    \draw[thick] (0,-1.6) -- (0,-4.8) (-0.5,-3.6) -- (0.5,-3.6) (-0.5,-4.4) -- (0.5,-4.4) (-0.5,-2.8) -- (0.5,-2.8) (-0.5,-2) -- (0.5,-2);
    \draw[thick,fill=white] (-0.25,-3.85) rectangle ++(0.5,0.5) (-0.25,-4.65) rectangle ++(0.5,0.5) (-0.25,-3.05) rectangle ++(0.5,0.5) (-0.25,-2.25) rectangle ++(0.5,0.5);
    \draw[thick] (0,-1.6) arc[start angle=180, end angle=0, x radius=0.05, y radius=0.2]
    (0,-4.8) arc[start angle=-180, end angle=0, x radius=0.05, y radius=0.2];
    \draw[line width=0.1cm,white] (0,-3.6) -- (-0.3,-3.3) (0,-4.4) -- (-0.3,-4.1) (0,-4.4) -- (-0.3,-4.1) (0,-2) -- (-0.3,-1.7) (0,-2.8) -- (-0.3,-2.5);
    \draw[ultra thick,purplish] (0,-3.6) -- (-0.3,-3.3) (0,-4.4) -- (-0.3,-4.1) (0,-2) -- (-0.3,-1.7) (0,-2.8) -- (-0.3,-2.5);
    \draw (-6.5,-3) node {$=$};
    \draw (-4.6,-3.6) node {$W^{\mathrm{Left}}(g^{[y]}f^{[y+1]})$} (-1.75,-3.6) node {$V^{-1}_{y+1}(g^{[y]}f^{[y+1]})$} (-1.25,-4.4) node {$V^{-1}_{y}(g^{[y]})$} (-1.55,-2.8) node {$V^{-1}_{y+2}(f^{[y+1]})$};
    
    \draw (4.6,-3.6) node {$W^{\mathrm{Right}}(g^{[y]}f^{[y+1]})$} (1.7,-3.6) node {$V_{y+1}(g^{[y]}f^{[y+1]})$} (1.15,-4.4) node {$V_{y}(g^{[y]})$} (1.45,-2.8) node {$V_{y+2}(f^{[y+1]})$};  \node[rotate=270] at (-3.1,-3.6) {$\underbrace{\hspace{2cm}}$};
    \node[rotate=-270] at (3,-3.6) {$\underbrace{\hspace{2cm}}$};
    \draw[red,thick,dashed] (0.5,-3.2) -- (2.9,-3.2) -- (2.9,-4) -- (0.5,-4) -- cycle;
\end{tikzpicture},
\end{equation}
which may also exhibit a $U(1)$ phase ambiguity on both boundaries. 
This yields the boundary symmetry condition for adjacent subsystems in the $y$-th and the $(y+1)$-th rows
\begin{equation}
    W^{\mathrm{Right}}(g^{[y]})W^{\mathrm{Right}}(f^{[y+1]}) = \omega(g^{[y]},f^{[y+1]}) W^{\mathrm{Right}}(g^{[y]}f^{[y+1]}).
\end{equation}
The \textit{mixed quantum anomaly} of $G_s^{[y]}\times G_s^{[y+1]}$ is calculated using the projected representations $V_{y+1}$ on a single virtual bond, as represented by the red dashed box. 
The anti-commutativity of adjacent subsystems is described by
\begin{equation}
    W^{\mathrm{Right}}(g^{[y]})W^{\mathrm{Right}}(f^{[y+1]}) = \phi(g^{[y]},f^{[y+1]})W^{\mathrm{Right}}(f^{[y+1]})W^{\mathrm{Right}}(g^{[y]}).\label{eq proj rep W}
\end{equation}
Given that $W^{\mathrm{Right}}(g^{[y]})$ can be decomposed into two adjacent local boundary Hilbert spaces
\begin{equation}
    W^{\mathrm{Right}}(g^{[y]}) = V_y(g^{[y]})V_{y+1}(g^{[y]}),
\end{equation}
we reformulate Eq. (\ref{eq proj rep W}) as
\begin{equation}
    V_{y}(g^{[y]})V_{y+1}(g^{[y]})V_{y+1}(f^{[y+1]})V_{y+2}(f^{[y+1]}) = \phi(g^{[y]},f^{[y+1]}) V_{y+1}(f^{[y+1]})V_{y+2}(f^{[y+1]})V_{y}(g^{[y]})V_{y+1}(g^{[y]}).
\end{equation}
Considering that $[V_{y_1},V_{y_2}] = 0$ for $y_1\neq y_2$, we conclude
\begin{equation}
    V_{y+1}(g^{[y]})V_{y+1}(f^{[y+1]}) = \phi(g^{[y]},f^{[y+1]}) V_{y+1}(f^{[y+1]})V_{y+1}(g^{[y]}).\label{eq single bond phi}
\end{equation}
Therefore, $\phi(g^{[y]},f^{[y+1]})$ is determined by analyzing the group representation $V_{y+1}(g^{[y]})$ within each local boundary space. 
For a strong SSPT phase with translational invariance, it is sufficient to explore the mixed anomaly between two adjacent subsystem symmetries $S^h(g^{[y]})$ and $S^h(f^{[y+1]})$.

\section{Numerical calculation of $Q(g^{[y]},f^{[y+1]})$}\label{app anomaly detection numerical}
In this appendix, we outline the numerical methods for computing the symmetry charge of the twisted sector state, defined as
\begin{align}
\begin{aligned}
    Q(g^{[y]},f^{[y+1]}) = \frac{\langle\psi_{f^{[y+1]}}|S^h(g^{[y]})|\psi_{f^{[y+1]}}\rangle}{\langle\psi_{f^{[y+1]}}|\psi_{f^{[y+1]}}\rangle}.
\end{aligned}
\end{align}
We denote the spectrum of the transfer matrix $\mathbb{T}(g^{[y]},f^{[y+1]})$ as
\begin{equation}
    \mathbb{T}(g^{[y]},f^{[y+1]}) = \sum_{\alpha=0}^{\chi-1} \lambda_\alpha(g^{[y]},f^{[y+1]}) |r_\alpha)(l_\alpha|,
\end{equation}
with eigenvalues ordered as $|\lambda_0(g^{[y]},e)|\geq|\lambda_1(g^{[y]},e)|\geq\cdots$ and satisfying the bi-orthogonality condition $(l_\alpha|r_\beta)=\delta_{\alpha\beta}$.
The symmetry charge of the twisted sector state is computed using
\begin{align}
\begin{aligned}
    Q(g^{[y]},f^{[y+1]})=&\frac{\mathrm{Tr}[\mathbb{T}(g^{[y]},e)\cdots\mathbb{T}(g^{[y]},e)\mathbb{T}(g^{[y]},f^{[y+1]})]}{\mathrm{Tr}[\mathbb{T}(e,e)\cdots\mathbb{T}(e,e)\mathbb{T}(e,f^{[y+1]})]}\\
    =&\frac{\sum_{\alpha=0}^{\chi-1} \lambda_\alpha^{L_x-1}(g^{[y]},e)(l_\alpha|\mathbb{T}(g^{[y]},f^{[y+1]})|r_\alpha)}{\sum_{\alpha=0}^{\chi-1} \lambda_\alpha^{L_x-1}(e,e)(l_\alpha'|\mathbb{T}(e,f^{[y+1]})|r_\alpha')},\label{eq norm phi}
\end{aligned}
\end{align}
where $\{\lambda_\alpha(e,e), |r_\alpha'),(l_\alpha'|\}$ labels the eigenspectrum of $\mathbb{T}(e,e)$. 
Since the transfer matrix of an SRE state features a unique leading eigenvalue, we simplify the expression for $Q(g^{[y]},f^{[y+1]})$ in the thermodynamic limit to
\begin{align}
\begin{aligned}
\lim_{L_x\rightarrow\infty}Q(g^{[y]},f^{[y+1]})=\lim_{L_x\rightarrow\infty}\biggl(\frac{\lambda_0(g^{[y]},e)}{\lambda_0(e,e)}\biggr)^{L_x-1}\cdot\frac{(l_0|\mathbb{T}(g^{[y]},f^{[y+1]})|r_0)}{(l_0'|\mathbb{T}(e,f^{[y+1]})|r_0')}
\end{aligned}.
\end{align}
Here, $\lambda_0(g^{[y]},e)/\lambda_0(e,e)$ measures the subsystem symmetry of the state, while $(l_0|\mathbb{T}(g^{[y]},f^{[y+1]})|r_0)/(l_0'|\mathbb{T}(e,f^{[y+1]})|r_0')$ evaluates the nontrivial phase factor of the quantum anomaly. 

\section{Tensor network representation of 2D cluster state}\label{app tn rep 2d cluster}
\begin{figure}
    \centering
    \begin{tikzpicture}[>=stealth]
        \draw[densely dotted,thick] 
        (-0.25,0.25) -- (0.25,-0.25)
        (-0.25,2.25) -- (2.25,-0.25)
        (-0.25,4.25) -- (4.25,-0.25)
        (1.75,4.25) -- (6.25,-0.25)
        (3.75,4.25) -- (6.25,1.75)
        (5.75,4.25) -- (6.25,3.75)
        (0.25,4.25) -- (-0.25,3.75)
        (2.25,4.25) -- (-0.25,1.75)
        (4.25,4.25) -- (-0.25,-0.25)
        (6.25,4.25) -- (1.75,-0.25)
        (6.25,2.25) -- (3.75,-0.25)
        (6.25,0.25) -- (5.75,-0.25);

        \filldraw[blue] 
        (1,1) circle[radius=0.125]
        (1,3) circle[radius=0.125]
        (3,1) circle[radius=0.125]
        (3,3) circle[radius=0.125]
        (5,1) circle[radius=0.125]
        (5,3) circle[radius=0.125];

        \filldraw[red] (-0.13,-0.13) -- (-0.13,0.13) -- (0.13,0.13) -- (0.13,-0.13) -- cycle
        (-0.13,1.87) -- (-0.13,2.13) -- (0.13,2.13) -- (0.13,1.87) -- cycle
        (-0.13,3.87) -- (-0.13,4.13) -- (0.13,4.13) -- (0.13,3.87) -- cycle
        (1.87,-0.13) -- (1.87,0.13) -- (2.13,0.13) -- (2.13,-0.13) -- cycle
        (3.87,-0.13) -- (3.87,0.13) -- (4.13,0.13) -- (4.13,-0.13) -- cycle
        (1.87,1.87) -- (1.87,2.13) -- (2.13,2.13) -- (2.13,1.87) -- cycle
        (1.87,3.87) -- (1.87,4.13) -- (2.13,4.13) -- (2.13,3.87) -- cycle
        (3.87,1.87) -- (3.87,2.13) -- (4.13,2.13) -- (4.13,1.87) -- cycle
        (3.87,3.87) -- (3.87,4.13) -- (4.13,4.13) -- (4.13,3.87) -- cycle
        (5.87,-0.13) -- (5.87,0.13) -- (6.13,0.13) -- (6.13,-0.13) -- cycle
        (5.87,1.87) -- (5.87,2.13) -- (6.13,2.13) -- (6.13,1.87) -- cycle
        (5.87,3.87) -- (5.87,4.13) -- (6.13,4.13) -- (6.13,3.87) -- cycle
        ;
        
        \draw[thick,rounded corners=5,rotate around={-45:(2.6,3.05)}] 
        (2.6,3.05) rectangle ++(2,0.5);
        \draw[thick,rounded corners=5,rotate around={-45:(2.6,1.05)}] (2.6,1.05) rectangle ++(2,0.5);
        \draw[thick,rounded corners=5,rotate around={-45:(0.6,3.05)}] (0.6,3.05) rectangle ++(2,0.5);
        \draw[thick,rounded corners=5,rotate around={-45:(0.6,1.05)}] (0.6,1.05) rectangle ++(2,0.5);

        \filldraw[blue,opacity=0.2] (0.75,-0.25) rectangle ++(0.5,4.5)
        (-0.25,2.75) rectangle ++(6.5,0.5);
        \filldraw[red,opacity=0.2] (1.75,-0.25) rectangle ++(0.5,4.5)
        (-0.25,1.75) rectangle ++(6.5,0.5);
        \draw
        (1.15,0.15) node[rotate=-45]  {$(x,y)$}
        (1.15,2.15) node[rotate=-45] {$(x,y+1)$}
        (3.15,0.15) node[rotate=-45]  {$(x+1,y)$}
        (3.15,2.15) node[rotate=-45] {$(x+1,y+1)$};

    \draw[densely dotted,thick] (7.75,0.25) -- (8.25,-0.25) (7.75,2.25) -- (10.25,-0.25) (7.75,4.25) -- (12.25,-0.25) (9.75,4.25) -- (14.25,-0.25) (11.75,4.25) -- (14.25,1.75) (13.75,4.25) -- (14.25,3.75)
    (8.25,4.25) -- (7.75,3.75) (10.25,4.25) -- (7.75,1.75) (12.25,4.25) -- (7.75,-0.25) (14.25,4.25) -- (9.75,-0.25) (14.25,2.25) -- (11.75,-0.25) (14.25,0.25) -- (13.75,-0.25);
    \draw[very thick,red,->] (8,4.3) -- (8,3.7);
    \draw[very thick,red,->] (8,2.3) -- (8,1.7);
    \draw[very thick,red,,->] (8,-0.3) -- (8,0.3);
    \draw[very thick,red,->] (10,4.3) -- (10,3.7);
    \draw[very thick,red,->] (10,2.3) -- (10,1.7);
    \draw[very thick,red,->] (10,-0.3) -- (10,0.3);
    \draw[very thick,red,->] (12,4.3) -- (12,3.7);
    \draw[very thick,red,->] (12,1.7) -- (12,2.3);
    \draw[very thick,red,->] (12,-0.3) -- (12,0.3);
    \draw[very thick,red,->] (14,4.3) -- (14,3.7);
    \draw[very thick,red,->] (14,1.7) -- (14,2.3);
    \draw[very thick,red,->] (14,-0.3) -- (14,0.3);
    \draw[red,fill=red,opacity=0.2] (7.75,1) -- (11,1) -- (11,3) -- (14.25,3) -- (14.25,-0.25) -- (7.75,-0.25) -- (7.75,1);
    \draw[very thick,blue,->] (11.3,1) -- (10.7,1);
    \draw[very thick,blue,->] (11.3,3) -- (10.7,3);
    \draw[very thick,blue,->] (8.7,1) -- (9.3,1);
    \draw[very thick,blue,->] (8.7,3) -- (9.3,3);
    \draw[very thick,blue,->] (12.7,1) -- (13.3,1);
    \draw[very thick,blue,->] (12.7,3) -- (13.3,3);
    \draw (3,-0.5) node {(a)}
    (11,-0.5) node {(b)};
    \draw[blue] (9,1.25) node {$\tau$}
    (9,3.25) node {$\tau$}
    (11,1.25) node {$\tau$}
    (11,3.25) node {$\tau$};
    \draw[red]
    (8.3,0) node {$\sigma$} 
    (8.3,2) node {$\sigma$} 
    (10.3,0) node {$\sigma$} 
    (10.3,2) node {$\sigma$};
    \end{tikzpicture}
    \caption{
    (a) Lattice model of 2D cluster state. 
    The blue dots and red squares denote the spin-$\frac{1}{2}$ degrees of freedom $i_\tau$ and $i_\sigma$ respectively. 
    Each unit cell (hollow rectangle) contains a faithful representation of the group $Z_2^\tau\times Z_2^\sigma$. 
    The subsystem symmetry and site index are defined on the rotated square grid. 
    (b) Ground state wave function with domain wall decoration pattern.
    }
    \label{2D cluster state}
\end{figure}
The Hamiltonian of the 2D cluster state is given by
\begin{align}
\begin{aligned}
    H_{\mathrm{cluster}} = -\sum_{x,y}\left(\tau_{x,y}^x\sigma^z_{x,y}\sigma^z_{x-1,y}\sigma^z_{x,y+1}\sigma^z_{x-1,y+1}+\sigma^x_{x,y}\tau^z_{x,y}\tau^z_{x+1,y}\tau^z_{x,y-1}\tau^z_{x+1,y-1} \right)
\end{aligned}.
\end{align}
The site index $(x,y)$ is defined on the rotated square lattice in Fig. \ref{2D cluster state}(a).
The ground-state wave function of $H_{\mathrm{cluster}}$ is understood as follows. 
The first term of the Hamiltonian ensures that the ground state comprises the decorated domain wall structure depicted in Fig. \ref{2D cluster state}(b). 
Here, the red arrows $\{\uparrow,\downarrow\}$ denote the configuration of the $Z^\sigma_2$ domain. 
The black dotted line represents the rotated square lattice, and the domain wall is placed at the boundary between the red and white regions. 
The blue arrows $\{\leftarrow,\rightarrow\}$ label the domain wall decoration $\{|-\rangle,|+\rangle\}$ of $\tau^x$. 
The decoration rule is given by the constraint $\tau_{x,y}^x=\sigma^z_{x,y}\sigma^z_{x-1,y}\sigma^z_{x,y+1}\sigma^z_{x-1,y+1}$. 
The second term of the Hamiltonian fluctuates between distinct domain configurations. 
Accordingly, the ground state of $H_{\mathrm{cluster}}$ is the superposition of all different $\sigma^z$ domain configurations equipped with $\tau^x$ decorations.

We construct the PEPS representation $(T_\tau,T_\sigma)$ of the 2D cluster state from these decoration rules. 
The $\sigma$ domain is fixed by the local tensor in the $\sigma^z$ basis
\begin{equation}
\begin{tikzpicture}
    \draw[ultra thick,red] (0.5,0.5) -- (0.5,0);
    \draw (-0.5,0) node {$T_\sigma=$}
    (2,-0.025) node {$=\delta_{i_\sigma j_1 \cdots j_4}$};
    \draw[thick] (0,0) -- (1,0) (0.25,-0.25) -- (0.75,0.25);

    \filldraw[red] (0.4,-0.1) -- (0.4,0.1) -- (0.6,0.1) -- (0.6,-0.1) -- cycle;
\end{tikzpicture}.
\end{equation}
Here, we introduce $i_\sigma$ and $j$ to label the physical and virtual indices in $T_{\sigma}$. 
On the other hand, $T_\tau$ is viewed as a map $T_{\tau}: \mathcal{H}_{\sigma^z}^{\otimes4} \rightarrow \mathcal{H}_{\tau^x}$, which is uniquely determined by the decoration rules
\begin{align}
\begin{aligned}
    T_{\tau}(\sigma^z_1,\sigma^z_2,\sigma^z_3,\sigma^z_4) = 
    \begin{cases}
        |+\rangle,\quad \sigma^z_1\sigma^z_2\sigma^z_3\sigma^z_4 = 1\\
        |-\rangle,\quad \sigma^z_1\sigma^z_2\sigma^z_3\sigma^z_4 = -1
    \end{cases}.
\end{aligned}\label{eq decoration rule}
\end{align}
Therefore, the nonzero elements of $T_\tau$ are given by
\begin{equation*}
\begin{tikzpicture}
    \draw (-0.75,0) node {$T_\tau=$}
    (1.5,0) node {$+$}
    (3.5,0) node {$+$}
    (5.5,0) node {$+$}
    (-0.5,-1.5) node {$+$}
    (1.5,-1.5) node {$+$}
    (3.5,-1.5) node {$+$}
    (5.5,-1.5) node {$+$}
    (-0.5,-3) node {$+$}
    (1.5,-3) node {$+$}
    (3.5,-3) node {$+$}
    (5.5,-3) node {$+$}
    (-0.5,-4.5) node {$+$}
    (1.5,-4.5) node {$+$}
    (3.5,-4.5) node {$+$}
    (5.5,-4.5) node {$+$}; 
    
    \draw[thick] (0,0) -- (1,0) (0.5,0.5) -- (0.5,-0.5)
    (2,0) -- (3,0) 
    (2.5,0) -- (2.5,-0.5)
    (4,0) -- (4.5,0)
    (4.5,0.5) -- (4.5,-0.5)
    (6,0) -- (7,0)
    (6.5,0.5) -- (6.5,0)
    (0.5,-1.5) -- (1,-1.5)
    (0.5,-1) -- (0.5,-2)
    (2,-1.5) -- (2.5,-1.5) 
    (2.5,-1.5) -- (2.5,-2)
    (4,-1.5) -- (5,-1.5)
    (6.5,-1.5) -- (7,-1.5)
    (6.5,-2) -- (6.5,-1.5)
    (0,-3) -- (0.5,-3) -- (0.5,-2.5)
    (2.5,-3.5) -- (2.5,-2.5)
    (4.5,-2.5) -- (4.5,-3) -- (5,-3)
    (6,-3) -- (6.5,-3)
    (0.5,-4) -- (0.5,-4.5)
    (2.5,-4.5) -- (3,-4.5)
    (4.5,-4.5) -- (4.5,-5);

    \draw[very thick,dotted] (2.5,0.5) -- (2.5,0) (4.5,0) -- (5,0) (6.5,-0.5) -- (6.5,0) (0,-1.5) -- (0.5,-1.5) (3,-1.5) -- (2.5,-1.5) -- (2.5,-1)   (4.5,-1) -- (4.5,-2) (6,-1.5) -- (6.5,-1.5) -- (6.5,-1) (1,-3) -- (0.5,-3) -- (0.5,-3.5) (2,-3) -- (3,-3) (4,-3) -- (4.5,-3) -- (4.5,-3.5) (6.5,-2.5) -- (6.5,-3.5) (6.5,-3) -- (7,-3) (0,-4.5) -- (1,-4.5) (0.5,-4.5) -- (0.5,-5) (2.5,-4) -- (2.5,-5) (2,-4.5) -- (2.5,-4.5) (4,-4.5) -- (5,-4.5) (4.5,-4) -- (4.5,-4.5) (6,-4.5) -- (7,-4.5) (6.5,-4) -- (6.5,-5);
    \draw[very thick,blue,fill=white] (0.5,0) circle[radius=0.1]
    (2.5,-1.5) circle[radius=0.1]
    (4.5,-1.5) circle[radius=0.1]
    (6.5,-1.5) circle[radius=0.1]
    (0.5,-3) circle[radius=0.1]
    (2.5,-3) circle[radius=0.1]
    (4.5,-3) circle[radius=0.1]
    (6.5,-4.5) circle[radius=0.1];
    \filldraw[blue] (2.5,0) circle[radius=0.1]
    (4.5,0) circle[radius=0.1]
    (6.5,0) circle[radius=0.1]
    (0.5,-1.5) circle[radius=0.1]
    (6.5,-3) circle[radius=0.1]
    (0.5,-4.5) circle[radius=0.1]
    (2.5,-4.5) circle[radius=0.1]
    (4.5,-4.5) circle[radius=0.1];
\end{tikzpicture},
\end{equation*}
where the solid and dotted legs represent the virtual degrees of freedom $|\uparrow\rangle$ and $|\downarrow\rangle$. 
The solid and hollow blue balls denote the physical states as $|-\rangle$ and $|+\rangle$ in the $\tau^x$ basis. 
This local tensor $T_\tau$ is transformed into the $\tau^z$ basis as
\begin{equation}
    T_\tau^\uparrow = \frac{1}{\sqrt{2}}\bigl( T_\tau^+ + T_\tau^- \bigr),\quad T_\tau^\downarrow = \frac{1}{\sqrt{2}}\bigl( T_\tau^+ - T_\tau^- \bigr),
\end{equation}
which is represented graphically as
\begin{equation}
\begin{tikzpicture}
    \draw[thick] (0,0) -- (2,0) (1.5,0.5) -- (0.5,-0.5) (3.5,0) -- (4.5,0) (3.75,-0.25) -- (4.25,0.25);
    \draw[blue,ultra thick] (1,0) -- (1,0.5) (4,0) -- (4,0.5);
    \draw[very thick,white,fill=white] 
    (0.5,0) circle[radius=0.15]
    (1.5,0) circle[radius=0.15]
    (1.25,0.25) circle[radius=0.175]
    (0.75,-0.25) circle[radius=0.175];
    \draw (0.5,0) node {\small $H$}
    (1.5,0) node {\small $H$}
    (1.25,0.25) node {\small $H$}
    (0.75,-0.25) node {\small $H$}
    (-0.5,0) node {$T_\tau=$}
    (2.75,0) node {$,$ where }
    (5.55,0) node {$=\delta_{i_\tau j_1 \cdots i_4}$.};
    \filldraw[blue] (1,0) circle[radius=0.1]
    (4,0) circle[radius=0.1];
\end{tikzpicture}\label{eq T_tau}
\end{equation}
By contracting $T_\tau$ with $T_\sigma$, we construct the PEPS representation of the 2D cluster state as illustrated in Fig. \ref{2D cluster PEPS}. 
Here, we denote $T_\tau$ and $T_\sigma$ as
\begin{equation}
\begin{tikzpicture}
    \draw[thick] 
    (3,1) -- (4,0)
    (4,1) -- (3,0)
    (-0.1,1.1) -- (1.1,-0.1)
    (1.1,1.1) -- (-0.1,-0.1);
    \draw[white,fill=white]
    (0.15,0.15) circle[radius=0.2]
    (0.85,0.15) circle[radius=0.2]
    (0.15,0.85) circle[radius=0.2]
    (0.85,0.85) circle[radius=0.2];
    \draw (4.5,0.5) node {$:T_\sigma$}
    (1.5,0.5) node {$:T_\tau$}
    (0.15,0.15) node {\small $H$}
    (0.85,0.15) node {\small $H$}
    (0.15,0.85) node {\small $H$}
    (0.85,0.85) node {\small $H$};

    \filldraw[red] (3.35,0.35) -- (3.35,0.65) -- (3.65,0.65) -- (3.65,0.35) -- cycle;
    \filldraw[blue] (0.5,0.5) circle[radius=0.15];
    \draw[very thick,white,fill=blue] (0.5,0.5) circle[radius=0.075];
    \draw[very thick,white,fill=red] 
    (3.5,0.5) circle[radius=0.075];
\end{tikzpicture}.
\end{equation}
To investigate the symmetry condition of a local tensor, we insert delta functions in the virtual spaces
\begin{equation}
\begin{tikzpicture}
    \draw[thick] (0,2) -- (2,0) (0,1) -- (1,2) (2,1) -- (1,0);
    \draw[thick] 
    (3.9,0.95) -- (4.6,0.25)
    (3.9,0.95) -- (3.9,1.35) 
    (4.6,0.25) -- (5,0.25)
    (3.9,0.95) -- (3.55,0.6)
    (5,0.25) -- (5.35,0.6)
    (5,0.25) -- (5.35,-0.1)
    (3.55,1.7) -- (3.9,1.35)
    (4.25,1.7) -- (3.9,1.35)
    (4.6,0.25) -- (4.25,-0.1);
    \draw[thick,white,fill=white] 
    (1,1) circle[radius=0.2]
    (0,1) circle[radius=0.2]
    (1,2) circle[radius=0.2]
    (0,2) circle[radius=0.2]
    (1,0) circle[radius=0.2]
    (2,1) circle[radius=0.2]
    (2,0) circle[radius=0.2]
    (4.25,0.6) circle[radius=0.2]
    (3.55,0.6) circle[radius=0.2]
    (4.25,1.7) circle[radius=0.2]
    (3.55,1.7) circle[radius=0.2]
    (5.35,0.6) circle[radius=0.2]
    (4.25,-0.1) circle[radius=0.2]
    (5.35,-0.1) circle[radius=0.2];
    \draw 
    (1,1) node {\small $H$}
    (0,1) node {\small $H$}
    (1,2) node {\small $H$}
    (0,2) node {\small $H$}
    (1,0) node {\small $H$}
    (2,1) node {\small $H$}
    (2,0) node {\small $H$}
    (2.75,1) node {$\Rightarrow$}
    (4.25,0.6) node {\small $H$}
    (3.55,0.6) node {\small $H$}
    (4.25,1.7) node {\small $H$}
    (3.55,1.7) node {\small $H$}
    (5.35,0.6) node {\small $H$}
    (4.25,-0.1) node {\small $H$}
    (5.35,-0.1) node {\small $H$};
    \filldraw[blue] (0.5,1.5) circle[radius=0.15]
    (3.9,0.95) circle[radius=0.15];

    \filldraw[red] (1.35,0.35) -- (1.35,0.65) -- (1.65,0.65) -- (1.65,0.35) -- cycle
    (4.45,0.1) -- (4.45,0.4) -- (4.73,0.4) -- (4.75,0.1) -- cycle;
    \draw[very thick,white,fill=blue] 
    (0.5,1.5) circle[radius=0.075]
    (3.9,0.95) circle[radius=0.075];
    \draw[very thick,white,fill=red] 
    (1.5,0.5) circle[radius=0.075]
    (4.6,0.25) circle[radius=0.075];
\end{tikzpicture}.\label{eq deformation}
\end{equation}
Consequently, the PEPS transforms into the structure shown in Fig. \ref{fig 2D cluster PEPS} under the $H$-gate deformation, which is decomposed into the following local tensor $T^{i_\tau i_\sigma}$
\begin{equation}
\begin{tikzpicture}
    \draw[thick] (1.25,-0.75) -- (1.25,0.75) (0.5,0) -- (2,0);
    \draw[thick,fill=white] (0.95,-0.3) -- (1.55,-0.3) -- (1.55,0.3) -- (0.95,0.3) -- cycle;
    \draw[white,line width=0.15cm] (0.75,0.5) -- (1.25,0);
    \draw[ultra thick,purplish] (0.75,0.5) -- (1.25,0);
    \draw[ultra thick,blue] (4.5,0.5) -- (4.5,1);
    \draw[ultra thick,red] (5,0) -- (5,0.5);

    \draw (2.25,0) node {$=$};
    \draw[purplish]
    (0.6,0.8) node {$(i_\tau,i_\sigma)$};
    \draw[blue]
    (4.5,1.25) node {$i_\tau$};
    \draw[red]
    (5,0.75) node {$i_\sigma$};
    \draw[thick] (2.5,0) -- (3,0) (5,0) -- (5.5,0) 
    (3.25,-0.75) -- (3.5,-0.5) (4.5,0.5) -- (4.75,0.75) (3,0) -- (4.5,0.5) -- (5,0) -- (3.5,-0.5) -- (3,0);
    \filldraw[very thick,white,fill=white] (3.75,0.25) circle[radius=0.15]
    (4.75,0.25) circle[radius=0.15]
    (4.25,-0.25) circle[radius=0.15]
    (3.25,-0.25) circle[radius=0.15];
    \draw (3.75,0.25) node {\small $H$}
    (4.75,0.25) node {\small $H$}
    (4.25,-0.25) node {\small $H$}
    (3.25,-0.25) node {\small $H$}
    (5.75,0) node {$=$};
    \draw[thick]
    (6,0) -- (6.4,0) -- (7.1,0.7) -- (7.8,0) -- (7.1,-0.7) -- (6.4,0)
    (7.1,0.7) -- (7.1,1.1) (7.1,-0.7) -- (7.1,-1.1) (7.8,0) -- (8.2,0);
    \draw[very thick,white,fill=white] (6.65,0.35) circle[radius=0.2]
    (7.45,0.35) circle[radius=0.2]
    (6.65,-0.35) circle[radius=0.2]
    (7.45,-0.35) circle[radius=0.2];
    \draw (6.65,0.35) node {\small $H$}
    (6.65,-0.35) node {\small $H$}
    (7.45,0.35) node {\small $H$}
    (7.45,-0.35) node {\small $H$};
    \filldraw[blue] (7.1,0.7) circle[radius=0.15]
    (4.5,0.5) circle[radius=0.1];
    \filldraw[red] (7.65,-0.15) -- (7.65,0.15) -- (7.95,0.15) -- (7.93,-0.15) -- cycle
    (4.9,-0.1) -- (4.9,0.1) -- (5.1,0.1) -- (5.1,-0.1) -- cycle;
    \draw[very thick,white,fill=blue]
    (7.1,0.7) circle[radius=0.075];
    \draw[very thick,white,fill=red] 
    (7.8,0) circle[radius=0.075];
\end{tikzpicture}.\label{eq local cluster state}
\end{equation}

\begin{figure}
    \centering
    \begin{tikzpicture}
        \draw[thick] 
        (-0.25,0.25) -- (0.25,-0.25)
        (-0.25,2.25) -- (2.25,-0.25)
        (-0.25,4.25) -- (4.25,-0.25)
        (1.75,4.25) -- (6.25,-0.25)
        (3.75,4.25) -- (6.25,1.75)
        (5.75,4.25) -- (6.25,3.75)
        (0.25,4.25) -- (-0.25,3.75)
        (2.25,4.25) -- (-0.25,1.75)
        (4.25,4.25) -- (-0.25,-0.25)
        (6.25,4.25) -- (1.75,-0.25)
        (6.25,2.25) -- (3.75,-0.25)
        (6.25,0.25) -- (5.75,-0.25);

        \filldraw[blue] 
        (1,1) circle[radius=0.15]
        (1,3) circle[radius=0.15]
        (3,1) circle[radius=0.15]
        (3,3) circle[radius=0.15]
        (5,1) circle[radius=0.15]
        (5,3) circle[radius=0.15];

        \filldraw[red] (-0.15,-0.15) -- (-0.15,0.15) -- (0.15,0.15) -- (0.15,-0.15) -- cycle
        (-0.15,1.85) -- (-0.15,2.15) -- (0.15,2.15) -- (0.15,1.85) -- cycle
        (-0.15,3.85) -- (-0.15,4.15) -- (0.15,4.15) -- (0.15,3.85) -- cycle
        (1.85,-0.15) -- (1.85,0.15) -- (2.15,0.15) -- (2.15,-0.15) -- cycle
        (3.85,-0.15) -- (3.85,0.15) -- (4.15,0.15) -- (4.15,-0.15) -- cycle
        (1.85,1.85) -- (1.85,2.15) -- (2.15,2.15) -- (2.15,1.85) -- cycle
        (1.85,3.85) -- (1.85,4.15) -- (2.15,4.15) -- (2.15,3.85) -- cycle
        (3.85,1.85) -- (3.85,2.15) -- (4.15,2.15) -- (4.15,1.85) -- cycle
        (3.85,3.85) -- (3.85,4.15) -- (4.15,4.15) -- (4.15,3.85) -- cycle
        (5.85,-0.15) -- (5.85,0.15) -- (6.15,0.15) -- (6.15,-0.15) -- cycle
        (5.85,1.85) -- (5.85,2.15) -- (6.15,2.15) -- (6.15,1.85) -- cycle
        (5.85,3.85) -- (5.85,4.15) -- (6.15,4.15) -- (6.15,3.85) -- cycle
        ;
        
        \draw[very thick,white,fill=blue]
        (1,1) circle[radius=0.075]
        (1,3) circle[radius=0.075]
        (3,1) circle[radius=0.075]
        (3,3) circle[radius=0.075]
        (5,1) circle[radius=0.075]
        (5,3) circle[radius=0.075];
        \draw[very thick,white,fill=red] 
        (0,0) circle[radius=0.075]
        (0,2) circle[radius=0.075]
        (0,4) circle[radius=0.075]
        (2,0) circle[radius=0.075]
        (2,2) circle[radius=0.075]
        (2,4) circle[radius=0.075]
        (4,0) circle[radius=0.075]
        (4,2) circle[radius=0.075]
        (4,4) circle[radius=0.075]
        (6,0) circle[radius=0.075]
        (6,2) circle[radius=0.075]
        (6,4) circle[radius=0.075];
        
    \draw[white,very thick,fill=white] 
        (0.5,0.5) circle[radius=0.175]
        (0.5,1.5) circle[radius=0.175]
        (0.5,2.5) circle[radius=0.175]
        (0.5,3.5) circle[radius=0.175]
        (1.5,0.5) circle[radius=0.175]
        (1.5,1.5) circle[radius=0.175]
        (1.5,2.5) circle[radius=0.175]
        (1.5,3.5) circle[radius=0.175]
        
        (2.5,0.5) circle[radius=0.175]
        (2.5,1.5) circle[radius=0.175]
        (2.5,2.5) circle[radius=0.175]
        (2.5,3.5) circle[radius=0.175]
        (3.5,0.5) circle[radius=0.175]
        (3.5,1.5) circle[radius=0.175]
        (3.5,2.5) circle[radius=0.175]
        (3.5,3.5) circle[radius=0.175]
        
        (4.5,0.5) circle[radius=0.175]
        (4.5,1.5) circle[radius=0.175]
        (4.5,2.5) circle[radius=0.175]
        (4.5,3.5) circle[radius=0.175]
        (5.5,0.5) circle[radius=0.175]
        (5.5,1.5) circle[radius=0.175]
        (5.5,2.5) circle[radius=0.175]
        (5.5,3.5) circle[radius=0.175];
    \draw 
        (0.5,0.5) node {\small $H$}
        (0.5,1.5) node {\small $H$}
        (0.5,2.5) node {\small $H$}
        (0.5,3.5) node {\small $H$}
        (1.5,0.5) node {\small $H$}
        (1.5,1.5) node {\small $H$}
        (1.5,2.5) node {\small $H$}
        (1.5,3.5) node {\small $H$}
        (2.5,0.5) node {\small $H$}
        (2.5,1.5) node {\small $H$}
        (2.5,2.5) node {\small $H$}
        (2.5,3.5) node {\small $H$}
        (3.5,0.5) node {\small $H$}
        (3.5,1.5) node {\small $H$}
        (3.5,2.5) node {\small $H$}
        (3.5,3.5) node {\small $H$}
        (4.5,0.5) node {\small $H$}
        (4.5,1.5) node {\small $H$}
        (4.5,2.5) node {\small $H$}
        (4.5,3.5) node {\small $H$}
        (5.5,0.5) node {\small $H$}
        (5.5,1.5) node {\small $H$}
        (5.5,2.5) node {\small $H$}
        (5.5,3.5) node {\small $H$};
    
    \end{tikzpicture}
    \caption{
    PEPS representation of 2D cluster state. 
    The blue ``$\odot$" and red ``$\boxdot$" denote the physical bond of $T_\tau$ and $T_\sigma$. 
    The vertex and link of the PEPS denote $\delta_{i j_1\cdots j_4}$ and the $H$-gate.
    }
    \label{2D cluster PEPS}
\end{figure}

For the subsystem symmetry action associated with $g_\tau$ and $g_\sigma$, we have the tensor equations
\begin{equation}
\begin{tikzpicture}
    \draw (-2,0) node {$g_\tau$:};
    \draw[ultra thick,blue] (0.5,0.5) -- (0.5,1);
    \draw[ultra thick,red] (1,0) -- (1,0.5);
    \draw[thick] (-1.5,0) -- (-1,0) (1,0) -- (1.5,0) 
    (-0.75,-0.75) -- (-0.5,-0.5) (0.5,0.5) -- (0.75,0.75) (-1,0) -- (0.5,0.5) -- (1,0) -- (-0.5,-0.5) -- (-1,0);
    \filldraw[very thick,white,fill=white] (-0.25,0.25) circle[radius=0.15]
    (0.75,0.25) circle[radius=0.15]
    (0.25,-0.25) circle[radius=0.15]
    (-0.75,-0.25) circle[radius=0.15];
    \draw (-0.25,0.25) node {\small $H$}
    (0.75,0.25) node {\small $H$}
    (0.25,-0.25) node {\small $H$}
    (-0.75,-0.25) node {\small $H$};
    \draw[blue] (0.5,1.25) node {$\tau^x$};
    \draw[red] (1,0.75) node {$I$};
    \draw (2.25,0) node {$Z$}
    (5.75,0) node {$Z$}
    (4.8,0.9) node {$X$};
    \draw (1.75,-0.05) node {$=$};
    \draw[ultra thick,blue] (4.5,0.5) -- (4.5,1);
    \draw[ultra thick,red] (5,0) -- (5,0.5);
    \draw[thick] (2.5,0) -- (3,0) (5,0) -- (5.5,0) 
    (3.25,-0.75) -- (3.5,-0.5) (4.5,0.5) -- (4.75,0.75) (3,0) -- (4.5,0.5) -- (5,0) -- (3.5,-0.5) -- (3,0);
    \filldraw[very thick,white,fill=white] (3.75,0.25) circle[radius=0.15]
    (4.75,0.25) circle[radius=0.15]
    (4.25,-0.25) circle[radius=0.15]
    (3.25,-0.25) circle[radius=0.15];
    \draw (3.75,0.25) node {\small $H$}
    (4.75,0.25) node {\small $H$}
    (4.25,-0.25) node {\small $H$}
    (3.25,-0.25) node {\small $H$};
    \filldraw[red] (0.9,-0.1) -- (0.9,0.1) -- (1.1,0.1) -- (1.1,-0.1) -- cycle
    (4.9,-0.1) -- (4.9,0.1) -- (5.1,0.1) -- (5.1,-0.1) -- cycle;
    \filldraw[blue] 
    (0.5,0.5) circle[radius=0.1]
    (4.5,0.5) circle[radius=0.1];
\end{tikzpicture},\quad\quad
\begin{tikzpicture}
    \draw (-2,0) node {$g_\sigma$:};
    \draw[ultra thick,blue] (0.5,0.5) -- (0.5,1);
    \draw[ultra thick,red] (1,0) -- (1,0.5);
    \draw[thick] (-1.5,0) -- (-1,0) (1,0) -- (1.5,0) 
    (-0.75,-0.75) -- (-0.5,-0.5) (0.5,0.5) -- (0.75,0.75) (-1,0) -- (0.5,0.5) -- (1,0) -- (-0.5,-0.5) -- (-1,0);
    \filldraw[very thick,white,fill=white] (-0.25,0.25) circle[radius=0.15]
    (0.75,0.25) circle[radius=0.15]
    (0.25,-0.25) circle[radius=0.15]
    (-0.75,-0.25) circle[radius=0.15];
    \draw (-0.25,0.25) node {\small $H$}
    (0.75,0.25) node {\small $H$}
    (0.25,-0.25) node {\small $H$}
    (-0.75,-0.25) node {\small $H$};
    \draw[blue] (0.5,1.25) node {$I$};
    \draw[red] (1.1,0.75) node {$\sigma^x$};
    \draw (2.25,0) node {$X$}
    (5.75,0) node {$X$};
    \draw (1.75,-0.05) node {$=$};
    \draw[thick] (2.5,0) -- (3,0) (5,0) -- (5.5,0) 
    (3.25,-0.75) -- (3.5,-0.5) (4.5,0.5) -- (4.75,0.75) (3,0) -- (4.5,0.5) -- (5,0) -- (3.5,-0.5) -- (3,0);
    \draw[ultra thick,blue] (4.5,0.5) -- (4.5,1);
    \draw[ultra thick,red] (5,0) -- (5,0.5);
    \filldraw[very thick,white,fill=white] (3.75,0.25) circle[radius=0.15]
    (4.75,0.25) circle[radius=0.15]
    (4.25,-0.25) circle[radius=0.15]
    (3.25,-0.25) circle[radius=0.15];
    \draw (3.75,0.25) node {\small $H$}
    (4.75,0.25) node {\small $H$}
    (4.25,-0.25) node {\small $H$}
    (3.25,-0.25) node {\small $H$};

    \filldraw[red] (0.9,-0.1) -- (0.9,0.1) -- (1.1,0.1) -- (1.1,-0.1) -- cycle
    (4.9,-0.1) -- (4.9,0.1) -- (5.1,0.1) -- (5.1,-0.1) -- cycle;
    \filldraw[blue] 
    (0.5,0.5) circle[radius=0.1]
    (4.5,0.5) circle[radius=0.1];
\end{tikzpicture}.\label{eq cluster tensor eq 1}
\end{equation}
Moreover, the $X$ operator on the bottom virtual index changes the tensor elements by
\begin{equation}
\begin{tikzpicture}
    \draw[ultra thick,blue] (0.5,0.5) -- (0.5,1);
    \draw[ultra thick,red] (1,0) -- (1,0.5);
    \draw[thick] (-1.5,0) -- (-1,0) (1,0) -- (1.5,0) 
    (-0.75,-0.75) -- (-0.5,-0.5) (0.5,0.5) -- (0.75,0.75) (-1,0) -- (0.5,0.5) -- (1,0) -- (-0.5,-0.5) -- (-1,0);
    \filldraw[very thick,white,fill=white] (-0.25,0.25) circle[radius=0.15]
    (0.75,0.25) circle[radius=0.15]
    (0.25,-0.25) circle[radius=0.15]
    (-0.75,-0.25) circle[radius=0.15];
    \draw (-0.25,0.25) node {\small $H$}
    (0.75,0.25) node {\small $H$}
    (0.25,-0.25) node {\small $H$}
    (-0.75,-0.25) node {\small $H$};
    \draw (2.25,0) node {$Z$}
    (5.75,0) node {$Z$}
    (-0.95,-0.95) node {$X$};
    \draw (1.75,-0.05) node {$=$};
    \draw[ultra thick,blue] (4.5,0.5) -- (4.5,1);
    \draw[ultra thick,red] (5,0) -- (5,0.5);
    \draw[thick] (2.5,0) -- (3,0) (5,0) -- (5.5,0) 
    (3.25,-0.75) -- (3.5,-0.5) (4.5,0.5) -- (4.75,0.75) (3,0) -- (4.5,0.5) -- (5,0) -- (3.5,-0.5) -- (3,0);
    \filldraw[very thick,white,fill=white] (3.75,0.25) circle[radius=0.15]
    (4.75,0.25) circle[radius=0.15]
    (4.25,-0.25) circle[radius=0.15]
    (3.25,-0.25) circle[radius=0.15];
    \draw (3.75,0.25) node {\small $H$}
    (4.75,0.25) node {\small $H$}
    (4.25,-0.25) node {\small $H$}
    (3.25,-0.25) node {\small $H$};
    \filldraw[red] (0.9,-0.1) -- (0.9,0.1) -- (1.1,0.1) -- (1.1,-0.1) -- cycle
    (4.9,-0.1) -- (4.9,0.1) -- (5.1,0.1) -- (5.1,-0.1) -- cycle;
    \filldraw[blue] 
    (0.5,0.5) circle[radius=0.1]
    (4.5,0.5) circle[radius=0.1];
\end{tikzpicture}.\label{eq cluster tensor eq 2}
\end{equation}
Considering a wave function on a cylindrical surface with circumference $L_y=2$, the above equations are transformed to the following MPS symmetry conditions
\begin{equation}
\begin{tikzpicture}
    \draw[thick] (0,0.8) -- (0,-0.8) (-0.5,0.4) -- (0.5,0.4) (-0.5,-0.4) -- (0.5,-0.4);
    \draw[thick,fill=white] (-0.25,0.15) rectangle ++(0.5,0.5) (-0.25,-0.65) rectangle ++(0.5,0.5);
    \draw[thick] (0,0.8) arc[start angle=180, end angle=0, x radius=0.075, y radius=0.3]
    (0,-0.8) arc[start angle=-180, end angle=0, x radius=0.075, y radius=0.3];
    \draw[line width=0.1cm,white] (0,0.4) -- (-0.3,0.7) (0,-0.4) -- (-0.3,-0.1);
    \draw[very thick,purplish] (0,0.4) -- (-0.3,0.7) (0,-0.4) -- (-0.3,-0.1);
    \draw (1,0) node {$=$};
    \draw[blue] (-0.5,0) node {$\tau^x$};
    \draw[purplish] (-0.5,0.75) node {$I$};
    \draw (1.6,-0.4) node {$Z$} (1.6,0.4) node {$Z$};
    \draw (3.2,-0.4) node {$Z$} (3.2,0.4) node {$Z$};
    
    \draw[thick] (2.4,0.8) -- (2.4,-0.8) (1.9,0.4) -- (2.9,0.4) (1.9,-0.4) -- (2.9,-0.4);
    \draw[thick,fill=white] (2.15,0.15) rectangle ++(0.5,0.5) (2.15,-0.65) rectangle ++(0.5,0.5);
    \draw[thick] (2.4,0.8) arc[start angle=180, end angle=0, x radius=0.075, y radius=0.3]
    (2.4,-0.8) arc[start angle=-180, end angle=0, x radius=0.075, y radius=0.3];
    \draw[line width=0.1cm,white] (2.4,0.4) -- (2.1,0.7) (2.4,-0.4) -- (2.1,-0.1);
    \draw[very thick,purplish] (2.4,0.4) -- (2.1,0.7) (2.4,-0.4) -- (2.1,-0.1);

    \draw[thick] (6,0.8) -- (6,-0.8) (5.5,0.4) -- (6.5,0.4) (5.5,-0.4) -- (6.5,-0.4);
    \draw[thick,fill=white] (5.75,0.15) rectangle ++(0.5,0.5) (5.75,-0.65) rectangle ++(0.5,0.5);
    \draw[thick] (6,0.8) arc[start angle=180, end angle=0, x radius=0.075, y radius=0.3]
    (6,-0.8) arc[start angle=-180, end angle=0, x radius=0.075, y radius=0.3];
    \draw[line width=0.1cm,white] (6,0.4) -- (5.7,0.7) (6,-0.4) -- (5.7,-0.1);
    \draw[ultra thick,purplish] (6,0.4) -- (5.7,0.7) (6,-0.4) -- (5.7,-0.1);
    \draw (7,0) node {$=$};
    \draw[red] (5.5,0) node {$\sigma^x$};
    \draw[purplish] (5.5,0.75) node {$I$};
    \draw (7.6,-0.4) node {$X$};
    \draw (9.2,-0.4) node {$X$};
    
    \draw[thick] (8.4,0.8) -- (8.4,-0.8) (7.9,0.4) -- (8.9,0.4) (7.9,-0.4) -- (8.9,-0.4);
    \draw[thick,fill=white] (8.15,0.15) rectangle ++(0.5,0.5) (8.15,-0.65) rectangle ++(0.5,0.5);
    \draw[thick] (8.4,0.8) arc[start angle=180, end angle=0, x radius=0.075, y radius=0.3]
    (8.4,-0.8) arc[start angle=-180, end angle=0, x radius=0.075, y radius=0.3];
    \draw[line width=0.1cm,white] (8.4,0.4) -- (8.1,0.7) (8.4,-0.4) -- (8.1,-0.1);
    \draw[ultra thick,purplish] (8.4,0.4) -- (8.1,0.7) (8.4,-0.4) -- (8.1,-0.1);
\end{tikzpicture}.
\end{equation}
Hence, the boundary operators of $G_s^{[y]}$ in $\mathcal{H}_{\mathrm{edge}}^{\mathrm{Right}}$ is given by
\begin{equation}
    W^{\mathrm{Right}}(g_\tau^{[y]}) = Z_{L_x,y}Z_{L_x,y+1},\quad W^{\mathrm{Right}}(g_\sigma^{[y]}) = X_{L_x,y}.
\end{equation}
The mixed anomaly $\phi(g_\tau^{[y]},g_\sigma^{[y+1]})=-1$ is encoded within the boundary symmetry condition
\begin{equation}
    W^{\mathrm{Right}}(g_\tau^{[y]})W^{\mathrm{Right}}(g_\sigma^{[y+1]}) = -W^{\mathrm{Right}}(g_\sigma^{[y+1]})W^{\mathrm{Right}}(g_\tau^{[y]})
\end{equation}

\section{Mixed-state anomaly detection from twisted sector density matrix}\label{app mixed-state anomaly detection}
In this appendix, we analyze the relationship between $Q^{\mathrm{ave}},Q^{\mathrm{exa}}$ and $\phi$ through tensor equations.
To explore the 't Hooft anomaly in mixed-state systems, we construct a twisted sector density matrix by inserting a zero-dimensional symmetry defect $V(\Tilde{g})$
\begin{equation}
\begin{tikzpicture}
    \draw (1.5,0) node {$\cdots$}
    (-0.9,0) node {$\rho_{\Tilde{g}}=$}
    (4.25,-0.35) node {$V(\Tilde{g})$};
    \draw[thick] (0,0) -- (1,0) (2,0) -- (4.5,0);
    \draw[ultra thick,purplish] (0.5,-0.6) -- (0.5,0.6) (2.5,-0.6) -- (2.5,0.6) (3.5,-0.6) -- (3.5,0.6);
    \draw[thick,fill=white] (0.25,-0.25) rectangle ++(0.5,0.5) (2.25,-0.25) rectangle ++(0.5,0.5) (3.25,-0.25) rectangle ++(0.5,0.5);
    \draw[thick] (0,0) arc[start angle=270, end angle=90, x radius=0.3, y radius=0.075]  (4.5,0) arc[start angle=-90, end angle=90, x radius=0.3, y radius=0.075];
    \filldraw[black] (4.25,0) circle[radius=0.075];
\end{tikzpicture},
\end{equation}
where $\Tilde{g}=(g,k)\in\Tilde{G}$ denotes a general group element. The 't Hooft anomaly between the exact symmetry operator $S(k)$ and the symmetry defect $V(\Tilde{g})$ are derived from the symmetry charge of the twisted sector density matrix
\begin{equation}
    Q^{\mathrm{exa}}(k,\Tilde{g}) = \frac{\mathrm{Tr}[S(k)\rho_{\Tilde{g}}]}{\mathrm{Tr}[\rho_{\Tilde{g}}]}.
\end{equation}
For the convenience of discussion, we introduce the following MPO transfer matrix
\begin{equation}
\begin{tikzpicture}
    \draw (-1.4,0) node {$\mathbb{T}_{1}(k,\Tilde{g})=$}
    (0.75,0.25) node {$V(\Tilde{g})$};
    \draw[blue] (-0.5,0.5) node {$U(k)$};
    \draw[thick] (-0.5,0) -- (0.75,0);
    \draw[ultra thick,purplish] (0,0.5) -- (0,-0.5) ;
    \draw[thick,fill=white] (-0.25,-0.25) rectangle ++(0.5,0.5);
    \draw[ultra thick,purplish] (0,0.5) arc[start angle=180, end angle=0, x radius=0.075, y radius=0.3] (0,-0.5) arc[start angle=-180, end angle=0, x radius=0.075, y radius=0.3];
    \filldraw[black] (0.5,0) circle[radius=0.075];
    \filldraw[blue] (0,0.5) circle[radius=0.075];
\end{tikzpicture}.
\end{equation}
For MPOs that are invariant under the exact symmetry transformation $S(k)$, we have the local tensor equation
\begin{equation}
\begin{tikzpicture}
    \draw (0.75,0.25) node {$V(\Tilde{g})$} (3.5,0.25) node {$V(\Tilde{g})$};
    \draw (1.25,0) node {$=$};
    \draw[blue] (3.3,-0.3) node {$V(k)$} (1.85,-0.3) node {$V^{-1}(k)$};
    \draw[blue] (-0.5,0.5) node {$U(k)$};
    \draw[thick] (-0.5,0) -- (0.75,0) (1.75,0) -- (3.5,0);
    \draw[ultra thick,purplish] (0,0.5) -- (0,-0.5) (2.625,0.5) -- (2.625,-0.5);
    \draw[thick,fill=white] (-0.25,-0.25) rectangle ++(0.5,0.5) (2.375,-0.25) rectangle ++(0.5,0.5);
    \draw[ultra thick,purplish] (0,0.5) arc[start angle=180, end angle=0, x radius=0.075, y radius=0.3] (0,-0.5) arc[start angle=-180, end angle=0, x radius=0.075, y radius=0.3] (2.625,0.5) arc[start angle=180, end angle=0, x radius=0.075, y radius=0.3](2.625,-0.5) arc[start angle=-180, end angle=0, x radius=0.075, y radius=0.3];
    \filldraw[black] (0.5,0) circle[radius=0.075] (3.325,0) circle[radius=0.075] ;
    \filldraw[blue] (0,0.5) circle[radius=0.075] (2.025,0) circle[radius=0.075] (3.025,0) circle[radius=0.075];
\end{tikzpicture},
\end{equation}
which is expressed as
\begin{equation}
    \mathbb{T}_1(k,\Tilde{g}) = 
    \phi(k,\Tilde{g})V^{-1}(k)\cdot\mathbb{T}_1(e,\Tilde{g})\cdot V(k)
\end{equation}
Using this, we rigorously calculate the exact symmetry charge of the system
\begin{align}
\begin{aligned}
    Q^{\mathrm{exa}}(k,\Tilde{g}) = \frac{\mathrm{Tr}[\mathbb{T}_1(k,e)\cdots\mathbb{T}_1(k,\Tilde{g})]}{\mathrm{Tr}[\mathbb{T}_1(e,e)\cdots\mathbb{T}_1(e,\Tilde{g})]}
    = \phi(k,\Tilde{g})\frac{\mathrm{Tr}[V^{-1}(k)\mathbb{T}_1(e,e)\cdots\mathbb{T}_1(e,\Tilde{g})V(k)]}{\mathrm{Tr}[\mathbb{T}_1(e,e)\cdots\mathbb{T}_1(e,\Tilde{g})]} = \phi(k,\Tilde{g})
\end{aligned}
\end{align}

For average symmetries $\mathcal{G}$, we construct the following quantity to detect the mixed-state anomaly between the average symmetry transformation and the symmetry defect $V(\Tilde{g})$
\begin{equation}
    Q^{\mathrm{ave}}(g,\Tilde{g}) = \frac{\mathrm{Tr}[S(g)\rho_{\Tilde{g}} S^\dagger(g) \rho_{\Tilde{g}}]}{\mathrm{Tr}[\rho_{\Tilde{g}}^2]}.
\end{equation}
We denote the transfer matrix of the double-layer MPO as
\begin{equation}
\begin{tikzpicture}
    \draw (-2,0.5) node {$\mathbb{T}_{2}(g,\Tilde{g})=$}
    (0.75,0.25) node {$V(\Tilde{g})$}
    (0.75,1.25) node {$V(\Tilde{g})$};
    \draw[red] (-0.6,0.5) node {$U^\dagger(g)$} (-0.6,1.5) node {$U(g)$};
    \draw[thick] (-0.5,0) -- (0.75,0) (-0.5,1) -- (0.75,1);
    \draw[ultra thick,purplish] (0,-0.5) -- (0,1.5) ;
    \draw[thick,fill=white] (-0.25,-0.25) rectangle ++(0.5,0.5) (-0.25,0.75) rectangle ++(0.5,0.5);
    \draw[ultra thick,purplish] (0,1.5) arc[start angle=180, end angle=0, x radius=0.075, y radius=0.3] (0,-0.5) arc[start angle=-180, end angle=0, x radius=0.075, y radius=0.3];
    \filldraw[black] (0.5,0) circle[radius=0.075] (0.5,1) circle[radius=0.075];
    \filldraw[red] (0,0.5) circle[radius=0.075] (0,1.5) circle[radius=0.075];
\end{tikzpicture}.
\end{equation}
The average symmetry condition of $\rho$ indicates that
\begin{equation}
\begin{tikzpicture}
    \draw (1.25,0.5) node {$=$}
    (0.75,0.25) node {$V(\Tilde{g})$}
    (0.75,1.25) node {$V(\Tilde{g})$}
    (3.5,0.25) node {$V(\Tilde{g})$}
    (3.5,1.25) node {$V(\Tilde{g})$};
    \draw[red] (-0.6,0.5) node {$U^\dagger(g)$} (-0.6,1.5) node {$U(g)$} (3.25,0.7) node {$V(g)$} (1.85,0.7) node {$V^{-1}(g)$};
    \draw[thick] (-0.5,0) -- (0.75,0) (-0.5,1) -- (0.75,1) (1.75,0) -- (3.5,0) (1.75,1) -- (3.5,1);
    \draw[ultra thick,purplish] (0,-0.5) -- (0,1.5) (2.625,-0.5) -- (2.625,1.5);
    \draw[thick,fill=white] (-0.25,-0.25) rectangle ++(0.5,0.5) (-0.25,0.75) rectangle ++(0.5,0.5)  (2.375,-0.25) rectangle ++(0.5,0.5) (2.375,0.75) rectangle ++(0.5,0.5);
    \draw[ultra thick,purplish] (0,1.5) arc[start angle=180, end angle=0, x radius=0.075, y radius=0.3] (0,-0.5) arc[start angle=-180, end angle=0, x radius=0.075, y radius=0.3] (2.625,1.5) arc[start angle=180, end angle=0, x radius=0.075, y radius=0.3] (2.625,-0.5) arc[start angle=-180, end angle=0, x radius=0.075, y radius=0.3];
    \filldraw[black] (0.5,0) circle[radius=0.075] (0.5,1) circle[radius=0.075] (3.35,0) circle[radius=0.075] (3.35,1) circle[radius=0.075];
    \filldraw[red] (0,0.5) circle[radius=0.075] (0,1.5) circle[radius=0.075] (3.05,1) circle[radius=0.075] (2.05,1) circle[radius=0.075];
\end{tikzpicture},
\end{equation}
which is expressed as 
\begin{equation}
    \mathbb{T}_2(g,\Tilde{g}) = 
    \phi(g,\Tilde{g})\biggl(V^{-1}(g)\otimes I\biggr)\cdot\mathbb{T}_2(e,\Tilde{g})\cdot \biggl(V(k)\otimes I\biggr)\label{eq ave transfer eq}
\end{equation}
According to Eq. (\ref{eq ave transfer eq}), we calculate the $Q^{\mathrm{ave}}$ of the density matrix with average symmetry 
\begin{align}
\begin{aligned}
    Q^{\mathrm{ave}}(g,\Tilde{g}) = \frac{\mathrm{Tr}[\mathbb{T}_2(g,e)\cdots\mathbb{T}_2(g,\Tilde{g})]}{\mathrm{Tr}[\mathbb{T}_2(e,e)\cdots\mathbb{T}_2(e,\Tilde{g})]}
    = \phi(g,\Tilde{g})\frac{\mathrm{Tr}[\biggl(V^{-1}(g)\otimes I\biggr)\mathbb{T}_2(e,e)\cdots\mathbb{T}_2(e,\Tilde{g})\biggl(V(k)\otimes I\biggr)]}{\mathrm{Tr}[\mathbb{T}_2(e,e)\cdots\mathbb{T}_2(e,\Tilde{g})]} = \phi(g,\Tilde{g})
\end{aligned}
\end{align}
Therefore $Q^{\mathrm{exa}}$ ($Q^{\mathrm{ave}}$) serves as a \textit{mixed-state anomaly indicator} between the exact (average) symmetry operator $k$ ($g$) and the zero-dimensional symmetry defect $V(\tilde{g})$ in open quantum systems.

\end{document}